\documentclass{emulateapj}
\usepackage{apjfonts}

\slugcomment{To appear in The Astronomical Journal}

\shorttitle{Catalog of Northern Stars With $\mu>0.15\arcsec$ yr$^{-1}$}
\shortauthors{L\'epine \& Shara}

\begin{document}

\title{A Catalog of Northern Stars With Annual Proper Motions Larger
Than 0.15 Seconds of Arc (LSPM catalog - North)\altaffilmark{1,2}}

\author{S\'ebastien L\'epine\altaffilmark{3}, and Michael
M. Shara}

\affil{Department of Astrophysics, Division of Physical Sciences,
American Museum of Natural History, Central Park West at 79th Street,
New York, NY 10024, USA}

\altaffiltext{1}{Based on data mining of the Digitized Sky Surveys,
developed and operated by the Catalogs and Surveys Branch of the Space
Telescope Science Institute, Baltimore, USA.}

\altaffiltext{2}{Developed with support from the National Science
Foundation, as part of the NASA/NSF NStars program.}

\altaffiltext{3}{American Museum of Natural History Kalbfleich
Research Fellow.}

\begin{abstract}
The LSPM-north catalog is a comprehensive list of 61,977 stars north of the
J2000 celestial equator that have proper motions larger than 0.15
seconds of arc per year (local-background-stars frame). The catalog
has been generated primarily as a result of our systematic search for
high proper motion stars in the Digitized Sky Surveys using our
SUPERBLINK software. At brighter magnitudes, the catalog incorporates
stars and data from the Tycho-2 Catalogue and also, to a lesser
extent, from the All-sky Compiled Catalogue of 2.5 million stars. The
LSPM catalog expands considerably over the old Luyten (LHS, NLTT)
catalogs, superseding them for northern declinations. Positions are
given with an accuracy $\lesssim$100 mas at the 2000.0 epoch, and
absolute proper motions are given with an accuracy of $\approx$8 mas
yr$^{-1}$. Corrections to the local-background-stars proper motions
have been calculated, and absolute proper motions in the extragalactic
frame are given. Whenever available, we also give optical $B_T$ and
$V_T$ magnitudes (from TYCHO-2, ASCC-2.5), photographic B$_{\rm J}$,
R$_{\rm F}$, I$_{\rm N}$ magnitudes (from USNO-B1 catalog) and
infrared J, H, K$_s$ magnitudes (from 2MASS). We also provide an
estimated $V$ magnitude and $V-J$ color for nearly all catalog
entries, useful for initial classification of the stars. The catalog
is estimated to be over 99\% complete at high Galactic latitudes
($|b>15|$), and over 90\% complete at low Galactic latitudes
($|b>15|$), down to a magnitude $V=19.0$, and has a limiting magnitude
$V=21.0$. All the northern stars listed in the LHS and NLTT catalogs
have been re-identified, and their positions, proper motions and
magnitudes re-evaluated. The catalog also lists a large number of
completely new objects, which promise to expand very significantly the
census of red dwarfs, subdwarfs, and white dwarfs in the vicinity of
the Sun.
\end{abstract}

\keywords{astrometry --- surveys --- stars: kinematics --- solar
neighborhood --- stars: white dwarfs --- stars: Population II}


\section{Introduction}

The search for and identification of stars with large proper motions
has traditionally been the dominant method for finding nearby
stars. Stellar distances are ultimately determined with parallax 
measurements, but these generally require substantial effort, and
until the HIPPARCOS mission (1991), large systematic parallax surveys
were impractical. One therefore had to rely on secondary diagnostics
of proximity that could be more easily measured, such as large
proper motions. It is a historical fact that the vast majority of the
stars within 25 parsecs of the Sun (the local volume of space known as
the ''Solar Neighborhood'') have first been identified as high proper
motion stars. These include some of our closest neighbors, such as
Proxima Centauri \citep{I15}, and Barnard's Star \citep{B16}, which
still is the star with the largest known proper motion
($\mu\simeq10.3\arcsec$ yr$^{-1}$).

Massive, large scale searches for high proper motion stars have been
performed, and continually improved, throughout the 20th century,
thanks to innovations in wide field astrophotography. Early
compilations of known stars with large proper motions \citep{VM15},
were expanded by catalogs such as Max Wolf's {\it Katalog von 1053
starker bewegten Fixternen} \citep{W19}, which he complemented over
the years with several new additions, naming a total of 1567 stars
after himself. Frank E. Ross also contributed numerous discoveries,
publishing lists of new high proper motion stars over a period of 14
years (1925-1939), discovering a total of 1080 nearby objects
\citep{R39}. These early investigations were made with the use of a
visual blink comparator; photographic plates obtained at different
epochs were blinked in succession, and examined by eye. The typical
motions of the stars detected (which were at that time simply called
``proper motion stars'') were $\approx0.2-1.0\arcsec$ yr$^{-1}$.

Over the years, the discovery of increasingly fainter stars having
large proper motions pointed to the existence of a significant
population of low luminosity stars. Some of the most extreme examples
were discovered as faint, common proper motion companions of brighter
objects \citep{VB61}. The very first free-floating brown dwarf was
also discovered in a survey of high proper motion stars
\citep{RLA97}. Besides being an extremely useful tool in the
identification of nearby populations of low luminosity objects, proper
motion surveys are also very sensitive to high-velocity stars. Because
high-velocity (e.g. thick disk, halo) stars are detected out to larger
distances than nearby disk stars in proper motion selected samples,
they are significantly over-represented in catalogs of high proper
motion stars. Far from being a problem, this makes proper motion
catalogs highly useful tools for the study of Galactic stellar
populations (including low-luminosity halo stars). This has motivated
intensive searches for faint high proper motion stars over the whole
sky.

The largest deep surveys of high proper motion stars were carried out
by the Lowell Observatory, and by Willem J. Luyten at the University
of Minnesota. Results from the Lowell Proper Motion Survey were
compiled and published in two large catalogs \citep{GBT71,GBT78}
listing a total of 11,749 stars with proper motion larger than
$0.2\arcsec$ yr$^{-1}$. Luyten's work, on the other hand, was
constantly updated over the decades, culminating in the publication of
two major catalogs: the {\it New Luyten Catalogue of stars with proper
motions larger than two tenths of an arcsecond} \citep{L79b} known in
short as the NLTT, listing 58,845 objects, and {\it A catalogue of
stars with proper motions exceeding 0".5 annually} \citep{L79a} known
as the LHS, and which is essentially a subset of the NLTT listing 4470
of the fastest-moving stars.

Most of Luyten's success stems from his use of the National Geographic
Palomar Sky Survey (POSS-I), completed in the 1950s. Luyten obtained second
epoch images in the late 50s and 60s using the same instrument and
setup (at the Palomar Schmidt telescope). Luyten also developed a
laser-scanning microdensitometer machine to process the northern sky
images, considerably improving over previous eye-blinking methods. The
depth of the Palomar plates allowed him to probe deeper than anyone
before, mapping the turnover in the local luminosity function at the
bottom of the main sequence \citep{L68}; he was however limited by the
depth of his second-epoch plates, which were about 1 magnitude
shallower (19th magnitude limit) that the POSS-I plates (20th
magnitude limit). Until today, the NLTT catalog remains the largest
and most complete list of high proper motion objects, at least for
declinations north of $-32.5^{\circ}$ (the southern limit of POSS-I).

Because the LHS and NLTT catalogs contain large numbers of
astrophysically significant objects, they have been used as a source
of targets in many follow-up programs, with the faster LHS stars
naturally taking precedence. Photometric studies have included the
search for nearby dwarfs \citep{W84,W96}, multiband studies of halo
stars \citep{R89} and of low-luminosity dwarfs and subdwarfs
\citep{B91}. Spectroscopic follow-up surveys have resulted in the
identification of new nearby stars \citep{GR97}, cool halo subdwarfs
\citep{RA93}, and white dwarfs \citep{H86,BRL92,VK03}. Despite numerous
studies and observing programs, hundreds of LHS stars and the majority
of NLTT stars are still lacking formal spectral classification. The
NLTT catalog in particular remains a goldmine of astrophysically
interesting but uncharacterized targets. Perhaps the most intensive
follow-up study to date is a recent program devoted to the
identification of nearby stars missing from the census of objects
within 25 parsecs of the Sun \citep{RC02,RKC02,CR02,Retal03}.

Modern astrometric techniques require much more accurate positions
and proper motions than initially recorded by Luyten. One example is
the identification of future possible microlensing events \citep{SG00}.
Motivated by these requirements, proper motions and positions of the
LHS stars have been recently recalculated by \citet{BSN02}, who
systematically searched for all the stars in the Digitized Sky
Surveys. A revision of the NLTT positions and proper motions was also
initiated by \citet{GS03} and \citet{SG03}, using the USNO-A2 catalog
\citep{Metal98} as a first epoch, and the 2MASS {\it Second
incremental Release} as a second epoch. These revisions revealed that
a significant fraction of LHS/NLTT entries contained large positional
errors, up to several minutes of arc in some cases. Indeed, a
comparison of NLTT and LHS positions, for stars appearing in both
catalogs, has revealed the existence of typographical errors in both
catalogs \citep{LSR02}. The uncertainties in Luyten's positions explain
in part the difficulty in carrying out follow-up observations of his
objects, and raises the possibility that background stars have been
mistaken for NLTT stars, resulting in erroneous spectral
classifications. This motivates a complete re-evaluation of the
positions of all NLTT objects.

The problem of the completeness of the NLTT has been much debated (see
Pokorny, Jones, \& Hambly 2003, and references therein). The reason is
that statistical studies of the Luyten stars (LHS stars in particular)
have been used in estimates of the local density of low-mass and
degenerate stars. In particular, the NLTT has been used to estimate
the local density and luminosity function of white dwarfs
\citep{LDM88,LRB98} and the density and luminosity function of the
local halo population \citep{D86,L91,G03a}. The accuracy of those
statistical studies, however, is entirely dependent on the
completeness of the underlying sample, or at least on their {\em
estimated} completeness. Hence the importance of being able to
estimate the completeness (as a function of position, magnitude,
proper motion, ...) of the NLTT catalog.

As initially noted by \citet{D86}, the NLTT and LHS catalogs are
notably incomplete in two distinct areas: (1) south of $-32.5^{\circ}$
in declination, and (2) in a band within $\pm10^{\circ}$ of the
Galactic plane. However, outside of these specific areas, in an area
referred to as the Completed Palomar Region (CPR), the catalogs do
appear to be significantly complete ($\gtrsim80\%$) at least down to
19th magnitude. This is significant because the magnitude distribution
of NLTT stars reflects the turnover at the faint end of the luminosity
function: the number density of NLTT stars peaks at $R\approx15$, well
before the limiting magnitude of the POSS-I plates on which they are
based. Because they probe the luminosity function turnover, and
because they contain significant samples of both disk and halo stars,
the LHS and NLTT catalogs are major tools in the determination of the
local luminosity function of both the Galactic disk and Galactic
halo. In any case, the accuracy of the luminosity function and of the
stellar density is dependent on a proper evaluation of the
completeness of these proper motion selected samples \citep{G03a}.

The completeness of a proper motion catalog can be determined either
{\em internally}, or {\em externally}. One internal test was devised
by \citet{Fetal2001}, and uses the fact that both the magnitude and
proper motion are a function of distance. For example, if one takes
subsamples with a lower proper motion limits $\mu_1$ and
$\mu_2=1.259\mu_1$, sample 1 should contain objects that are on
average more distant by a distance modulus of 0.5. The completeness at
a magnitude $V_1$ in sample 1 can thus be determined by comparing the
relative number of stars in bins $V_1$ and $V_2=V_1-0.5$ in sample
2. One can then work iteratively, starting from a bin $V_2$ assumed to
be complete. Applying the technique to the NLTT, \citet{Fetal2001}
estimated that the completeness in the CPR (for stars with
$\mu>0.2\arcsec$ yr$^{-1}$) falls linearly from 100\% at $V=13$ to
60\% at $V=18$, and then breaks down to 0\% at $V=20$. The method was,
however, put in doubt by \citet{MFLCHR00} who noted that at high
galactic latitudes the density of stars decreases with distance, and
sample 2 is thus not equivalent to the more distant sample 1; the
method of \citet{Fetal2001} would thus tend to underestimate the
completeness. \citet{G03a} discusses this problem (see his appendix),
and concludes that there is little evidence to suggest the
\citet{Fetal2001} result to be in error. The main problem is a
lack of reliable external completeness estimates.

External completeness tests are based on a direct comparison of the
proper motion catalog (the NLTT) to a deeper or more sensitive proper
motion survey conducted over a selected area. The main caveat of that
method is that the results then become dependent on the completeness
of the new survey itself, which has to be estimated by other means. Of
course, if the new survey is conducted over a sufficiently large
area, it can simply supersede the former catalog.

From a small sample of only 100 square degrees from the APM Proper
Motion Project, \citet{E92} concluded that the NLTT has an
incompleteness of $\approx16\%$, most of it probably due to random
measurement errors at the lower proper motion limit of
$\mu=0.18\arcsec$ yr$^{-1}$. This completeness level is larger than
suggested by the internal test of \citet{Fetal2001}, but the small
area and limited number of objects puts this external test in doubt. A
much larger survey of 1,378 square degrees in the northern sky
\citep{MFLCHR00} was performed using pairs of plates from the Second
Palomar Sky Survey \citep{R91}. Because of the short timespan between
pairs of plates, this survey was limited to the detection of stars
with very large proper motions ($0.4\arcsec$ yr$^{-1}$). Nevertheless,
fifteen stars were found that had been missed by Luyten, of which 6
have proper motions $\mu>0.5\arcsec$ yr$^{-1}$, suggesting a completeness
$\gtrsim90\%$ for the LHS catalog. (N.B. while the Monet {\it et al.}
paper cites 17 ''new'' objects, two of them were subsequently found by
\citet{G03a} to be the NLTT catalog stars 58785 and 52890.)
Unfortunately, their survey did not cover the $0.18\arcsec$
yr$^{-1}<\mu<0.4\arcsec$ yr$^{-1}$ range, where most of the NLTT stars
are found, leaving open the question of the NLTT completeness.

Many more attempts have been made at conducting more sensitive proper
motion surveys in the south. Because the NLTT is well known to be much
less complete south of $-32.5^{\circ}$, the potential pay-off is much
higher. In the Cal\'an-ESO Proper Motion Survey \--- famous for its
discovery of Kelu-1, the first free floating brown dwarf \citep{RLA97}
\--- 14 pairs of ESO Schmidt plates were used, covering an area of 350
square degrees \citep{RWRG}. Fourteen new stars with $\mu>0.5''$
yr$^{-1}$ were found, suggesting a completeness of only $\approx60\%$
for the LHS in the south. But since several of their areas are south
of -32.5$^\circ$, this overestimates the incompleteness of the NLTT
in the CPR. The larger survey conducted by
\citet{WT89,WT91,WT94,WT96,WT97} and continued by
\citet{WC99,WC01}, now covers a total of 3,275 square degrees in 131
scattered areas. This survey is performed by direct visual inspection
of photographic plates, using a Zeiss-Jena plate comparator. The
survey now has 2,495 cataloged objects, all of which are new
(i.e., not in the NLTT). Within the limits of the CPR, they typically
discover $\approx$50 new stars with $\mu>0.2\arcsec$ yr$^{-1}$ in
every 100 square degree area. When compared to the NLTT density of
roughly $90$ stars per 100 square degrees in the same areas, this
yields an overall estimated completeness of $\approx65\%$ for the
NLTT, but a more detailed analysis would be warranted.

Other large southern surveys are built from lists of objects generated
from machine scans of photographic Schmidt plates. A survey of 2,000
square degrees near the south polar galactic cap was made with lists of
objects from UK Schmidt plates, scanned with the APM machine
\citep{SIIJM00}. This survey is ongoing, and may eventually cover much
larger areas of the southern sky. The Liverpool-Edinburgh high proper
motion survey \citep{PJH03} is based on lists of objects from ESO
Schmidt and UK Schmidt plates, scanned with the SuperCOSMOS machine,
and has a lower proper motion limit of 0.18$\arcsec$ yr$^{-1}$; the
survey is currently limited to a moderately large area of the southern
sky, covering $\approx3,000$ square degrees around the south Galactic
cap. More recently, \citet{Hetal04} have used SuperCOSMOS data too, to
search $\approx3,000$ square degrees south of -57.5$^{\circ}$ for
stars with proper motion $\mu>0.4$$\arcsec$ yr$^{-1}$ (although they only
list those they found having $\mu>1.0$$\arcsec$ yr$^{-1}$). What is clear
from these southern sky surveys is that the LHS/NLTT catalogs are
definitely incomplete south of $-32.5^{\circ}$. The number of
new high proper motion stars discovered is a sizable fraction of all
the high proper motion stars identified. The consensus is that in the
southern sky, the NLTT is no more than $\approx70\%$ complete down to
R=19, and possibly much less complete in some areas.

More massive astrometric catalogs based on lists of objects from
scanned Schmidt plates are in preparation, including the second Guide
Star Catalog (GSC-II). By combining lists of objects from multiple
epochs, these will attempt to provide proper motions for all stars
detected. The already available USNO-B1.0 catalog \citep{Metal03} is a
first attempt at such a deep, all-sky, astrometric catalog with proper
motions. For the northern sky, the plate material includes the Oschin
Schmidt plates from the first and second epoch Palomar Sky Surveys
(POSS-I, POSS-II), providing a large temporal baseline and deep (R=20)
limiting magnitude; in the southern sky scans of the ESO schmidt and
UK schmidt were also used. Unfortunately, the huge number of objects
involved (1 billion sources), and the large number of false detections
(plate defects, mismatches from multiple epoch detections) makes the
identification of high proper motion stars very difficult. An analysis
of the USNO-B1.0 catalog by \citet{G03b} shows a huge contamination
rate (200 to 1) in bogus high proper motion objects. Recent efforts
\citep{GK04,Metal04} have however shown that it is possible to
eliminate most of the false detections by cross-correlating the
USNO-B1.0 catalog with other large area surveys like the Sloan Digital
Sky Survey (SDSS). Unfortunately, the completeness of the USNO-B1.0
for high proper motion stars (based on the recovery of NLTT stars)
does not exceed $90\%$ over the whole sky, and drops to $70\%$ at low
galactic latitudes.

Very high completeness levels have however been achieved by us, in a
novel approach to surveys of high proper motion stars
\citep{LSR02,LSR03}. Instead of working with lists of objects
identified from plate scans, we work directly with the pixel data,
using an image subtraction algorithm. With the help of a specialized
software (SUPERBLINK), we have performed a very successful proper
motion survey based on a massive re-analysis of all image scans of the
Oschin Schmidt plates (POSS-I and POSS-II) made at STScI for the
Digitized Sky Surveys (DSS). The image subtraction method is
significantly more efficient in densely populated fields, where
examination with the blink comparator is difficult, and identification
of point stellar sources by scanning machines considerably less
efficient because of crowding. To massively apply plate subtraction
methods to the DSS data, we have developed the SUPERBLINK software,
which works like an automated blink comparator. Our initial survey was
a search for stars with very large proper motions $\mu>0.5\arcsec$
yr$^{-1}$ over the whole 20,000 square degrees of northern sky. We
showed our method to be extremely successful, by recovering
essentially all LHS stars, and with the discovery of 198 new stars
with $\mu>0.5\arcsec$ yr$^{-1}$. This yielded the definitive
completeness measure of the LHS catalog in the northern sky
($1662/1866=89.1\%$) and confirmed the results of \citet{MFLCHR00}
about the northern sky completeness of the LHS. However, it left open
the question of the completeness of the NLTT. Our new proper motions
catalog directly addresses this problem since it is an extension of
our survey to smaller proper motions ($\mu>0.15\arcsec$ yr$^{-1}$),
with a completeness level exceeding 99\% down to R$_{\rm F}=19$. The
executive summary is: {\em in the northern hemisphere, we find and
report 40,843 stars with $\mu>0.18\arcsec$ yr$^{-1}$, of which
28,486 are NLTT objects; the LSPM catalog also lists an additional
21,133 stars with $0.15\arcsec$ yr$^{-1}<\mu<0.18\arcsec$ yr$^{-1}$,
only 2,875 of which are in the NLTT catalog.} Our analysis
demonstrates that {\em in the CPR, the completeness of the NLTT is
85\% at $V=18$, and breaks down at $V=19$, while at low galactic
latitudes ($|b|<10$) its completeness falls from 90\% at $V=15$ to
only 30\% at $V=17$.} This suggests that the internal test of
\citet{Fetal2001} underestimates the completeness of the NLTT in the
CPR.

Note that other major proper motion surveys, which have been extremely
successful in determining highly accurate proper motions of selected
stars, are not very helpful in increasing the completeness of our
proper motion catalogs. These include the astrometric survey conducted
with the HIPPARCOS satellite, whose data are now compiled in the
TYCHO-2 catalog \citep{H00}. Only stars brighter than V=9 were
observed systematically, and an input catalog was used for stars down
to the limiting magnitude (V=13). Bright NLTT stars were included in
the survey, but very few new high proper motion stars were discovered
above Luyten's cutoff. Also limited as tools for finding new high
proper motions stars are the Lick Northern Proper Motion Program
\citep{KJH87} and its southern extension, the Yale/San Juan Proper
Motion Survey \citep{VL91}. Both programs aim at very precise
astrometric measurements of {\em selected} stars, and largely rely on
an input catalog, although a subset of stars were picked at
random. Another highly accurate astrometric survey is the US Naval
Observatory CCD Astrograph program \citep{Z00}, from which an all-sky
astrometric catalog (UCAC) is being assembled \citep{Z03}. While
extremely promising as an expansion to the TYCHO-2 catalog, the UCAC
will have a limiting magnitude of R=16, and it is unclear how
sensitive the survey will be for stars with very large proper
motions. At this time, it appears that our image subtraction method
holds the best promise for generating an all-sky replacement to the
LHS/NLTT catalogs.

As part of the NASA/NSF NStars initiative, we have been expanding our
DSS-based survey, aiming at the systematic detection and verification
of all stars in the northern sky with proper motions larger than
0.05$\arcsec$ yr$^{-1}$. Our goals are to achieve optimal detection
rates, with completeness exceeding 99\% down to $R=19.0$ over most of
the sky, with minimal contamination from false detections. This paper
presents our first major data release: a catalog of all known stars in
the northern sky with Proper Motions larger than 0.15 $\arcsec$
yr$^{-1}$. This catalog which we refer to as the LSPM catalog,
includes improved astrometry and photometry for more than 31,000 high
proper motion stars previously listed in the LHS and NLTT
catalogs. The LSPM catalog also incorporates bright high proper motion
stars from the TYCHO-2 catalog. Finally, the LSPM catalog contains over
28,000 newly discovered high proper motion stars.

The LSPM catalog represents a major improvement over the NLTT
catalog. Not only is it much more complete, but the positions and
proper motions are also much more accurate. In effect, the LSPM
supersedes the NLTT for the sky north of the celestial equator, and
should now be used in its place for all applications. This paper
provides information that is essential in understanding how the LSPM
catalog was built, and what are its strengths and limitations. A
description of the SUPERBLINK code, used to find LSPM candidates in
the DSS, follows in \S2. The detailed procedure for the inclusion of a
star in the LSPM catalog is detailed in \S3. The sources used for the
photometry are presented in \S4, while the catalog astrometric
accuracy is discussed in \S5. The format of the catalog is explained
in \S6. The completeness of the LSPM is discussed in \S7. A
preliminary analysis of the stellar contents of the catalog is given
in \S8. Plans for future expansion and improvement of the catalog are
summarized in the conclusion (\S9).


\section{A new survey for high proper motion stars}

\subsection{The SUPERBLINK software}

SUPERBLINK is an automated blink comparator developed by SL, and first
described in \citet{LSR02}. Given two different images of the same
patch of sky on input (up to 2k$\times$2k pixels in size), SUPERBLINK
automatically identifies any object that has moved between the two
epochs, such as a star with a large proper motion. On output, the
software generates a list of possible moving objects in the field,
with their positions, proper motions, image magnitude, and a
probability index that estimates the likelihood of the object being
real. The software also generates identification charts
($151\times151$ pixels in size) centered on each object. These charts
are dual-epoch, and can be blinked on the computer screen for easy
examination of the moving object. The two core elements of SUPERBLINK
are an image superposition and subtraction algorithm (SUPER), and a
shift-and-match search algorithm (BLINK).

The current version of the code has been optimized for use with
Digitized Sky Survey images (first epoch DSS and second epoch XDSS) to
look for stars with large proper motions. Pairs of images are provided
to the code on input, each pair consisting of one
$17\arcmin\times17\arcmin$ field extracted from the DSS at a specified
position, and a second $17\arcmin\times17\arcmin$ extracted from the
XDSS and centered on the same position. In the northern sky, the DSS
image invariably consists of a POSS-I scan, with a typical resolution
of $1.7\arcsec$ per pixel, while the XDSS image is a POSS-II scan with
a resolution of $1.01\arcsec$ per pixel. The POSS-II image is
generally of higher quality than the POSS-I image: the background
noise is lower, the image reaches about one magnitude deeper, and the
astronomical resolution (seeing) is better. 

The SUPER procedure performs image transformations to make the two
images in the pair look as similar as possible. The procedure
uses the first image (DSS) as a template, and attempts to modify/degrade
the second image (XDSS) in such a way that it can be subtracted from
the first with the smallest residuals possible. The SUPER procedure
follows a series of steps described below.

{\em Rescaling.} The two images, on input, can have different
resolutions; the SUPER procedure resets the two images to the same
angular scale. Typically, the higher resolution second epoch image
(XDSS) is remapped onto a grid that matches the resolution of the lower
resolution first epoch (DSS).

{\em Rectification.} Images are rectified so that their background
levels (sky) are uniform and set to a value of 1. The code uses a
procedure that marks each pixel as either ``sky'' or ``object'' (based
on the statistics of intensity values). A two-dimensional linear fit
is then performed on the ``sky'' pixels. The image is then divided by
this fit, setting background levels to unity. While the background
level is never strictly linear on a photographic plate (edge effects
are important), it is a good approximation on the scale of the images
provided on input ($17\arcmin\times17\arcmin$), which are much smaller
than the typical size of the POSS plates
($384\arcmin\times384\arcmin$).

{\em Normalization.} The total flux from all ``object'' pixels is
determined for each image. The second image is then normalized so
that the total flux (above background) from ``object'' pixels is equal
to that in the first image. Note that this normalization might be
inaccurate if there are bright objects showing up in only one of the
two images. This does happen, particularly if the two images are not
exactly aligned initially. It may then happen that e.g. a bright
star near the edge of one image does not show up in the other, and
vice versa. The renormalization procedure (see below) will generally
correct for any normalization error.

{\em Shift and rotate.} The second image is shifted vertically and
horizontally ($\Delta$X, $\Delta$Y), and rotated ($\Delta\theta$),
before being subtracted from the first image. The procedure is repeated
recursively, first using small incremental values of $\Delta$X,
$\Delta$Y, and $\Delta\theta$ until a good match is found, i.e. until
the residuals are significantly smaller than the total flux in each
image. More precise values for the shift and rotation are then
determined using a multidimensional downhill simplex minimization
routine, which identifies a strong minimum in the residuals in
[$\Delta$X, $\Delta$Y, $\Delta\theta$] space. Note that this
procedure accurately superposes the two images using the assumption
that most stars in the field are ``fixed''. Any systematic motion in
the background ``fixed'' stars will be eliminated. This means that all
proper motions calculated by SUPERBLINK will be proper motions {\em
relative to the background of ``fixed'' stars, and not absolute proper
motions.}

{\em Renormalization.} The second image is normalized again, as
described above, but this time using only ``object'' pixels that are
common to both images in the pair. These can now be easily determined
since we know from the preceding shift and rotate procedure which part
of the field is common to both images.

{\em Convolution.} The second image is then degraded so that its PSF
matches that of the first one. A convolution profile of variable width
is applied to the second image, which is then subtracted from the
first image. The width of the profile is increased until a minimum
value in the residuals is found. The shape of the convolution profile
has been determined by trial and error. Several different shapes have
been considered; a simple profile generated by a sum of two gaussians
of different widths was found to yield the best results. This same
general profile was applied to all our fields. After this final
procedure, the first and second images generally look extremely
similar. In the best of cases, it is very difficult to tell the two
images apart just by looking at them. The only obvious differences are
variations in the noise patterns, or the presence of a variable star, an
asteroid track, or a high proper motion star.

The BLINK procedure starts with one pair of images that have first
been superposed with SUPER, and proceeds to identify any object that
has moved between the first and second epoch. Stars with very large
proper motions essentially appear as pairs of objects, one at each
epoch, that do not cancel out after image subtraction; these are
fairly easy to find. On the other hand, stars whose total motion
between the two epochs is less than their apparent sizes on the POSS
plates show a more complex pattern in the residuals, having been
partially canceled out. On scans of POSS-I plates, typical sizes
(full width at half maximum) of stars range from $\approx 3$ pixels
(5.1$\arcsec$) for unsaturated stars (R$_{J}>15$), to $\approx 15$ pixels
(25.5$\arcsec$) for the brightest, saturated objects detectable by SUPERBLINK
(R$_{J}\approx10$). The minimum motion of stars in our catalog is
about 6$\arcsec$ between the POSS-I and POSS-II plates (0.15$\arcsec$ yr$^{-1}$ in
40 yr). This means that while the fainter stars are always
well-separated after plate subtraction, the images of many of the
brighter proper motion stars will overlap, and will partially cancel
out after subtraction. The following procedures in BLINK allow for a
correct treatment of all moving objects, whether or not they partially
cancel out on plate subtraction.

{\em Subtraction and cataloging of residuals.} The two images
processed by SUPER are subtracted from each other. Any object that has
moved significantly between the two epochs induces a large, local
maximum/minimum on the residual image. All the minima/maxima are
mapped, cataloged, and matched to their source on the first or second
image.

{\em Removal of the candidate moving objects from the first epoch
image.} Each object on the first epoch image that is associated with a
large residual is removed from that image, with all pixels set to the
sky level. A new residual image is then calculated by subtracting the
second image from the first one (which now lacks the profile of the
candidate moving star). Because the moving object has now been removed
from the first image, its second epoch profile now shows up in its
entirety in the residuals. This allows for an easy identification of
the slow-moving objects, which would otherwise be canceling
themselves out partially on the residual image.

{\em Search, and match the moving star on the second epoch.} 
The profile of the object that was removed from the first image is
then recursively shifted in X and Y, and added to the residual
image. If the star is indeed a moving object, there will be a
near exact replica of it on the residual image. Because the second epoch
has been subtracted from the first, the replica will be a negative
source of similar flux. Hence, the object will cancel out its second
epoch replica when it is shifted by the $\Delta$X and $\Delta$Y which
corresponds to its motion on the plate between the first and second
epoch images.

{\em Calculate likelihood.} For each candidate moving object selected
on the first epoch image, the code identifies the best possible
match for that star on the second epoch image, within a radius of
$1.5\arcmin$. The software distinguishes between actual moving
objects, and accidental matches of unrelated features based on the
quality of the match. A probability index is calculated for each
candidate moving object which is function of: (1) the difference in
the object magnitude between the first and second epoch, and (2) the
difference in the magnitude density of the object between the first
and second epoch. A detection thus has a high likelihood if the object
has the same magnitude on both images, provided that the type of
object (compact/extended) is also the same. Criterion (2) essentially
prevents stars from being matched with galaxies, and vice-versa.

{\em Repeat, for candidate moving objects from the second epoch image.}
The procedure is repeated, but this time only the second
epoch counterparts are considered. The object is subtracted from the
second epoch image, residuals are recalculated, and the object is
shifted until it cancels out its first epoch profile, now showing in
its entirety in the residuals. Note that this means that most high
proper motion stars are identified {\em twice} by the code, once from
their first epoch location, and once from their second epoch
location. This redundancy is necessary for the identification of stars
whose profile is superposed on the profiles of other stars at either
epoch (especially in fields with significant crowding). In effect, this
increases the chance of detection for blended stars; the star will be
detected even if it blended with another source in either of the two
epochs. This is especially useful for faint stars moving in the
vicinity of brighter objects or in crowded fields.

Once the image has been completely analyzed and searched, the code
generates a list of all candidate moving objects along with
their positions, relative proper motions (in pixels per year),
integrated plate magnitudes, and likelihood index. The code uses the
plate solutions and epochs (found in the image headers) to determine
the local scale and orientation of the plates, and calculate the
magnitude (in seconds of arc per year) and direction of the proper
motion vector on the sky. The software also extracts $151\times151$
pixels$^{2}$ finder charts from the superposed images. These charts
are extremely useful, as they are used to subsequently verify each and
every detection by eye, on the computer screen.

\subsection{Application to the Digitized Sky Surveys}

The first epoch of the Digitized Sky Surveys (DSS) in the northern sky
consist of scans of photographic plates from the National Geographic
Palomar Observatory Sky Survey (POSS-I), obtained circa 1950. The
scans were performed with the GAMMA machine by the Catalogs and
Surveys Branch at the Space Telescope Science Institute. Only the red
plates (xx103aE emulsion + plexi) have been scanned. The second epoch
of the Digitized Sky Surveys (XDSS) consists, for the northern sky, of
scans of the Second Palomar Observatory Sky Survey \citep{R91}. Images
from the POSS-II include scans of plates from all three photographic
bands of the survey: the blue (IIIaJ emulsion with GG385 filter), red
(IIIaF emulsion with RG610 filter), and near infrared (IVN emulsion
with RG9 filter). The data from both the DSS and XDSS are publicly
available from a variety of on-line databases \footnote{Including
http://archive.stsci.edu/.}.

We divided the northern sky into 615,800 areas distributed
on a grid with a separation of 12$\arcmin$ in DEC and a mean
separation of 10$\arcmin$ in RA. At every grid point, we extracted
from the DSS (our first epoch) and XDSS (our second epoch \--- red
band only) pairs of images each $17\arcmin\times17\arcmin$ on a
side. We deliberately extracted images that are much larger than the
grid point separation, thus allowing for a significant overlap between
neighboring image pairs. 

We allowed for a large overlap between neighboring subfields in
part because of the required rotation of one of the images in the
superposition process. Square subfields extracted from the DSS are
generally not oriented with the Y axis pointing toward the celestial
north pole; rather they follow the local XY coordinates {\em of the
scanned POSS plates}. As a result, pairs of images extracted from the
DSS and XDSS are generally not aligned, and the XDSS image is rotated
(by up to 30 degrees at high latitudes) by the SUPER procedure. Areas
near the corners of the square subfields are thus cut out. We
therefore allow for a band $1\arcmin$ wide running along the edge of
each subfield so that no gaps in sky coverage occur.

A large overlap is also required for completeness because a high
proper motion star, to be detected by SUPERBLINK, must be present in a
given subfield at each of the two epochs. A star that has moved
from one subfield to another would not be detected as a moving object
but rather as two distinct ``variable'' stars. A star with a proper
motion $\mu\leq2.0\arcsec$ yr$^{-1}$ can move up to $1.5\arcmin$
between the two epochs of the POSS-I and POSS-II. This is why we also
allocated an additional band $1.5\arcmin$ wide running along the edge
of each field, to help in the detection of stars with very large proper
motions.

In summary, the different angular scales, scanning resolution,
non-alignment of subfields, different pixellation grids and offsets
between scans, and different image quality and limiting magnitude are
all accounted for and corrected by SUPERBLINK.

All of our DSS scans were extracted from {\em The Digitized
Sky Surveys} series of CD-ROMs, published by the Space Telescope
Science Institute (STScI). All the XDSS scans were downloaded off the
Internet directly from the STScI archive (where they are stored on a
CD-ROM jukebox), with kind permission of the STScI {\em Catalogs and
Surveys Branch}. All subfields were processed as they were
downloaded. Complete uploading/downloading and analysis of all 615,800
subfields was performed over a period of 11 months, from May 2001
through March 2002. Computations were performed on a dual Pentium-III
processor machine running Linux. Scripts were used to automate the
procedure, and the downloading and processing of the whole northern
sky with SUPERBLINK was completed with minimal user interaction. Most
of our human effort went into the quality control phase, described in
the next section.

\subsection{Visual confirmation of candidates}

False detections are inevitable when one is looking for high proper
motion stars on photographic plates. The POSS plates are filled with
plate defects of different sorts, such as grains and bubbles in the
emulsion, dust specks, and scratches. The plates also contains
transient images left by solar system bodies (asteroid tracks), and
the occasional meteor trail, narrow artificial satellite track
(POSS-II only), or wide airplane track. A combination of plate defects
and/or space junk may conspire to create the illusion, on a DSS/XDSS
pair, of an object moving at a rate within our detection limits
($0.15\arcsec$ yr$^{-1}<\mu<2.00\arcsec$ yr$^{-1}$). Of course,
SUPERBLINK automatically eliminates most such bogus detections with the
requirement that candidate high proper motion stars must have
comparable fluxes and flux densities on both plates. However, it is
not uncommon to see plate defects of the same magnitude within seconds
of arc, or minutes of arc of each other on two different epochs,
mimicking the behavior of a high proper motion star. This is
especially true for faint features near the detection limit of the
plate, which tend to be very numerous.

Another major source of false detections is the long diffraction
spikes associated with the brighter stars. Because fields from the DSS
and XDSS often do not have the same orientation on the sky, the
position angles of the diffraction spikes change from the first to the
second epoch. Once the two images are superposed and subtracted out in
SUPERBLINK, diffraction spikes systematically show up as intense
residual features. When the superposed images are blinked on the
computer screen, the spikes display a remarkable rotating motion
between the two epochs. This motion is of course recorded by
SUPERBLINK which systematically lists moving spikes as possible proper
motion objects. One solution that was considered at first was to
reject any detection of a moving object within a certain distance of a
bright star. However, after the detection of several faint high proper
motion stars in the vicinity of bright diffraction spikes, we decided
to investigate them all, to maximize the detection rate of genuine
high proper motion stars.

The most direct and reliable way to eliminate false detections
is by visually inspecting each and every candidate high proper motion
star, using a blink comparator. A trained eye can easily distinguish
real stars from plate defects, for objects down to a magnitude of
$r\approx19$. The second epoch of the Digitized Sky Surveys (the XDSS)
also contains images in the B$_{\rm J}$ and I$_{\rm N}$ band, which
can be used as a third epoch for confirmation of ambiguous objects.

Blinking each and every object identified by SUPERBLINK is a daunting
task. However, the task is actually made easy (if only time consuming)
thanks to the convenience of the finding charts generated by
SUPERBLINK. The SUPERBLINK charts are more than just pairs of
DSS/XDSS scans. While the first epoch of the chart is essentially the
DSS image centered on the candidate moving object, the second epoch of
the chart (as explained in \S2.1 above) is an XDSS image that has been
processed and modified by SUPERBLINK to match the appearance and
quality of the DSS image. Using simple software (designed by SL), it
is possible to blink sequentially large numbers of finder charts,
accepting and rejecting stars with a single keystroke, and
automatically updating the list of confirmed high proper motion
stars. With a little training, it is possible to quickly sift through
hundreds of candidates, at a rate of about 1 star per second. False
detections typically outnumber real objects by a factor of 3 to
4. The visual confirmation of $\approx$60,000 high proper motion stars
carried out for this catalog thus represents a total of about 75 hours
of intensive human inspection.

An interesting benefit of the 60,000 individual visual inspections was
the identification of close proper motion pairs. The SUPERBLINK
software does not discriminate between point sources and extended
objects, and in the course of the survey, several extended objects
were found to be moving. Many of these turned out to be double stars
with small separations ($\approx1-5\arcsec$). Close pairs, on the POSS
plates, produce images that are sometimes elliptical, if the two stars
are of equal magnitude, and sometimes ``pear-shaped'', if the stars
have different magnitudes. While these were all identified as single
moving objects by SUPERBLINK, they were flagged as probable multiple
systems during visual inspection. All those that could be confirmed
were then listed as distinct objects in the catalog (see \S3.2 below).

\section{Building the LSPM Catalog}

\subsection{Stars identified with SUPERBLINK in the Digitized Sky Surveys}

Using SUPERBLINK, we have successfully analyzed DSS/XDSS fields
covering 99.23\% of the northern sky (20,460 square degrees). Areas
that were not analyzed include a small patch of sky north of 87 degree
in declination ($\approx 30$ square degrees), which we avoided because
of problems associated with the very large rotation angles required in
the superposition of the first and second epoch image. SUPERBLINK also
failed to analyze some 4,766 scattered subfields, covering a total
area of 295 square degrees; these were rejected after SUPERBLINK was
unable to superpose the two images because of the presence of a very
bright, saturated object in the field. Rejected fields include all
those containing a star brighter than 5th magnitude, fields containing
cores of bright globular clusters, parts of M31, and of a few other
extended and saturated objects.

\begin{figure}
\epsscale{1.5}
\plotone{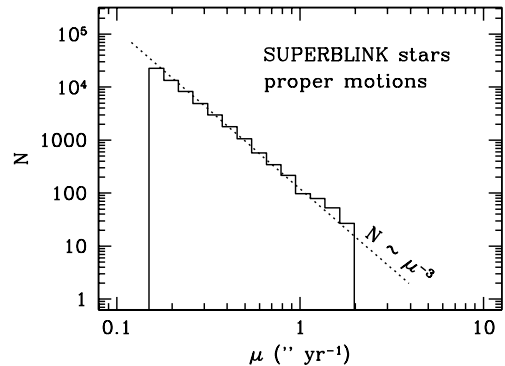}
\caption{\label{fig1}
Distribution as a function of the proper motion for stars found with
SUPERBLINK. The number density very closely follows a $N\sim\mu^{-3}$
law (dashed line, {\it this is not a fit}).}
\end{figure}

\placefigure{fig1}

Stars identified by SUPERBLINK with proper motions in the
$0.15\arcsec$ yr$^{-1}<\mu<2.0\arcsec$ yr$^{-1}$ range and confirmed
by visual inspection, are found to disproportionately consist of
slower-moving objects. As the proper motion of a star is inversely
proportional to the distance, a uniform density distribution of stars
in the volume around the Sun is expected to result in a cumulative
distribution inversely proportional to the cube of the proper
motion. The objects identified with SUPERBLINK very closely follow
such a $N\sim\mu^{-3}$ relationship, as illustrated in Figure 1. The
sharp drop in objects above $\mu=2.0\arcsec$ yr$^{-1}$ and below
$\mu=0.15\arcsec$ yr$^{-1}$ simply results from the detection limits
imposed on SUPERBLINK. While the upper limit was firmly set into the
software, the lower limit has been set only for the purpose of the
present catalog. We allowed SUPERBLINK to identify stars with proper
motions as small as $\mu=0.04\arcsec$ yr$^{-1}$ (totaling nearly one
million objects), but only those with $\mu<0.15\arcsec$ yr$^{-1}$,
considered the most valuable, were examined visually and retained for
the present analysis. The much more numerous slower-moving objects are
only now being examined, and their publication is planned for a future
release.

\begin{figure*}
\plotone{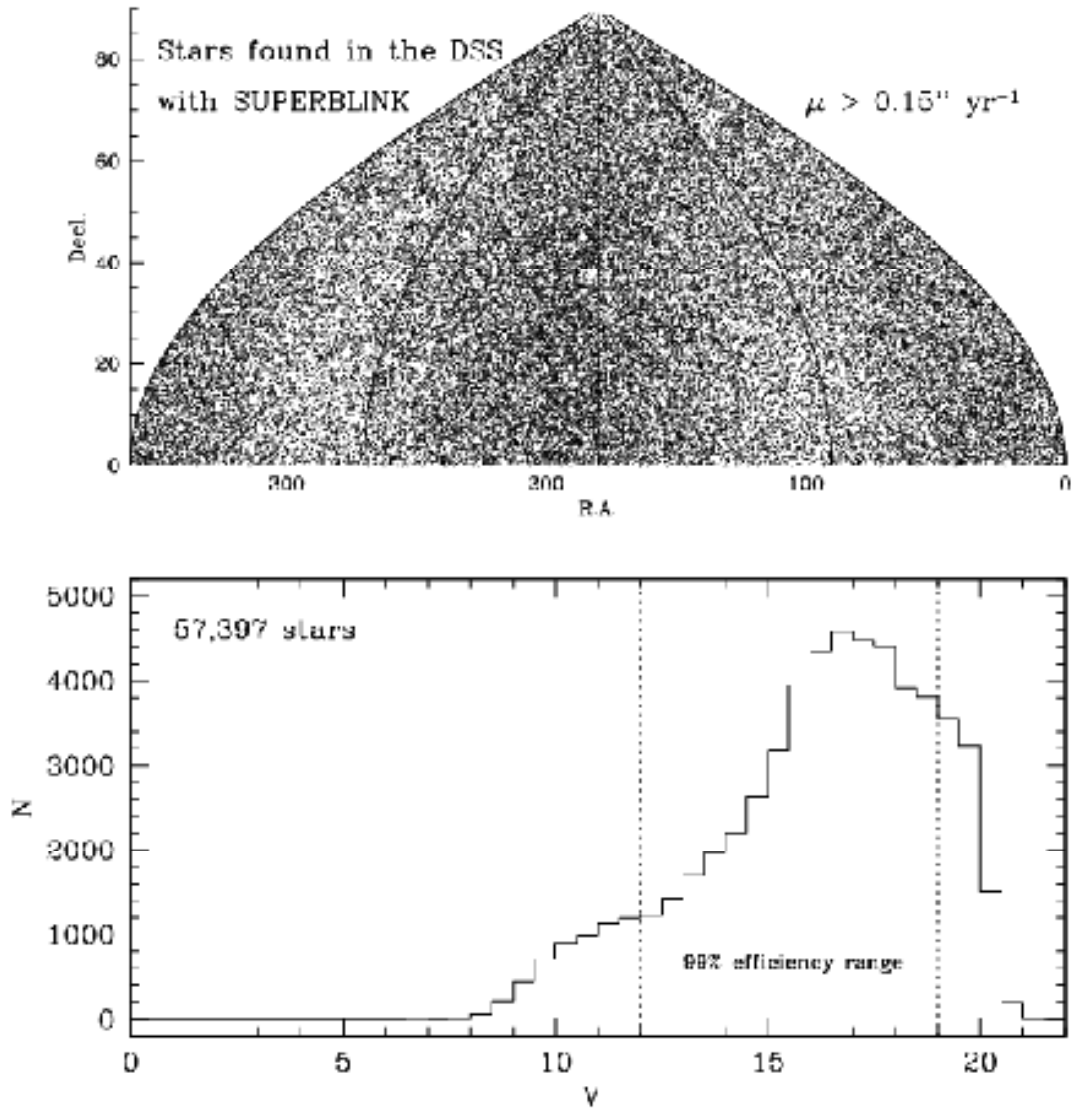}
\caption{\label{fig2}
Top: distribution on the north celestial hemisphere of the 57,763 high
proper motion stars identified in the Digitized Sky Surveys with the
SUPERBLINK software. Bottom: distribution as a function of
optical $V$ magnitude. The detection efficiency of SUPERBLINK
exceeds 99\% in the magnitude range $12.0<V<19.0$. The
efficiency drops for brighter ($V<12$) stars as the stellar images
become saturated on the POSS plates, and at fainter magnitudes
($V>19$) as one reaches the POSS plate limit. The turnover in the
distribution beyond $V=16$ is real, and is not a result of a declining
detection rate. It occurs because proper motion selected samples
survey a limited volume, combined with the fact that field stars also
have a turnover in their luminosity function.}
\end{figure*}

\placefigure{fig2}

In the area analyzed by SUPERBLINK, the software identified a total of
56,238 objects with proper motion exceeding 0.15$\arcsec$
yr$^{-1}$. Among these, a total of 1,159 objects were subsequently
found to be double stars (see \S3.2 below), and are listed as two
distinct objects in the LSPM catalog. This makes a total of 57,397
individual high proper motion objects identified with SUPERBLINK. The
distribution on the sky is displayed in Figure 2, along with their
distribution as a function of optical $V$ magnitude (see \S4.4 for a
discussion on how $V$ magnitudes are derived for SUPERBLINK
detections).

At magnitudes fainter than $V=19$, we observe a sharp drop in the
number of high proper motion stars detected with SUPERBLINK. This
reflects the limited capabilities of SUPERBLINK to detect stars near
the magnitude limit of the POSS-I plates. While it is true that the
POSS-II plates are marginally more sensitive, SUPERBLINK demands a
detection at both epochs in order to identify the object as moving,
and thus the detection threshold of SUPERBLINK is determined by the
sensitivity of the POSS-I plates.

At the bright end of the distribution ($V<12.0$), we observe a
steady decline in the number of stars detected, falling to zero for
$V<8$. While one naturally expects to find fewer high proper
motion stars with very bright magnitudes, many more high proper motion
stars are known with ($V<12.0$) than have been detected by
SUPERBLINK. The lack of stars detected at bright magnitudes reflects
the inability of SUPERBLINK to deal with stars that have a strong
saturated core on the photographic POSS-I and POSS-II plates. Tests
made with fields containing known, bright high proper motion stars
showed that SUPERBLINK generally did well with stars fainter than
$V=10.0$, correctly identifying them. However, tests revealed that
SUPERBLINK has much trouble identifying brighter stars. What generally
happens is that the BLINK procedure typically fails to determine the
centroid of bright saturated stars, and is thus unable to calculate a
proper motion; such objects are simply rejected by the code. In the most
extreme cases, i.e., for stars brighter than $V=4$, a complete failure
occurs in the SUPER procedure: the code is unable to superpose the two
$17\arcmin\times17\arcmin$ subfields that display extended,
saturated patches from the bright star. These fields are basically not
processed by the code. Unfortunately, this also guarantees that any
fainter high proper motion star that would normally be detected by
SUPERBLINK, but happens to be in a subfield occupied by a very bright
star, will also be missed by the code.

We identified all possible counterparts of the SUPERBLINK objects in
the HIPPARCOS and TYCHO-2 catalogs (see \S3.3 below), and adopted for
those stars the more precise proper motion value from the TYCHO-2
catalog. As a result, 91 SUPERBLINK objects that were initially above
our proper motion threshold were found to have TYCHO-2 proper motions
below $\mu=0.15\arcsec$ yr$^{-1}$. These rejected objects are not
considered in the current analysis, and are not counted among the
57,397 stars officially identified with SUPERBLINK.

The 57,397 SUPERBLINK stars form the core of our new LSPM catalog, and
include thousands of newly identified high proper motion stars.

\subsection{Resolved common proper motion doubles}

Common proper motion doubles with separations in the range
$\approx1-10\arcsec$ are not uncommon in the field. These objects
usually show up on the POSS plates as elongated, or oddly shaped
objects, and can be mistaken for distant galaxies or short asteroid
tracks. Our SUPERBLINK software identifies all moving objects,
regardless of their shape, and so it picks out barely resolved common
proper motion doubles just as well as single stars. Upon visual
inspection, any moving object with an odd shape is flagged for further
analysis. 

Objects flagged as possible common proper motion doubles are searched
for in the 2MASS All Sky Point Source Catalog to see whether they are
featured as pairs of stars. The resolution of the 2MASS infrared CCD
images is significantly better than the POSS plates, and pairs of
objects with separations smaller than $1\arcsec$ are often resolved. We
found that the vast majority of the stars that were initially flagged
by us as possible doubles indeed did show up as pairs of resolved
stars in the 2MASS catalog. All the objects found by SUPERBLINK were
eventually searched in the 2MASS catalog, in order to determine their
infrared (J,H,K$_s$) magnitudes (see \S4.2 below). In this process,
several more objects identified by SUPERBLINK, and not initially
flagged by us as candidate doubles, were also found to be resolved
into pairs in 2MASS. The individual components of those close common
proper motion doubles are included as separate entries in the catalog.

Using all available images from the DSS (POSS-I, POSS-II in three
colors) plus the 2MASS Quicklook Images obtained from the NASA/IPAC
Infrared Data Archive
\footnote{http:irsa.ipac.caltech.edu/applications/2MASS/QL/interactive.html},
we examined all the pairs to determine whether they were actual common
proper motion doubles, or chance alignments. We found about equal
numbers of each. In areas with significant crowding (low-galactic
latitude fields) there were an abundance of cases in which the high
proper motion star happened to be in the vicinity of a background
source; these were easily filtered out by noticing that the background
source had not moved between the POSS-I/POSS-II and 2MASS epochs. In
most other cases, it was clear from the POSS scans and 2MASS images
that we were dealing with a moving pair. There were only a few
ambiguous cases, for which we conservatively assumed the star to be
single.

In the end, we resolved 1,159 SUPERBLINK objects into common proper
motion pairs; each component is included in the LSPM catalog as a
separate entry. However, since it was generally not possible to obtain
proper motions for each component individually (SUPERBLINK only gave a
proper motion for the pair) the two stars are listed as having exactly
the same proper motion. This, of course, is only approximate, and one
should not conclude that the two stars do not show any significant
relative proper motion. 

A number of common proper motion doubles were already listed as such
in the NLTT catalog. In those instances, we have tried to assign the
correct NLTT numbers for each star of the pair. To check our
assignments, we have first used the coordinates listed in the NLTT to
determine which star of the pair was to the north or east of the
other. In many cases, the two stars were listed in the NLTT as having
exactly the same position. In those cases, we looked for notes to the
NLTT catalog, which usually specified the position angle of the
secondary.

A separate paper (L\'epine {\it et al.}, in preparation) will provide
a detailed analysis of all the common proper motion doubles identified
in our survey.

\subsection{Additional Stars from the TYCHO-2 and ASCC-2.5 catalogs}

Our SUPERBLINK survey of the POSS plates has a bright magnitude limit
that limited our identification of very bright ($V<12$) high proper
motion stars. In order to build a catalog that is the most complete
possible, we need to complement the SUPERBLINK stars with lists of
known, bright high proper motion stars.

The two sources we used to complement our catalog are the Tycho-2
Catalogue of the 2.5 Million Brightest Stars (TYCHO-2), and the All-sky
Compiled Catalogue of 2.5 million stars (ASCC-2.5). The TYCHO-2
catalogue \citep{H00} is the product of a re-analysis of data from the
ESA Hipparcos satellite, and combines space-determined positions and
proper motions for 2.5 million of the brightest stars in the sky (the
catalog is complete down to about $V_T=12$) with ground based astrometry
from a variety of sources. The ASCC-2.5 \citep{K01} is a catalog
largely built from the TYCHO and HIPPARCOS catalogs, and providing
essentially similar information on positions and proper
motions. However, the ASCC-2.5 includes complete data on a number of
stars whose proper motions and/or photometric data were missing in the
TYCHO-2 (including stars from the TYCHO-2 supplement-1, which 
contains all HIPPARCOS stars not listed in the TYCHO-2 catalog). The
ASCC-2.5 also includes astrometric information (including proper
motions) on an additional number of fainter stars, obtained from
various ground-based astrometric surveys. The ASCC-2.5 extends the
TYCHO-2 catalog down to slightly fainter magnitudes.

\begin{figure}
\plotone{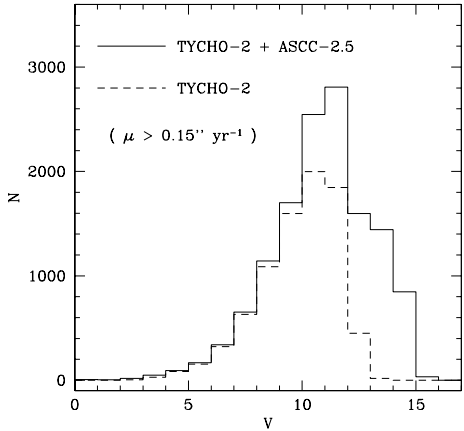}
\caption{\label{fig3}
Distribution as function of magnitude for stars listed in the TYCHO-2
and ASCC-2.5 catalogs with proper motions larger than 0.15$\arcsec$
yr$^{-1}$. The dashed lines show stars listed only in TYCHO-2, while
the full line shows the full set of TYCHO-2 stars complemented by
additional objects from the ASCC-2.5 catalog. The complete set is most
probably complete down to $V=10$, but has a sharp drop in
completeness fainter than $V=12$ at which point, fortunately, the
detection efficiency of SUPERBLINK reaches high levels (see Figure
2).}
\end{figure}

\placefigure{fig3}

The TYCHO-2 catalog was used as our primary source of bright, high
proper motion stars, while the ASCC-2.5 was used as a complement to
the TYCHO-2. We extracted from TYCHO-2 all stars listed with proper
motions exceeding $\mu=0.15\arcsec$ yr$^{-1}$. We found 8,225 objects
in the northern sky spanning a range in magnitude
$2.1<V<13.6$ (see \S4.4 for our derivation of $V$ magnitudes from the
TYCHO-2 $V_T$ and $B_T$ magnitudes), with 93\% of the stars brighter
than $V=12.0$ (Figure 3). Scanning the ASCC-2.5 catalog for additional
objects, we identified 5,239 stars stars listed with a proper motion
$\mu>0.15\arcsec$ yr$^{-1}$ that were not listed in the TYCHO-2, or
whose proper motion data were unavailable in TYCHO-2. The additional
ASCC-2.5 stars spanned a range in magnitude $0.0<V<15.5$, but with
95\% of the objects fainter than $V=10.0$. The vast majority of the
$V<10$ stars in the ASCC-2.5 catalog are listed in the TYCHO-2.

A comparison of the list of bright TYCHO-2/ASCC-2.5 high proper motion
stars with the list of SUPERBLINK detections indicated that 4,252 of
the TYCHO-2 stars, and 3,603 of the ASCC-2.5 stars were already in the
list of $\mu>0.15\arcsec$ yr$^{-1}$ stars detected by
SUPERBLINK. Among the stars that were not found in the SUPERBLINK
list, there were significant numbers of objects with a
TYCHO-2/ASCC-2.5 proper motion close to the $0.15\arcsec$ yr$^{-1}$
limit of our initial list of SUPERBLINK detections. We thus surmised
that that some might have been detected by SUPERBLINK but ranked in a
lower proper motion range. This is especially true of brighter
($V<10$) stars that are strongly saturated on the POSS plates and thus
have larger SUPERBLINK proper motion errors. We searched for
possible matches in a preliminary list of stars found by SUPERBLINK
with calculated proper motion $0.10\arcsec$ yr$^{-1}<\mu<0.15\arcsec$
yr$^{-1}$. We found matches to an additional 846 TYCHO-2 and 21
ASCC-2.5 stars. Because our TYCHO-2 sample contains brighter stars on
average than our ASCC-2.5 sample, it comes as no surprise that most of
the additional matches were from TYCHO-2 objects (whose proper motion
errors in SUPERBLINK are larger).

This left us with 3,127 stars from TYCHO-2 and 1,615 stars from
ASCC-2.5 that had not been detected by SUPERBLINK. Upon visual
inspection of DSS/XDSS images centered on those objects, we
discovered that {\em a significant fraction of them do not show any
detectable proper motion}. This suggests that some of the
TYCHO-2/ASCC-2.5 stars had their proper motions overestimated, which
would explain why they were not identified by SUPERBLINK.

In order to verify the proper motions quoted in the TYCHO-2 and
ASCC-2.5 catalogs, we identified counterparts for all these objects in
the 2MASS All-Sky catalog. We then recalculated their proper motions
using the 2MASS positions (epochs 1997-2001) and the Hipparcos-based
positions from the TYCHO-2 and ASCC-2.5 catalogs (epoch 1991.25). The
2MASS positions for these bright stars are accurate to about 120
mas, which means that it should be possible to derive proper motions
to an accuracy of 12 mas yr$^{-1}$ (for a 10 years baseline) to 20 mas
yr$^{-1}$ (for a 6 years baseline). This assumes, of course, that the
TYCHO-2 and ASCC-2.5 positions are significantly more accurate than
the 2MASS positions.

Results showed that a few hundred of the TYCHO-2 stars, and over a
thousand of the ASCC-2.5 stars had been missed by SUPERBLINK for a
good reason: their actual proper motion is definitely below our
adopted threshold of $0.15\arcsec$ yr $^{-1}$. Figure 4 compares the
proper motions quoted in the TYCHO-2/ASCC-2.5 catalog to the proper
motions determined from the differences in the 2MASS and
TYCHO-2/ASCC-2.5 positions. We plot the results separately for 4
groups of objects: (1) stars from the TYCHO-2 catalog that were also
recovered by SUPERBLINK (top left), (2) stars not listed in the
TYCHO-2 catalog but listed in the ASCC-2.5, and that were identified
by SUPERBLINK (bottom left), (3) stars listed in TYCHO-2 that were
not recovered by SUPERBLINK (top right), and (4) stars in the
ASCC-2.5 but not in TYCHO-2, and that were not identified by
SUPERBLINK (bottom left). First of all, the top left and lower left
plots show that there is a good correlation between the 2MASS-derived
proper motions and those quoted in TYCHO-2 and ASCC-2.5, at least for
stars that had their proper motions confirmed by SUPERBLINK. We find a
dispersion of $15 mas$ yr$^{-1}$ in the difference between the
TYCHO-2 and 2MASS-derived proper motions, and $18 mas$ yr$^{-1}$ in
the difference between the ASCC-2.5 and 2MASS-derived proper motions,
in good agreement with the predicted values (see above). There are
very few outliers in the distribution, with perhaps a few dozen stars
(out of several thousand) whose 2MASS-derived proper motion appears to
be clearly  overestimated, which can be accounted for by an
occasional, large error in the 2MASS, or TYCHO-2/ASCC-2.5, position.

\begin{figure*}
\plotone{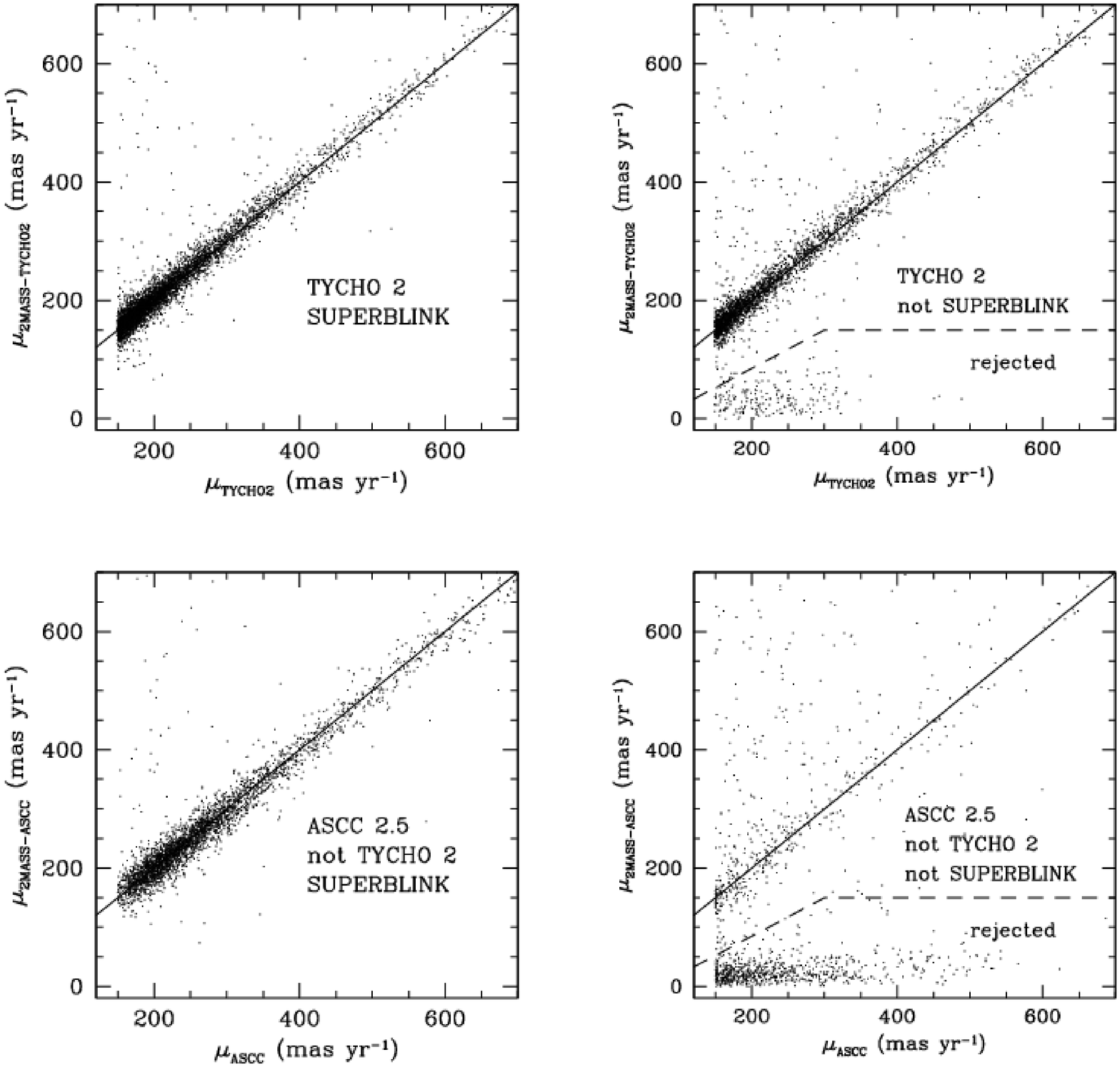}
\caption{\label{fig4} Comparison between the proper motions quoted in
the TYCHO-2/ASCC-2.5 catalogs (abscissa) and the proper motions
calculated from the difference between the 2MASS position (epoch
1997-2001) and the TYCHO-2/ASCC-2.5 position (epoch 1991.25). Top
left: 5,098 high proper motion stars from TYCHO-2 that have been
recovered by SUPERBLINK. Bottom left: 3,624 stars from the ASCC-2.5
(but not listed in TYCHO-2) that have been recovered by
SUPERBLINK. Top right: 3,127 stars from the TYCHO-2 catalog that have
not been recovered by SUPERBLINK (mostly because they are too bright
for SUPERBLINK to handle). Bottom right: 1,615 stars from the ASCC-2.5
catalog (but not in TYCHO-2) that were not recovered by SUPERBLINK. A
significant fraction of the TYCHO-2/ASCC-2.5 stars that were not
recovered by SUPERBLINK are found to have 2MASS-derived proper motions
inconsistent with their quoted values (areas below dashed
lines). These have not been included in the LSPM catalog.}
\end{figure*}

\placefigure{fig4}

The upper right and lower right plots, on the other hand, tell quite a
different story. A significant number of stars are found to have
2MASS-derived proper motions around and below 50 mas yr$^{-1}$. Given
the positional errors on the 2MASS positions, these values are
consistent with the stars having no detectable proper motions. We have
visually inspected DSS/XDSS pairs of images for over a hundred stars
rejected in the procedure, and confirmed that indeed none of these
stars showed any significant proper motion. The dashed lines in Figure
4 shows where we have set the limits under which a star is considered
to have no measurable proper motion, in which case the quoted
TYCHO-2/ASCC-2.5 proper motion is in error. A total of 230 TYCHO-2 and
917 ASCC-2.5 presumed high proper motion stars were thus determined to
be actual low proper motion objects.

\begin{figure*}
\plotone{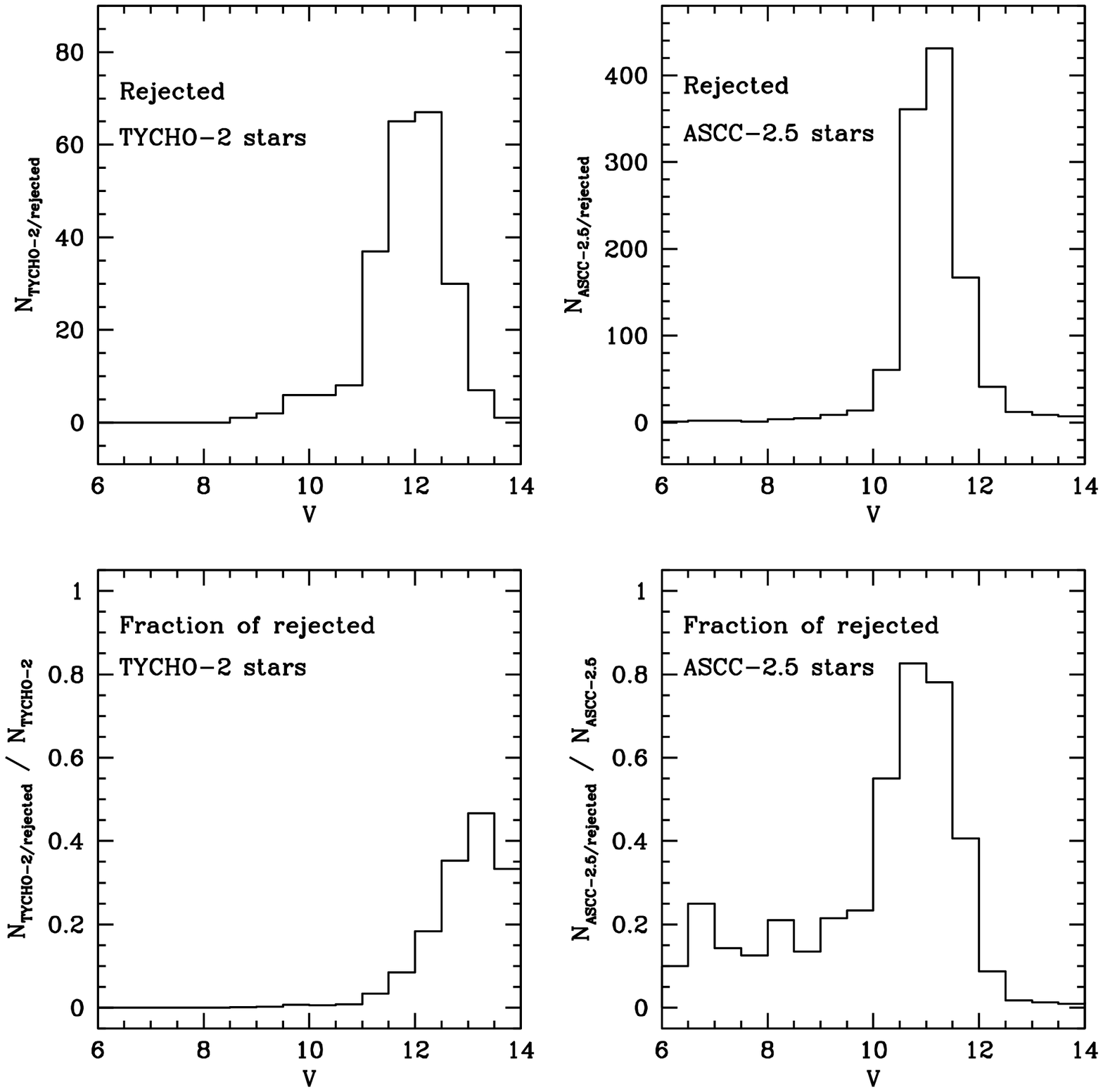}
\caption{\label{fig5} Top left: distribution as a function of $V$
magnitude of TYCHO-2 stars that have not been included in the LSPM
catalog (see Figure 4). Top right: the same, for additional stars from
the ASCC-2.5 catalog, i.e., stars in the ASCC-2.5 catalog listed with
$\mu>0.15\arcsec$ yr$^{-1}$ and which have no TYCHO-2
counterparts. The fractional contribution of these misidentified
high-$\mu$ stars to the full TYCHO-2/ASCC-2.5 samples is shown in the
bottom plots. Up to $\approx40\%$ of the TYCHO-2 stars listed with
$\mu>0.15\arcsec$ do not actually have large proper motions. While a
very significant fraction ($\approx65\%$) of the $10<V<12$
ASCC-2.5(non-TYCHO-2) stars are bogus, almost all of the $V>12$ object
are correctly identified (note that most of the latter have also been
identified with SUPERBLINK).}
\end{figure*}

\placefigure{fig5}

Erroneous proper motion entries in the TYCHO-2 catalog are certainly
cause for concern. We note that most of them are associated with stars
near the faint end of the TYCHO-2 catalog (Figure 5); most erroneous
entries have $V\approx12$. We plot in Figure 5 the fraction of TYCHO-2
stars with quoted proper motion $\mu>150$mas yr$^{-1}$ that  actually
made it into the LSPM catalog. One can see that fully 20\% of $V>12$
stars were found to be low proper motion objects.

Furthermore, we could not find any 2MASS counterpart for 230 of the
ASCC-2.5 stars. Visual inspection of DSS/XDSS images showed no trace
of these stars at the position quoted in the ASCC-2.5. Furthermore, no
high proper motion star was found within $2\arcmin$ of the quoted
position. We thus assume these entries to be bogus, although we cannot
rule out the possibility of a very large error (several arcmin) in the
quoted ASCC-2.5 position.

In addition, there were 66 TYCHO-2 stars and 239 ASCC-2.5 stars that
are listed as close visual companions of brighter TYCHO-2
objects. These are not resolved on the 2MASS images, and thus have no
2MASS catalog counterparts. They are, of course, not resolved on the
DSS/XDSS images either, and we thus cannot obtain an independent
confirmation of their existence (we also found no mention of
them in the Luyten catalogs). We refrain from including them in the
LSPM catalog at this point, while we are still investigating their
status. We do point out that at least all the primary components are
in the LSPM, which should make these secondaries (and probably many
more unsuspected ones) easy to recover eventually.

\begin{figure*}
\plotone{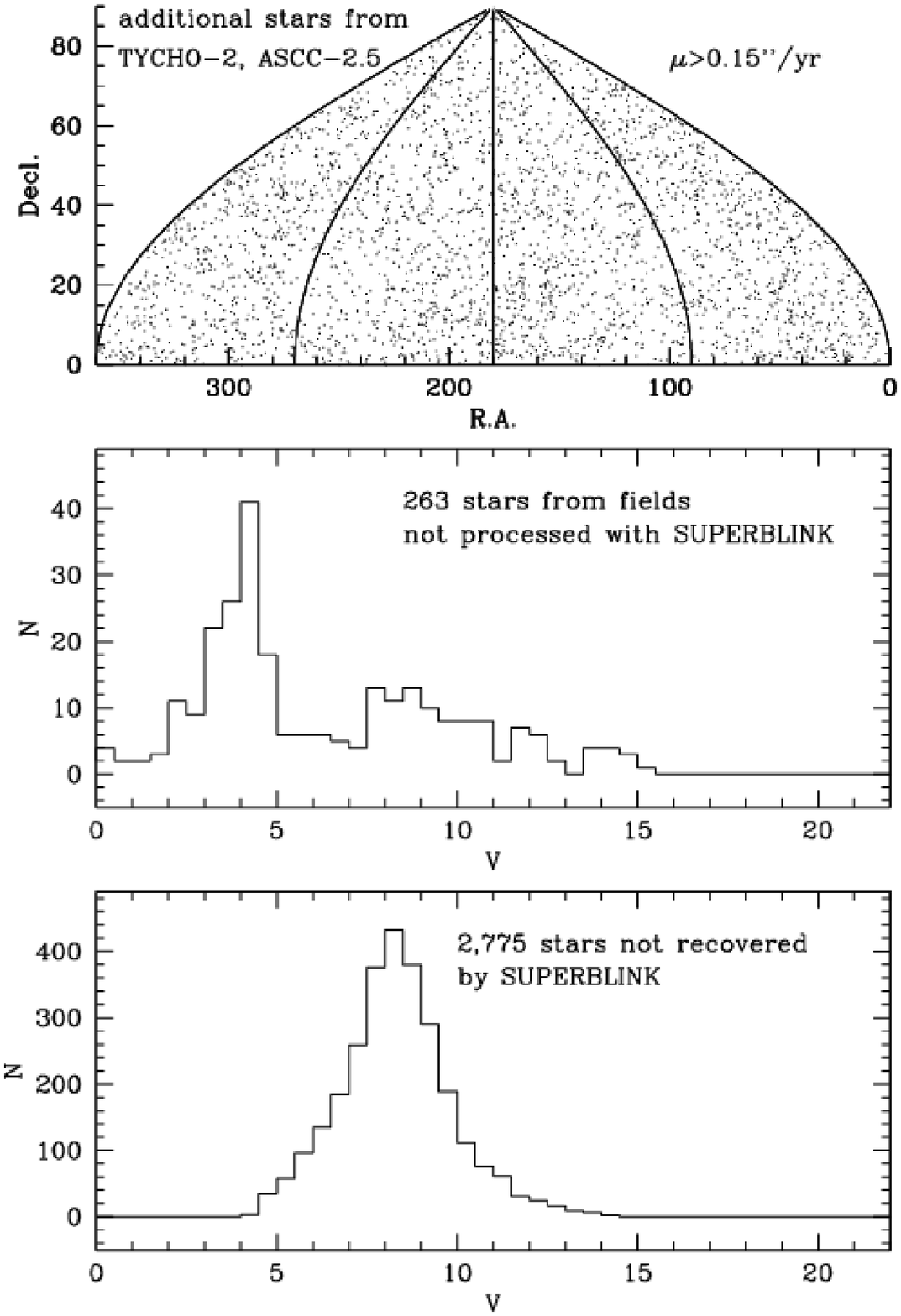}
\caption{\label{fig6}
High proper motion stars listed in the Tycho-2/ASCC-2.5 catalogs
that were not recovered by SUPERBLINK. Top: distribution on the
northern celestial hemisphere. Center: distribution as a function of
magnitude of the 278 stars that were in areas of the northern sky not
processed with SUPERBLINK (325 square degrees). Bottom: distribution
as a function of magnitude of the 3,787 stars missed by SUPERBLINK
because they were too bright, and their images saturated on the DSS
scans. These additional bright high proper motion stars have all
been incorporated into the LSPM catalog.}
\end{figure*}

\placefigure{fig6}

In the end, we were left with 2,831 stars from TYCHO-2, and 229 stars
from ASCC-2.5 which are {\it bona fide} high proper motion stars. Each
of these stars has been included in the LSPM catalog. The vast majority
of those additions are stars brighter than $V=10$. Their distribution
on the celestial sphere is shown in Figure 6. We find that 271 of the
additional TYCHO-2/ASCC-2.5 stars are located in areas that were not
processed by SUPERBLINK. These areas include the north polar cap, and
all areas containing very bright stars that could not be overlapped
properly with the SUPER procedure. The distribution of those stars as
a function of $V$ magnitude is plotted separately in Figure
6. It is obvious that many of the rejected areas are coincident with
very bright stars, as the rejected fields contain a disproportionate
number of $V<5$ objects. We finally find 2,775 stars with proper
motion $\mu>0.15\arcsec$ yr$^{-1}$ in fields that were analyzed by
SUPERBLINK but that were missed by the code. Their distribution peaks
at a magnitude $V\simeq8.5$, and spans a range in magnitude where the
efficiency of the SUPERBLINK software is limited because of saturation
on the POSS plates.

\subsection{Additional stars from the LHS and NLTT catalogs}

A comparison with the LHS and NLTT catalogs reveals a small number of
stars that are absent from our list of SUPERBLINK detection, and that
are too faint to be in the TYCHO-2 and ASCC-2.5 catalogs. All the
objects were investigated individually in order to determine whether
they are real, and whether they should be added to the LSPM catalog.

First of all, the LHS includes a list of 13 faint stars with
proper motions exceeding $\mu=2.0\arcsec$ yr$^{-1}$; much too fast to
have been detected with SUPERBLINK. Generally, LHS stars with very
large proper motions also happen to be relatively bright, and are thus
also present in the TYCHO-2 catalog. But those 13 LHS stars are
fainter than $V=14$, which explains why they are not in the
TYCHO-2/ASCC-2.5 catalogs either. In any case, they were easily
re-identified by direct inspection of DSS scans, and have been added
to the LSPM catalog.

Secondly, and most importantly, a significant number of NLTT stars
that are not in the LSPM were actually recovered by SUPERBLINK, but
found to have proper motions below the $0.15\arcsec$ yr$^{-1}$ cutoff
of the current version of the LSPM catalog. There are also 214 bright
NLTT stars that are listed in TYCHO-2 as $\mu<0.15\arcsec$ yr$^{-1}$
stars. From a search of a preliminary list of $0.10\arcsec$
yr$^{-1}<\mu<0.15\arcsec$ yr$^{-1}$ high proper motion stars
identified with SUPERBLINK, we also recovered an additional 1,204
stars from the NLTT catalog. A small fraction of these stars were
already listed in the NLTT catalog as having low proper motions, but
most were listed with having proper motions $0.18\arcsec$
yr$^{-1}<\mu<0.25\arcsec$ yr$^{-1}$.

Thirdly, we found 161 additional NLTT stars that are actually listed
in the NLTT as close companions ($<10\arcsec$) of brighter NLTT
objects. These companions are not resolved on the POSS scans, and
neither are they resolved on the 2MASS images, which explains why
they didn't show up in our initial search. However, Luyten provided
separations and position angles for most common proper motion doubles
in the NLTT in a comment line, which is available in the electronic
version of the catalog. Using this information, we rederived the
locations of those companions (using the revised positions of the
primaries) and included the companions as separate LSPM
entries. Exceptionally for those secondaries, the quoted optical $b$
and $r$ magnitudes (in our LSPM catalog) are directly recopied from
the NLTT catalog.

While investigating potential NLTT secondaries that might have been
missed in our initial search, we also found 39 duplicate
entries. These stars are apparently objects whose positions and/or
proper motions have been remeasured at some point, but for which the
initial, erroneous entry had been kept in the NLTT catalog by
mistake. They can be easily identified in that while the two stars
were supposed to be two objects of equal magnitude and colors within
$20-60\arcsec$ of each other, only one object of the pair was found on
the DSS scans. The fact that the two entries are listed in the NLTT
catalog with {\it exactly} the same magnitude and color betrays the
fact that they are indeed one and the same.

A more time-consuming job was to investigate the existence of the
remaining 900 or so NLTT stars that remained unaccounted for. Our
methodology was simple: we retrieved pairs of
$17\arcmin\times17\arcmin$ DSS/XDSS scans centered on the quoted
positions of the NLTT stars. The pairs were aligned
(shift-rotated) using SUPERBLINK subroutines, and examined by eye, by
blinking them on the computer screen. We searched for the presence of
a moving object within a $5\arcmin\times5\arcmin$ area centered on the
presumed location of the NLTT star. In 395 fields, no moving object
could be found whatsoever. In those instances, we must assume that
the NLTT entry is bogus. In 7 more cases, we did recover a moving
object, but its proper motion was very clearly smaller than our
$0.15\arcsec$ yr$^{-1}$ limit. All of these NLTT stars, none of which
were included in the LSPM catalog, are listed in a separate table in
the appendix.

\placefigure{fig7}
\begin{figure*}
\plotone{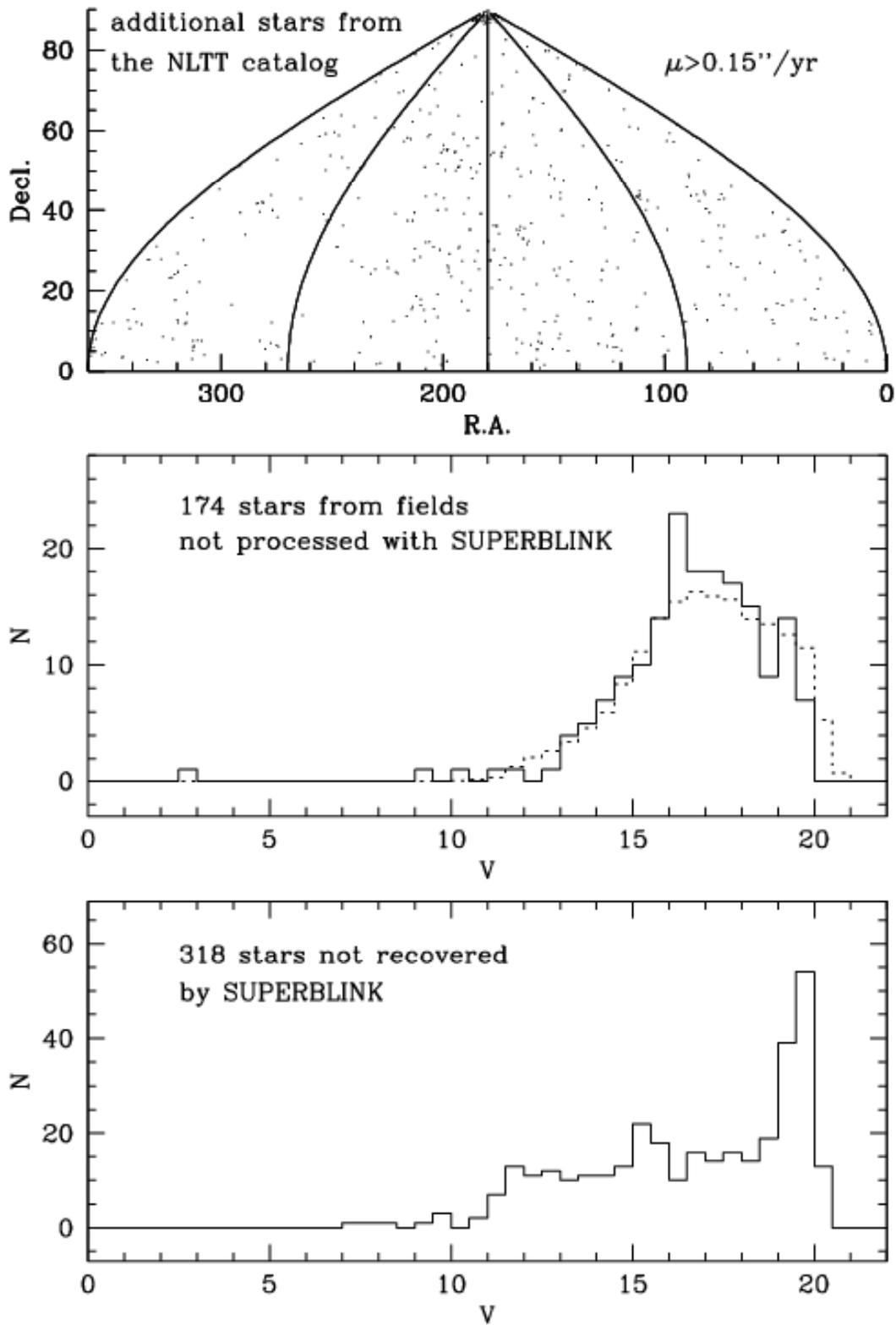}
\caption{\label{fig7}
High proper motion stars listed in the LHS and NLTT catalogs of high
proper motion stars, but that were not recovered by SUPERBLINK, and are
not listed in the Tycho-2 catalog either. Top: distribution on the northern
celestial hemisphere (Aitkins projection). Center: distribution as a
function of magnitude of the 174 stars that were in areas of the
northern sky not processed with SUPERBLINK (325 square
degrees). The normalized distribution of SUPERBLINK stars found in the
rest of the sky and that are not in the TYCHO-2/ASCC-2.5 catalog is
shown for comparison (dotted line). Bottom: distribution as a function
of magnitude of the 318 stars missed by SUPERBLINK. These stars have
all been added to our catalog.}
\end{figure*}

Finally, we positively identified a total of 492 NLTT stars from the
DSS/XDSS scans, with proper motions in the $0.15\arcsec$
yr$^{-1}<\mu<0.4\arcsec$ yr$^{-1}$ range (Figure 7). Of those genuine
high proper motion stars, a total of 174 were found to be located in
areas that were initially mishandled and rejected by SUPERBLINK, and
that were thus not part of the survey. In particular, there were 40
stars located north of $\delta=86.5$, the area near the north
celestial pole that was not properly processed. 

The distribution of stars in areas not analyzed by SUPERBLINK very
closely follows the distribution of high proper motion stars detected by
SUPERBLINK in the rest of the sky, if TYCHO-2/ASCC-2.5 stars are
excluded. In Figure 7, we superpose (dotted line) the distribution of
SUPERBLINK objects that do not have a TYCHO-2 or ASCC-2.5 counterpart
over the distribution of additional NLTT stars found in the areas not
analyzed by SUPERBLINK. The two curves are in very close agreement,
with a peak around $V=17$. (The main difference is that the SUPERBLINK
distribution extends to slightly fainter magnitudes, which is
consistent with a high completeness of SUPERBLINK over the NLTT beyond
$V=19$.) This is exactly what is expected for stars that are missing
because they are outside the SUPERBLINK survey areas: they should
follow the same general magnitude distribution as the stars extracted
inside the SUPERBLINK survey areas.

The remaining 318 NLTT stars are in areas that were processed by
SUPERBLINK, but that were nevertheless missed by the code. Several of
these elusive objects were within $1\arcmin$ of very bright stars at
one of the POSS-I or POSS-II epochs, or were close to plate edges, or
were coincident with local plate defects, making their identification
difficult. The distribution of these stars with $V$ magnitude is also
skewed toward very faint objects (see Figure 7 bottom panel), with a
peak at magnitude $V=19.5$. This marks the range at which SUPERBLINK
is beginning to suffer from incompleteness, as it reaches the
magnitude limit of the POSS-I plates.

\subsection{Other additional objects}

Two more stars were included in our catalog that are objects with
very large proper motions that were discovered in the past two
years. Both are too faint to have been in the TYCHO-2 or ASCC-2.5
catalogs, and because they are very recent additions, they are, of
course, not in the Luyten catalogs either.

The first star is LSPM 1826+3014, discovered by \citet{L02}. The
star has a proper motion $\mu=2.38\arcsec$ yr$^{-1}$ and a magnitude
$V=19.4$. In our catalog, it bears the name LSPM J1826+3014. The star
was actually discovered in the course of our own survey, but is
regarded as a serendipitous discovery: it was not initially identified
as a high proper motion star by SUPERBLINK, but rather as a pair of
variable stars within 2$\arcmin$ of each other, which we examined
further out of curiosity. This makes one seriously consider the
possibility that there still exist faint stars with very large proper
motions waiting to be discovered.

The second star is the extremely high proper motion object SO
025300.5+165258 discovered by \citet{T03}. The star has a
proper motion $\mu=5.05\arcsec$ yr$^{-1}$ and a magnitude
$V=15.4$, and is identified as LSPM J0253+1652 in our
catalog. It is believed to be a very nearby star. Its extremely large
proper motion is beyond the detection limit (2.0$\arcsec$ yr$^{-1}$)
of the SUPERBLINK. What is interesting is that the survey by
\citet{T03} that led to its discovery was initially aimed at the
identification of solar system objects, and uses a temporal baseline
of months to a few years. It thus appears that a pair of all sky
surveys with a short separation in time (e.g. 1-2 years) might well
lead the way to locating any possible remaining faint stars with
proper motions in excess of $2.0\arcsec$ yr$^{-1}$.

We note that most of the L dwarfs and T dwarfs discovered in recent
years \citep{K99,K00,H02,C03} very probably have proper motions within
the range of our catalog; however no effort was made to include any
high proper motion brown dwarf at this point. Proper motions have so
far been determined only for a small number of L and T dwarfs
\citep{D02}, and additional work would be required to obtain accurate
proper motions for most of them. For now, we have limited ourselves to
the very few L dwarfs that do show up on the POSS-I and POSS-II plates
and that were recovered by SUPERBLINK, although we do plan to add high
proper motion brown dwarfs to the LSPM catalog in the near future.

Adding up all the stars found by SUPERBLINK, those retrieved from the
TYCHO-2 and ASCC-2.5 catalog, the LHS/NLTT stars missed by our code,
and the two additional objects discussed in this section, we come to a
total and final tally of 61,618 stars in the LSPM catalog.

\subsection{Counterparts in the UCAC2 astrometric catalog}

The Second U.S. Naval Observatory CCD Astrograph Catalog (UCAC2) is
the second release in an all-sky astrometric survey of stars in the
magnitude range $7.5<$ R$_{\rm F}<16.0$ \citep{Z03}. The current
version lists 48,330,571 stars in the declination range
$-90<$Decl.$<+50$, and gives positions with an accuracy of $20-70$mas
(depending on magnitude). It also provides proper motions for all
cataloged stars with an accuracy $\sim5$mas yr$^{-1}$.

We found UCAC2 counterparts for 9,151 of the LSPM stars. All the
counterparts are south of Decl.=+53.23, reflecting the current,
limited sky coverage of the UCAC2 (Figure 8). Fewer stars are found
below a Decl. of 10 degrees, where the UCAC2 apparently has a brighter
magnitude limit ($V<12$). Overall, the UCAC2 lists stars
down to a magnitude $V\simeq16.0$ but appears to be significantly
incomplete for high proper motion stars at all magnitudes and
positions.

\begin{figure*}
\plotone{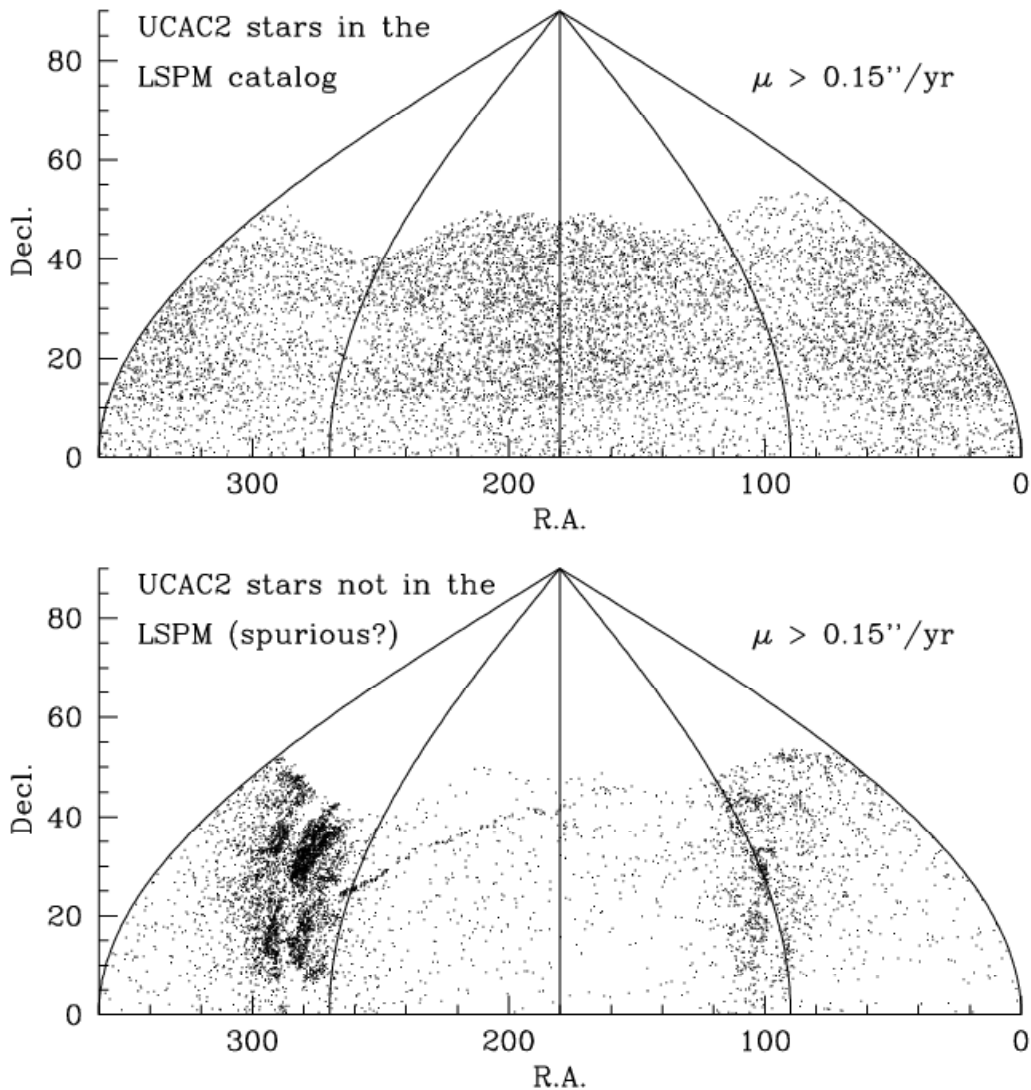}
\caption{\label{fig8}
High proper stars in the UCAC2 astrometric catalog. Top: positions of
9,151 stars with $\mu>0.15\arcsec$ yr$^{-1}$ that are also listed in
our LSPM catalog. Bottom: positions of 7,370 stars listed in the UCAC2
as having proper motions $\mu>0.15\arcsec$ yr$^{-1}$ but that are not
in the LSPM; these appear to be spurious entries.}
\end{figure*}

\placefigure{fig8}

While we were initially hoping to use the UCAC2 as a an additional
source of bright stars with large proper motions for the LSPM catalog,
we found this to be impractical at this point. The main reason is that
it appears that the UCAC2 is plagued with a large number of spurious
high proper motion entries. We retrieved all the stars in the UCAC2
that are in the northern sky and listed as having a proper motion
exceeding $0.15\arcsec$ yr$^{-1}$. Apart from the 9,168 objects also listed
in our LSPM catalog, we found an additional 7,370 entries (see Figure
6) with a very non-uniform distribution. The vast majority of the
additional entries are located at low galactic latitudes, and they
have cataloged proper motions in the range $0.15\arcsec$
yr$^{-1}<\mu<0.25\arcsec$ yr$^{-1}$. Our examination of several
DSS/XDSS scans centered on the presumed location of those stars failed
to reveal any of them as high proper motion objects.

Good UCAC2 counterparts do however prove useful as a means to estimate
the astrometric accuracy of the LSPM from an independent source. A
comparison of the UCAC2 and LSPM positions and proper motions for the
stars common to both catalogs is presented in \S4.3 and \S 4.4 below.

\section{Photometry}

\subsection{Optical magnitudes from TYCHO-2 and ASCC-2.5: B and V}

Optical B$_T$ and V$_T$ magnitudes are obtained from TYCHO-2 catalog
counterparts (see \S3.3 above). Photometric errors are 0.013 mag for
$V_T<9$, and 0.1 mag for $9<V_T<12$. A simple conversion transforms
these into Johnson B and V magnitudes:
\begin{equation}
B = B_T - 0.24 ( B_T - V_T ) ,
\end{equation}
\begin{equation}
V = V_T - 0.09 ( B_T - V_T ) ,
\end{equation}
following the prescription in the introduction to the HIPPARCOS and
TYCHO catalogs.

The ASCC-2.5 catalog provides both B and V magnitudes (converted
from $V_T$ and $B_T$), and can also be used as a relatively reliable
source of optical magnitudes for bright stars. In particular, we are
using it to obtain magnitudes of bright stars that are not listed in
the TYCHO-2 catalog. For stars fainter than about $V=12.0$, ASCC-2.5
magnitudes are derived from a variety of sources, and may not be as
accurate as the TYCHO-2 magnitudes, but since they were obtained from
photoelectric or CCD measurements, they should be relatively reliable.

\begin{figure}
\plotone{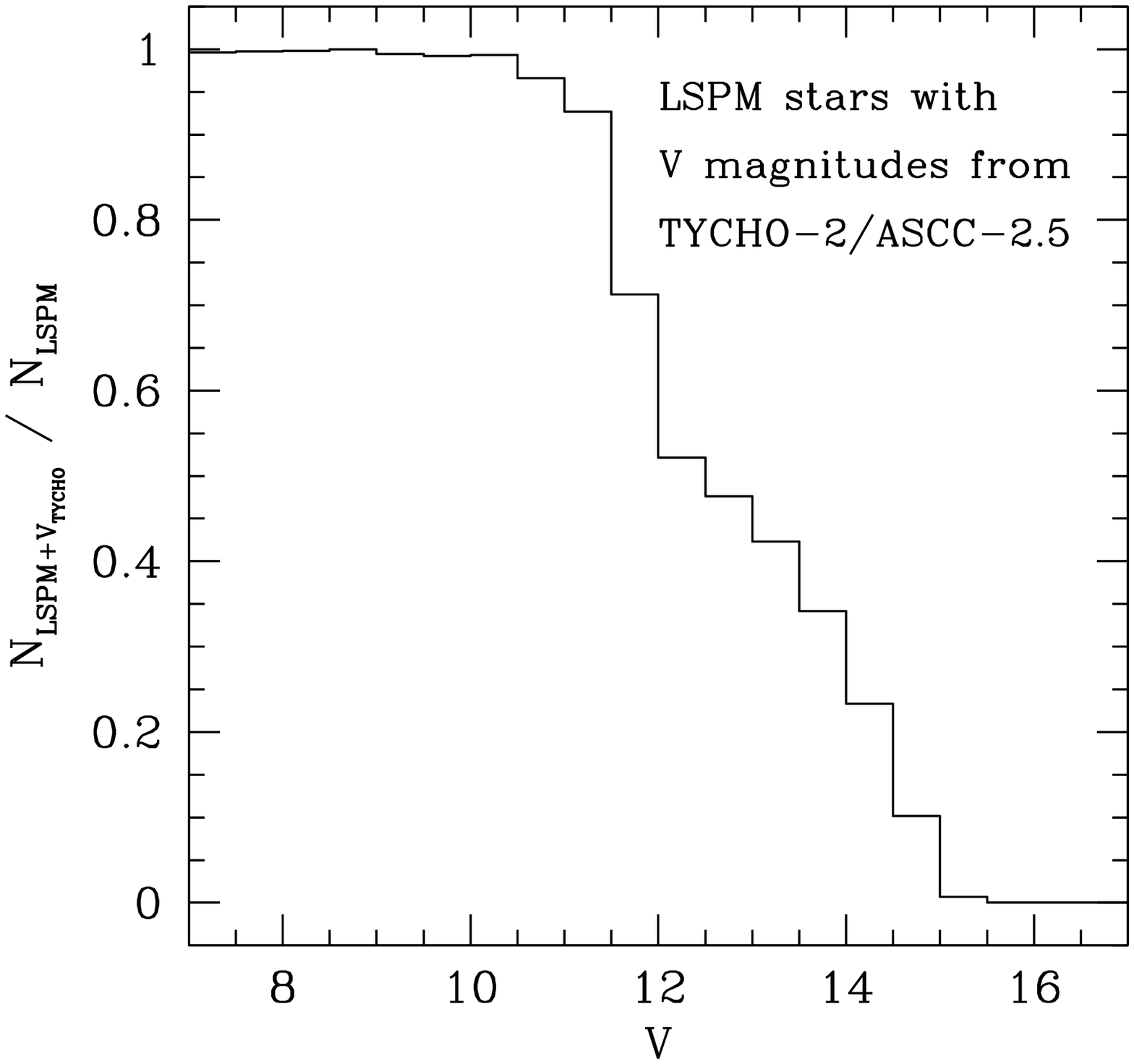}
\caption{\label{fig9}
Fraction of LSPM catalog stars with $B$ and $V$ magnitudes obtained
from the TYCHO-2 and ASCC-2.5, plotted as a function of the visual
magnitude. This indicates that we have reliable optical photometry for
essentially all stars brighter than $V=12.0$. For fainter objects, we
have to rely on photographic magnitudes to estimate $V$.}
\end{figure}

\placefigure{fig9}

We have gathered B and V magnitudes from TYCHO-2/ASCC-2.5 for a total
of 11,719 LSPM stars. The fraction of stars with TYCHO-2/ASCC-2.5
optical photometry is plotted in Figure 9 as a function of $V$. It
shows that the LSPM contains reliable optical photometry for
essentially all stars brighter than $V=12.0$. This is fortunate,
because these are stars for which photographic magnitudes are
subject to large errors, because of saturation on the POSS plates. For
fainter stars, and especially those with $V>14.0$, we do need to rely
mainly on photographic magnitudes to cover the optical regime: the
only existing all-sky catalogs of faint optical stars are based on
photographic plate material.

\subsection{USNO-B1.0 photographic magnitudes: $B_J$, $R_F$, $I_N$}

The USNO-B1.0 Catalog \citep{Metal03} is an all-sky catalog made from
scans of several photographic sky survey, including the POSS-I and
POSS-II. Astronomical objects have been identified using the PMM
scanning machine. The catalog gives positions, proper motions,
photographic magnitudes in 5 passbands, and star/galaxy estimators for
1,042,618,261 objects. While the USNO-B1.0 provides proper motions for
all objects detected in the POSS plates, it is not a reliable source
for stars with large proper motions. The main difficulty with the
USNO-B1.0 is the exceedingly large number of spurious entries
\citep{G03b}; at high galactic latitudes, up to 99\% of objects listed
with $\mu>0.18\arcsec$ yr$^{-1}$ are not real. The catalog also
suffers from serious incompleteness for high proper motion stars at
low galactic latitude (up to 30\%), as estimated from its recovery of
NLTT stars.

The USNO-B1.0 is however an extremely valuable complement to the LSPM
catalog because it provides reasonably accurate photographic B$_{\rm
J}$, R$_{\rm F}$, and I$_{\rm N}$ magnitudes (respectively IIIa-J,
IIIa-F, and IV-N) derived from the POSS-II scans. Because the
USNO-B1.0 is based on some of the same plate material as the DSS, it
also provides a very useful check for our SUPERBLINK detections.

We have succeeded in finding USNO-B1.0 counterparts for 60,396 of our
LSPM stars. Searching the USNO-B1.0 for high proper motion objects
however turned out to be a difficult and time-consuming
problem. Cross-correlation of the two catalogs yielded only
$\approx75\%$ of unambiguous matches. The large number of ambiguous
cases fell into three broad classes: (1) stars with erroneous USNO-B1.0
proper motions, (2) moving stars not identified as such in the
USNO-B1.0 and listed as separate stars, one for each epoch in which
the stars was detected, and (3) confusion with background stars at the
detection epoch.

One common problem was USNO-B1.0 entries with large errors in their
quoted proper motions ($\mu_{USNO-B1.0}$). Since their quoted RAJ2000
and DEJ2000 are calculated by extrapolating the position from the mean
epoch of observation with their estimated proper motions, some stars
have quoted positions incorrect by up to several seconds of arc. In
most cases, we were able to recover the star by extrapolating back to
the mean epoch of observations and recalculating the RAJ2000 and
DEJ2000 positions using the proper motion determined by SUPERBLINK.

Another major source of complication was high proper motion stars
listed as two or more distinct entries in the USNO-B1.0. Each entry
typically corresponds to a detection of the star in a distinct
photographic survey. This problem particularly affected stars with
larger proper motions. A typical example is a moving star listed
as three separate entries, one with the position of the star at
the epoch of the POSS-I survey, one with the position of the star at
the epoch of the POSS-II red and blue surveys, and one with the
position of the star at the epoch of the POSS-II near infrared
survey. The confusion was such that most of these cases had to be
dealt with individually.

Additional complications occurred because of confusion with background
sources. This problem was common especially in low galactic latitude
fields, where crowding is significant. All these cases had to be
examined one by one. Overall, we had to visually inspect
$\approx12,000$ LSPM objects in order to determine their correct
USNO-B1.0 counterpart. Again we made use of our interactive software,
this time overlaying the USNO-B1.0 catalog over our SUPERBLINK finder
charts. Whenever an LSPM star appeared as two or more separate
USNO-B1.0 entries, a choice had to be made as to which one should be
used as the ``official'' counterpart, in order to keep the LSPM
catalog simple. In general, we picked the entry having the most
complete photometric data, or in some cases the one least likely to be
contaminated by blending with background sources.

Not all USNO-B1.0 entries have magnitude information in all three
bands. Whenever possible, we tried to combine magnitude data if a
USNO-B1.0 star appeared as more than one entry. For example, if one
high proper motion star was listed as two distinct USNO-B1 entries,
one giving only B$_{\rm J}$ and R$_{\rm F}$ magnitudes, the other
giving only an I$_{\rm N}$ magnitude, we would combine the information
to obtain complete B$_{\rm J}$R$_{\rm F}$I$_{\rm N}$ photometry. For
practical purposes, however, we list the counterpart ID only for one
of the two entries. As a result, LSPM magnitudes are often more
complete than magnitudes extracted from the USNO-B1 catalog for the
listed counterpart.

Finally, we note that no USNO-B1 counterparts could be found for a
total of 1,580 LSPM stars. The majority of these are in close proper
motion double systems or in very crowded fields, and are simply not
resolved in the USNO-B1.0, although most of them are resolved in
2MASS. We note that 267 of the LSPM stars have neither a 2MASS nor
a USNO-B1 identifier; a majority of these are faint proper motion
companions that are not resolved in the USNO-B1 and are too faint to
have been detected by 2MASS. However, 229 of them are NLTT stars, and
the others all clearly show up on DSS/XDSS scans and/or 2MASS images.

\subsection{2MASS infrared magnitudes: J, H, and K$_s$}

The 2MASS All-Sky Point Source Catalog is an all-sky catalog of
sources detected in the Two Micron All-Sky Survey \citep{C03}. The
catalog covers the whole sky and is complete down to
$J\simeq16.5$. Infrared J, H, and K$_s$ magnitudes are provided, and are
accurate to 0.02 mag down to 15th magnitude. Positions are given for
the epoch of observation (1997-2001) and are accurate to 70-80 mas for
the fainter sources ($J>9$), and 120 mas for the brighter ones.

The vast majority of the LSPM catalog stars are found to have
counterparts in the 2MASS All-Sky Point Source Catalog: we have
reliable matches for 59,684 of our high proper motion stars. Finding
2MASS counterparts was straightforward for $\approx90\%$ of the LSPM
objects, and was obtained by a simple cross-correlation of the
SUPERBLINK and TYCHO-2/ASCC-2.5 high proper motion star positions with
the 2MASS source positions. To make the search more effective, all
positions were locally extrapolated to the epoch of the 2MASS
observations. Most 2MASS counterparts were found within $1\arcsec$ of
their predicted position. There were multiple possible matches for
$\approx3,000$ of the LSPM objects, most of them in very crowded, low
galactic latitude fields. We used again a software package developed
by SL, which overlays 2MASS catalog entries on the
$3.25\arcmin\times3.25\arcmin$ charts generated by SUPERBLINK. All
matches were then made interactively, by direct visual inspection. All
common proper motion doubles (see \S3.2 above) were also examined and
their 2MASS identification verified with the same software. No 2MASS
counterparts were found for 2,292 LSPM stars; in the majority of
cases, these are simply too faint in the infrared to have been detected
in the 2MASS survey.

Because of its relatively high astrometric accuracy, we have adopted
the 2MASS catalog as the {\em primary source of positional
information} for stars that are not listed in the TYCHO-2 catalog (see
\S 5.3 below). We believe the 2MASS is the best possible choice to
pinpoint the positions of those stars because: 1) it contains the
majority of the LSPM objects, 2) its positions are given in the ICRS
system and are reasonably accurate, and 3) the 2MASS observation
epochs are very close to 2000.0. The advantage of having all the 2MASS
positions to within a few years of the 2000.0 epoch is that it
minimizes positional errors introduced by the errors in the proper
motions, when one extrapolates the position of a high proper motion
star to the 2000.0 epoch.

\subsection{Estimated V magnitudes and V-J colors}

We have attempted to place all our stars on a simple, uniform
magnitude/color system, that includes both an optical and an infrared
magnitude. We do have uniform, reliable measurements of infrared
magnitudes (from 2MASS) for the vast majority of LSPM stars. However,
some of our objects do not have 2MASS counterparts. Furthermore, we are
in a situation where our optical magnitudes are in two different
systems. On the one hand, we have $V$ and $B$ magnitudes accurate to
$0.1$ mag, but only for a small fraction of our stars (those with
TYCHO-2 counterparts). On the other hand, we have much less reliable
photographic magnitudes ($B_J$, $R_F$, $I_N$), accurate to
$\approx$0.3-0.5 mags \citep{Metal03}, but most likely affected by
systematics errors \citep{Setal04}. Additional complications include
the fact that one or more of the photographic magnitudes are sometimes
missing. It is claimed by \citet{SG03} that photographic $R$
magnitudes from the USNO-A2.0 are accurate to 0.25 mag, which is
significantly better than USNO-B1.0. On the other hand, the
USNO-B1.0 catalog is more complete than the USNO-A2.0, particularly
for faint ($V>19$) stars, and also at low Galactic latitudes.
Nevertheless, we are currently trying to find USNO-A2.0 counterparts
of LSPM stars, and USNO-A2.0 magnitudes will be included in future
versions of the catalog.

In any case, it is desirable to provide an immediate means to classify
the stars in our catalog according to color and magnitude (even roughly).
The general idea is thus to get estimates of the optical $V$ magnitude
and of the optical-to-infrared $V-J$ color {\em for all the stars in the
catalog}. We already have reliable $V$ magnitudes for LSPM stars with
TYCHO-2 (see equation 2) or ASCC-2.5 counterparts, and $J$ magnitudes
for all stars with a 2MASS counterpart. What we need is a
transformation system to obtain estimates of $V$ and $V-J$ using the
photographic magnitudes from the USNO-B1.0 catalog. While one might be
tempted to use the excellent 2MASS infrared magnitudes to obtain
estimates of optical $V$, this cannot be done reliably. The reason is
that M dwarfs, which constitute the vast majority of the LSPM stars
with no TYCHO-2 counterparts, are largely degenerate in their infrared
colors: all M dwarfs (except the very coolest) have
$J-K_s\simeq0.7\pm0.2$ for $3<V-J<6$.

The $V$ band is located halfway between the photographic $B_J$ and
$R_F$ bands. We estimate $V$ using:
\begin{equation}
V = B_J - 0.46 ( B_J - R_F ) ,
\end{equation}
a relationship which is verified for all TYCHO-2 stars with USNO-B1.0
counterparts. For stars with $B_J$, but no $R_F$ magnitudes, we find
it is possible to estimate $V$ from the following set of
transformations:
\begin{displaymath}
V = B_J - 0.23 ( B_J - J ) - 0.10  \ \ \ \ [ B_J-J < 4 ]
\end{displaymath}
\begin{equation}
V = B_J - 0.05 ( B_J - J) - 0.72 \ \ \ \ [ B_J-J > 4 ]
\end{equation}
Likewise, for stars with $R_F$, but no $B_J$, we can estimate $V$
using:
\begin{displaymath}
V = R_F + 0.6 ( R_F - J ) - 0.10 \ \ \ \ [ R_F-J < 2 ]
\end{displaymath}
\begin{equation}
V = R_F + 1.10 \ \ \ \  [ R_F-J > 2 ]
\end{equation}
The V magnitudes estimated from the relationships given above are
generally accurate to about $\pm0.5$ mag. Following these simple
transformations, we calculate $V$ magnitude estimates for 61,550 LSPM
stars. Of the 427 LSPM stars for which we do not provide a $V$
magnitude estimates, 355 are close common proper motion doubles that
are resolved in 2MASS but not on the POSS plates (see \S3.2). At this
point, we refrain from trying to obtain a $V$ magnitudes using only
$J$, $H$, and $K_s$. The remaining 72 stars with no $V$ estimates are
stars that are not in the 2MASS of USNO-B1.0 catalogs, and for which
we only have $R_F$ magnitudes estimates from SUPERBLINK.

Prospective catalog users should be warned that these $V$ magnitude
estimates are generally not very accurate, and may be subject to
systematic errors and other effects. The LSPM catalog $V$ magnitudes
should only be trusted for stars brighter than $V=12.0$, whose $V$ are
from the TYCHO-2 catalog. At fainter magnitudes, there may be errors
of 0.5 mag or larger.

Since the majority of LSPM objects do have 2MASS counterparts,
calculations of V-J colors are straightforward. For stars that do
not have 2MASS counterparts, we use the photographic $I_N$ magnitudes
and use the following transformation:
\begin{equation}
V-J = 1.3 ( V - I_N ) + 0.3.
\end{equation}
From the 2MASS counterparts and the transformation above, we obtain
$V-J$ colors for all but 814 entries in the LSPM catalog. These
include the 427 stars for which we have no $V$ magnitude estimates
(see above), and 387 stars that are not in the 2MASS catalog and for
which we do have $I_N$ magnitudes. Note that the $V-J$ colors are only
as accurate as the $V$ magnitudes are. Since stars fainter than
$V=12.0$ have errors of up to 0.5 mag or even larger, the $V-J$ color
estimates should be used with extreme caution.

We emphasize again that the primary goal of the LSPM catalog is {\em to
provide the most complete list possible of high proper motion stars},
and is not intended to be a photometric catalog. The photometry that
we do provide for LSPM stars should be regarded as very preliminary,
and is given only as a help in identifying interesting classes of
objects for follow-up observations. Future efforts will be devoted to
obtaining more accurate optical magnitude estimates for all LSPM
stars.

\section{Astrometry}

\subsection{Conversion to absolute proper motions}

The SUPERBLINK software was largely designed to achieve the highest
possible recovery rate for high proper motion stars on photographic
plates. A such, it was optimized for raw detection, and not for
accurate astrometric measurements of detected objects. The main caveat
is that {\em proper motions are calculated relative to local
background sources}. This means, typically, all objects within
$\approx4\arcmin$ of the moving target. Because of this, the proper
motions calculated by SUPERBLINK are {\em local, relative proper
motions}. These are usually offset by up to several milliarcseconds 
(mas) per year relative to {\em absolute proper motions}, which are
proper motions measured in a fixed reference frame (defined e.g. by
the positions of distant quasars, such as for the ICRS reference
frame). It must be realized that most ``background'' sources used by
SUPERBLINK as a local reference system are Galactic objects, and they
all have significant proper motions at the mas level. The
local frames used by SUPERBLINK are thus moving frames, and this
potentially introduces both random and systematic offsets in the 
SUPERBLINK proper motions. The random offsets arise because of the
limited number of stars that locally define the frame, if these stars
are all moving in random directions, then their mean proper motion
will generally not add-up to zero; it will however converge to zero if
the number of reference stars is large enough. These random errors
affect most the fields at high Galactic latitudes, where the object
density is low and the local SUPERBLINK reference frames are defined
by very few stars (sometime $\lesssim100$). In any case, field
background stars generally have random proper motions smaller than 10
mas yr$^{-1}$, which means that local random offsets will be less
than 1 mas yr$^{-1}$ in frames defined by at least 100
stars. Systematic offsets, however, are more of a problem. If all
local stars participate in some local bulk motion, then the local
frame used by SUPERBLINK will definitely be moving at the bulk motion
rate, no matter how many stars define the frame.

Fortunately, we do have a means to estimate some of the 
systemic motions of the background stars, and correct for them in
order to obtain absolute proper motions. To that purpose, we can use
all TYCHO-2 stars that have also been measured with SUPERBLINK, and
compare their absolute and relative proper motions. The random errors
on the SUPERBLINK proper motion are on the order of, or larger than
the local systemic motions, but we can average out the residuals over
appropriate-sized areas, and calculate {\em zonal corrections}.

It is true that the averaging procedure may even out some of the local
fluctuations. However, background Galactic stars are expected to
display mainly {\em global} patterns of systemic motions. The main
sources are the rotation of the Galaxy, the systemic motion of the
local standard of rest (LSR) relative to other Galactic stellar
populations (old disk, halo), and the motion of the Sun within the
LSR. The resulting systemic absolute proper motions of the background
stars are dependent on their position on the sky, but on a global
scale. For instance, the systemic drift of old disk and halo stars is
largest in a direction perpendicular to the Galactic rotation (toward
the Galactic pole), and slowly decreases as one looks more toward the
direction of rotation.

The current version of the LSPM catalog only lists a few thousand
stars (with $\mu>0.15\arcsec$ yr$^{-1}$) for which we have both TYCHO-2
and SUPERBLINK proper motions, leaving very few objects to calculate
zonal corrections on a scale of less than a few tens of
degrees. This tends to make the map too coarse, and local values too
inaccurate. But the {\it complete} SUPERBLINK database actually
includes detections from the DSS of stars with proper motions down to
0.04$\arcsec$ yr$^{-1}$. While most of those lower proper motion
detections are still being processed and analyzed, we have recently
compiled a preliminary list of objects, which include over 30,000
TYCHO-2 stars with proper motions $0.05\arcsec$
yr$^{-1}<\mu<0.15\arcsec$ yr$^{-1}$, all of which have a relative
proper motion value measured by SUPERBLINK.

\begin{figure*}
\plotone{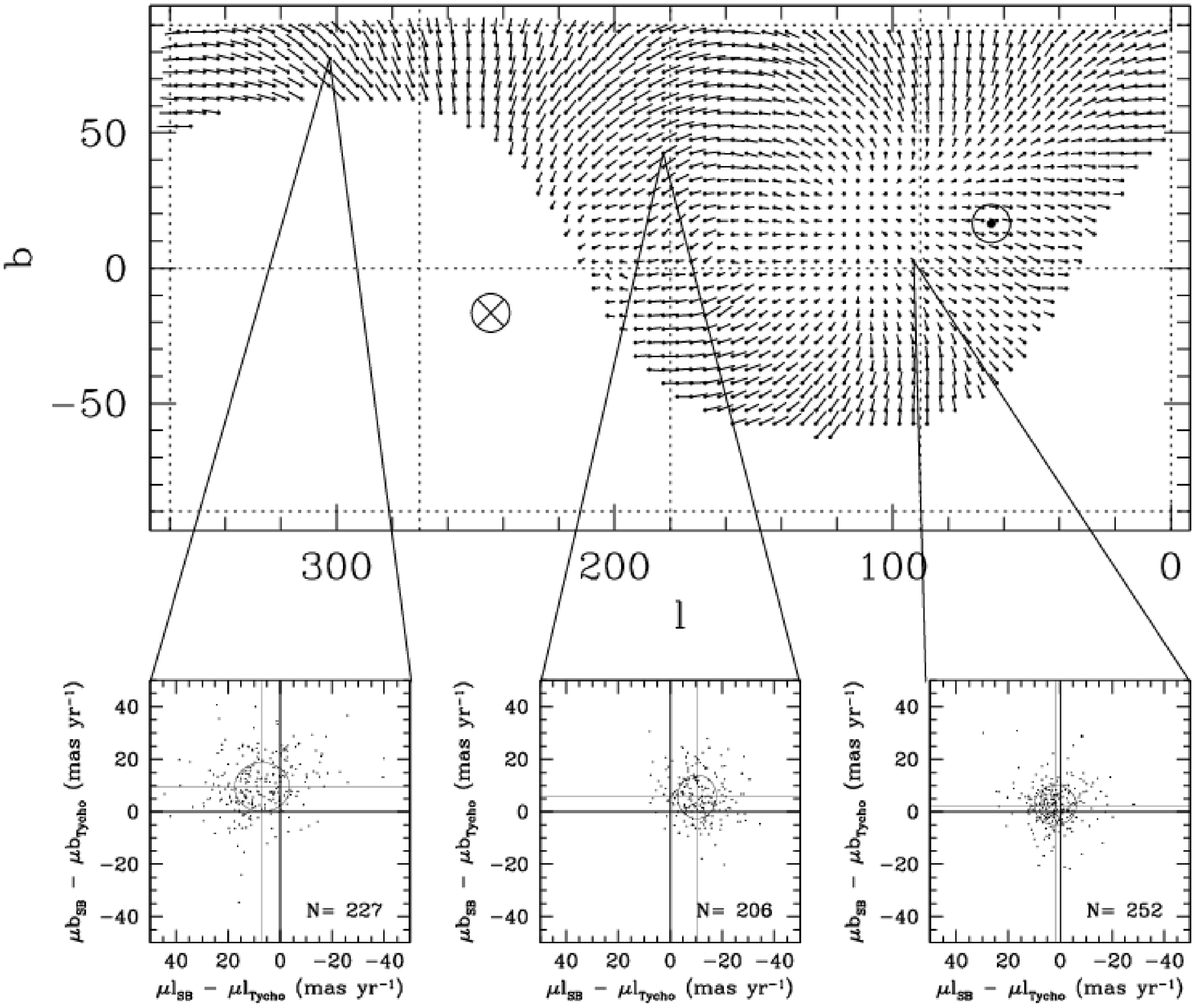}
\caption{\label{fig10} Local offsets between SUPERBLINK relative
proper motions and TYCHO-2 absolute proper motions. The offsets are
calculated using 33,300 TYCHO-2 stars with proper motions
$0.04\arcsec$ yr$^{-1}<\mu<0.50\arcsec$ yr$^{-1}$ whose relative
proper motions have been independently calculated with SUPERBLINK. The
three bottom panels shows local differences between the relative and
absolute proper motions for stars located within 7 degrees of the
specified location. Proper motion differences are calculated in the
local plane of the sky. The crosshairs mark the mean value of the
offset, while the ellipse shows the mean standard deviation
($1\sigma$). The top panel plots proper motion difference vectors as a
function of position on the sky, in galactic coordinates. The apex
($\odot$) and antapex ($\otimes$) of the Sun's motion are noted. These
offsets effectively map out the local mean proper motion of background
field stars in the TYCHO-2 (ICRS) reference system. Offsets are
largest at high galactic latitude, where there is a significant
drifting motion of old disk and halo stars relative to the local
standard of rest.} 
\end{figure*}

\placefigure{fig10}

Zonal corrections have thus been calculated using a list of 33,312
stars with magnitudes $10<V<13$ and proper motions
$0.05\arcsec$ yr$^{-1}<\mu<0.50\arcsec$ yr$^{-1}$. For each position
of the celestial sphere, the zonal correction is calculated from the
mean of the offsets between the relative (SUPERBLINK) and absolute
(TYCHO-2) proper motions of all stars {\em within a radius of 7
degrees}. Every position on the sky uses $\approx290$ stars on
average. All offsets are calculated in the local plane of the
specified location. Outliers, with offsets more than 3-sigma away
from the mean, are removed from the final calculation. 

Depending on the position on the sky, the mean
offsets vary from $-9.1$ to $+12.2$ mas yr$^{-1}$ in $\mu RA$, and
from $-0.3$ to $+13.9$ mas yr$^{-1}$ in $\mu DE$. A map of the zonal
corrections for the northern sky is shown in Figure 10, where it is
plotted in Galactic coordinates. The local distribution of offsets
between the relative and absolute proper motions is also shown for
three positions on the sky. The number of stars $N$ used in
calculating the local offset is noted.

From these zonal corrections, absolute proper motions are calculated
from the relative SUPERBLINK proper motions. We thus obtain absolute
proper motions for all LSPM stars. In the LSPM catalog, we list both
the relative proper motion determined with SUPERBLINK, and the
absolute proper motion obtained after applying the zonal corrections.

In effect, Figure 10 plots the local mean value of the absolute proper
motion of background stars (N.B.: the zonal correction vectors plotted
in Figure 10 all point in the direction opposite to the mean relative
proper motion). Our plot should be compared to Figure 13 in
\citet{Metal04}, which plots local mean values of the proper motion of
background stars, calculated by combining astrometry from the
USNO-B1.0 and SDSS Data Release I. Exactly the same pattern is
displayed in both figures. Our Figure 10, however, covers a much large
area on the sky (20,000 square degrees, compared to the 3,000 square
degrees of the Munn {\it et al.} survey), and conveys a more global
picture of the kinematics of the local Galactic stars.

The patterns observed in Figure 10 are most probably a combination of
three effects. The dominant pattern appears to be the drift of the
LSR relative to the stars in the old disk and halo. The mean motion of
the standard of rest relative to the centroid
of the velocity distribution of all local Galactic stars is pointing
in the direction of galactic rotation (l=90,b=0), and there should
thus be a general drift of the background stars in a direction
opposite to the Galactic rotation (l=270,b=0). This is largely what
is observed in Figure 10, although the relative proper motions diverge
from a point that does not exactly coincide with (l=90,b=0). But other
systematic motions should be producing drifting motions in the
background stars. One is the motion of the Sun itself relative to the
stars in the local standard of rest. It should be producing a general
drifting motion pointing away from the apex of the Sun's motion. For
comparison, we plot in Figure 10 the positions of the Solar
apex/antapex, as determined by \citet{FDB01}. From Figure 10, it is
unclear how much weight this effect has. Another effect is the
rotation of the Galaxy as a whole, which should result in a rotating
motion around the north and south Galactic poles. 

A detailed model that would account for all those possible effect
would be required for a more complete interpretation of Figure 10. We
note that the information summarized in our Figure 10 can all be
recovered (possibly with even greater detail) from a simple
subtraction of the relative and absolute proper motions, which are
listed in separate columns in the LSPM catalog.

\subsection{Accuracy of LSPM proper motions}

A check on the accuracy of our proper motions is obtained from LSPM
stars that have counterparts in the UCAC2 catalog. We separate these
objects into two groups: stars that have their proper motions directly
transcribed from the TYCHO-2, and stars that have their proper
motions determined by SUPERBLINK. 

The first group includes 4,181 TYCHO-2 stars (all listed in the LSPM
catalog) that also have UCAC2 counterparts. As expected, proper
motions from the TYCHO-2 catalog are in very close agreement with the
UCAC2 proper motions. The dispersion in the difference is $2.8$
mas yr$^{-1}$ in R.A. and $2.6$ mas yr$^{-1}$ in Decl. This is
comparable to the quoted rms errors on the UCAC2 proper motions.

\begin{figure*}
\plotone{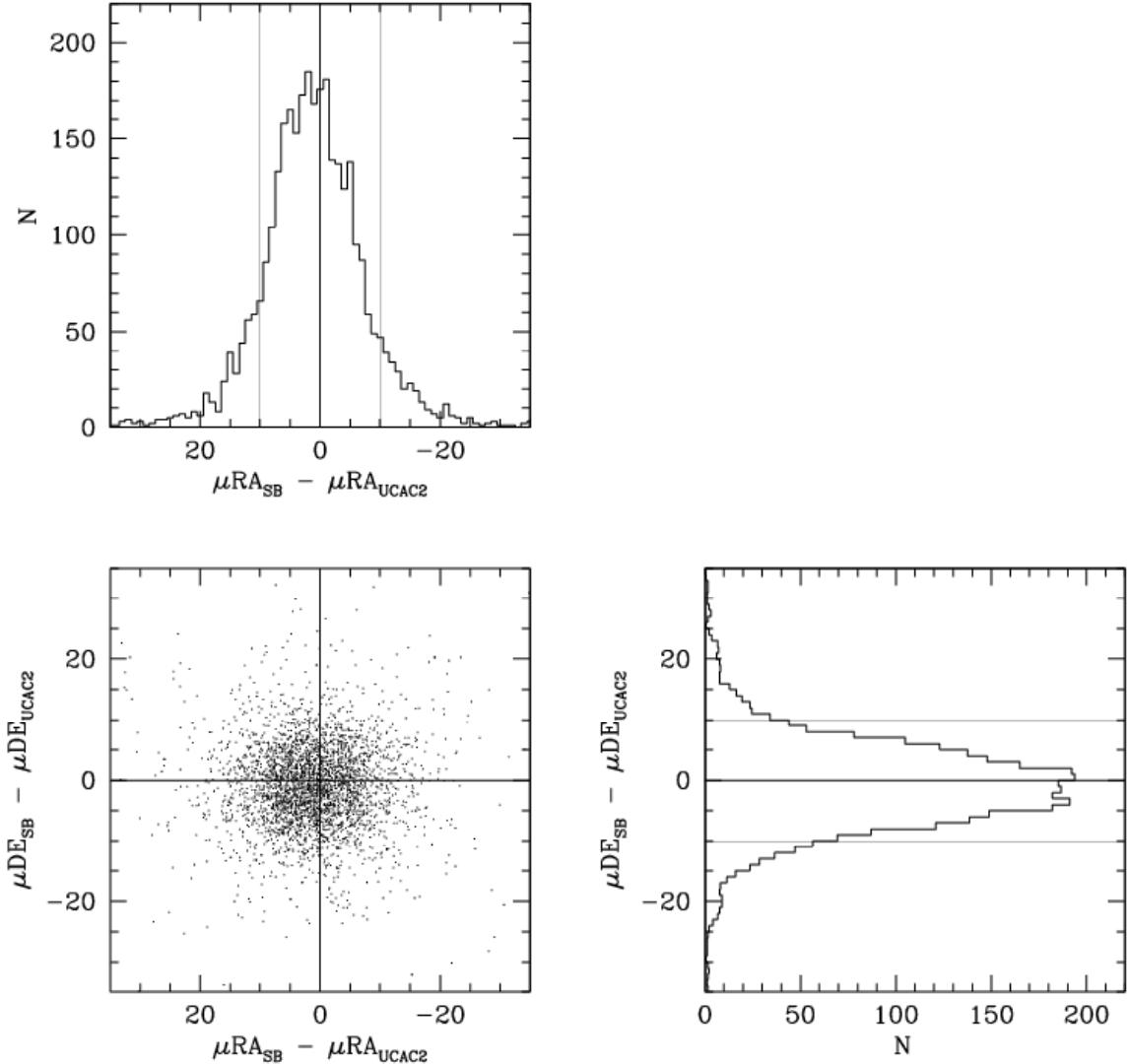}
\caption{\label{fig11} 
Difference between the SUPERBLINK and UCAC2 absolute proper
motions. The zonal-correction has been applied to the SUPERBLINK proper
motions. Only stars that have no TYCHO-2 counterparts are shown. There
is a dispersion of 7.9 mas yr$^{-1}$ in $\mu RA$ and 7.1 mas
yr$^{-1}$ in $\mu DE$. This provides an estimate of the rms errors in
the SUPERBLINK proper motions.}
\end{figure*}

\placefigure{fig11}

The second group comprises 4,380 stars from the LSPM in the magnitude
range $12<V<16$ that have UCAC2, but no TYCHO-2,
counterparts. Overall, the difference between SUPERBLINK and UCAC2
proper motions has a dispersion of $[7.5,6.7]$ mas yr$^{-1}$ in
RA and Decl. after removal of 3$\sigma$ outliers (Figure 11). There is
also a small offset of $[0.9,-0.9]$ mas yr$^{-1}$, which might
indicate a problem in the zonal corrections procedure. Our zonal
corrections are based on motions from relatively bright (TYCHO-2)
stars, while the local frames used by SUPERBLINK are largely
defined with respect to the more numerous fainter stars. Perhaps the
zonal corrections are dependent both position of the sky {\em and} 
magnitude. If bright stars are on average closer to us than fainter
background stars, then there could very well be a difference in their
systemic proper motions. In any case, since the nominal errors on the
UCAC2 proper motions are very small (1-3 mas yr$^{-1}$), the measured
$\approx$7 mas yr$^{-1}$ dispersion is a good estimate of the
SUPERBLINK astrometric errors on the proper motions.

The 3$\sigma$ outliers comprise $6\%$ of the stars in the sample, and
$2.5\%$ of the objects are beyond the $6\sigma$ limit. This
means there is an extended tail to the distribution. Indeed, 90 stars
have a difference in proper motion $>100$ mas yr$^{-1}$. Whether the
large difference arises from a faulty LSPM or UCAC2 proper motion
remains to be determined, although we do suspect that in a significant
number of cases it is the UCAC2 proper motion that is in error.

The SUPERBLINK proper motion error appears to be independent of
magnitude for stars fainter than $V=11$ (Figure 12). The proper
motions of fainter stars are perhaps marginally better, and we measure
a dispersion of $[7.3,6.3]$ mas yr$^{-1}$ at $V>15$. Astrometric
errors increase significantly for brighter ($V<11$) stars, as expected
from the fact that these are saturated on the DSS scans. This,
however, is of little consequence for the proper motions quoted in the
LSPM catalog since we use the more accurate TYCHO-2 astrometry for the
vast majority of the LSPM stars with $V<11$. The accuracy of the
SUPERBLINK astrometry is also largely independent of the proper motion
(Figure 13).

\begin{figure}
\epsscale{1.5}
\plotone{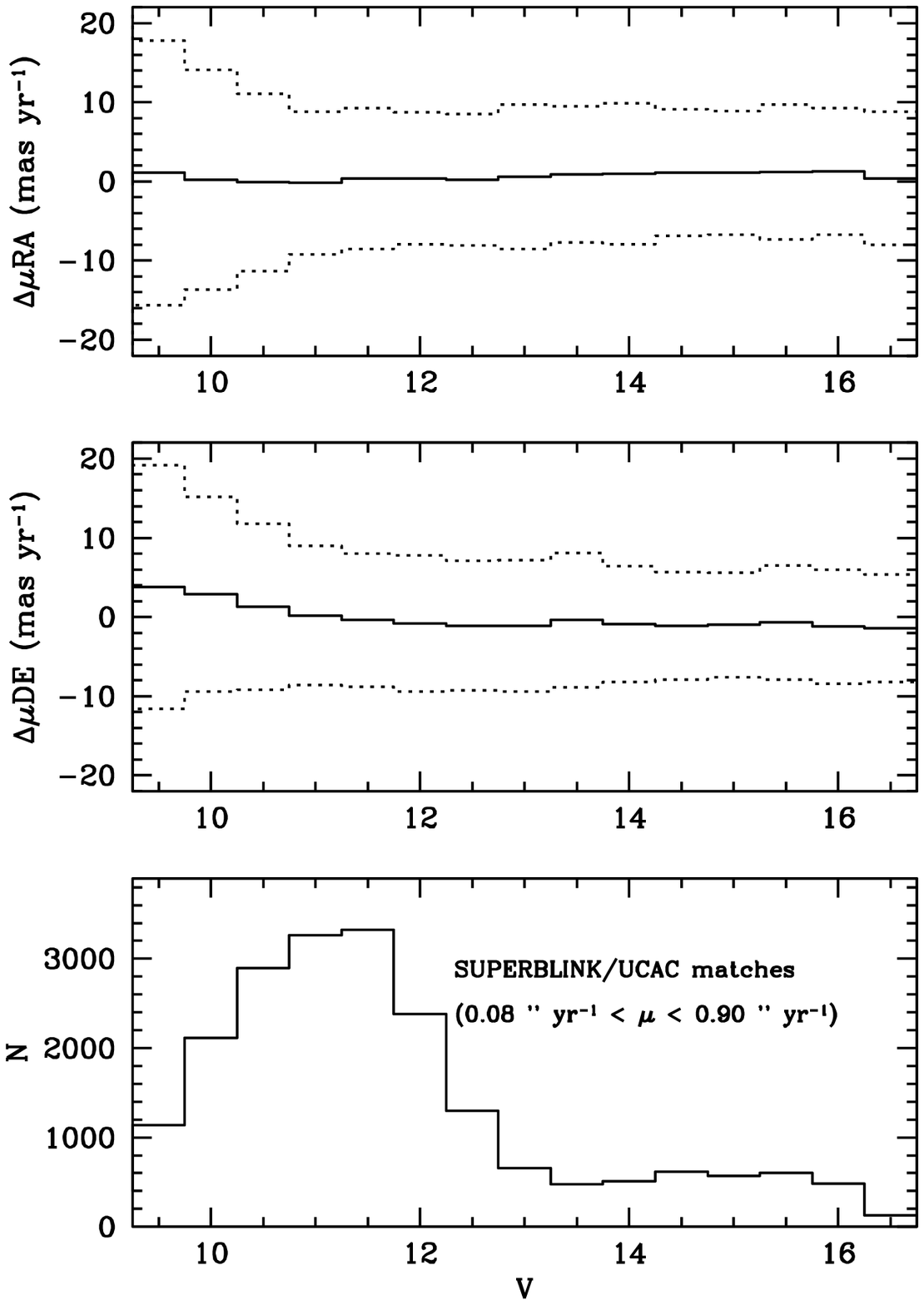}
\caption{\label{fig12} Dispersion in the difference between
SUPERBLINK and UCAC2 proper motions, as a function of magnitude. The
continuous line shows the mean of the difference, and the dotted lines
show the 1$\sigma$ dispersion. All stars with SUPERBLINK and UCAC2
proper motions have been included here, including those which also
have TYCHO-2 counterparts (to increase the sample of objects at bright
magnitudes). For stars brighter than $V=11.0$ the accuracy of the
SUPERBLINK proper motions degrades as the stars gets brighter, but the
mean errors are uniform in the $11.0<V<17.0$ range.}
\end{figure}

\placefigure{fig12}

\begin{figure}
\epsscale{1.5}
\plotone{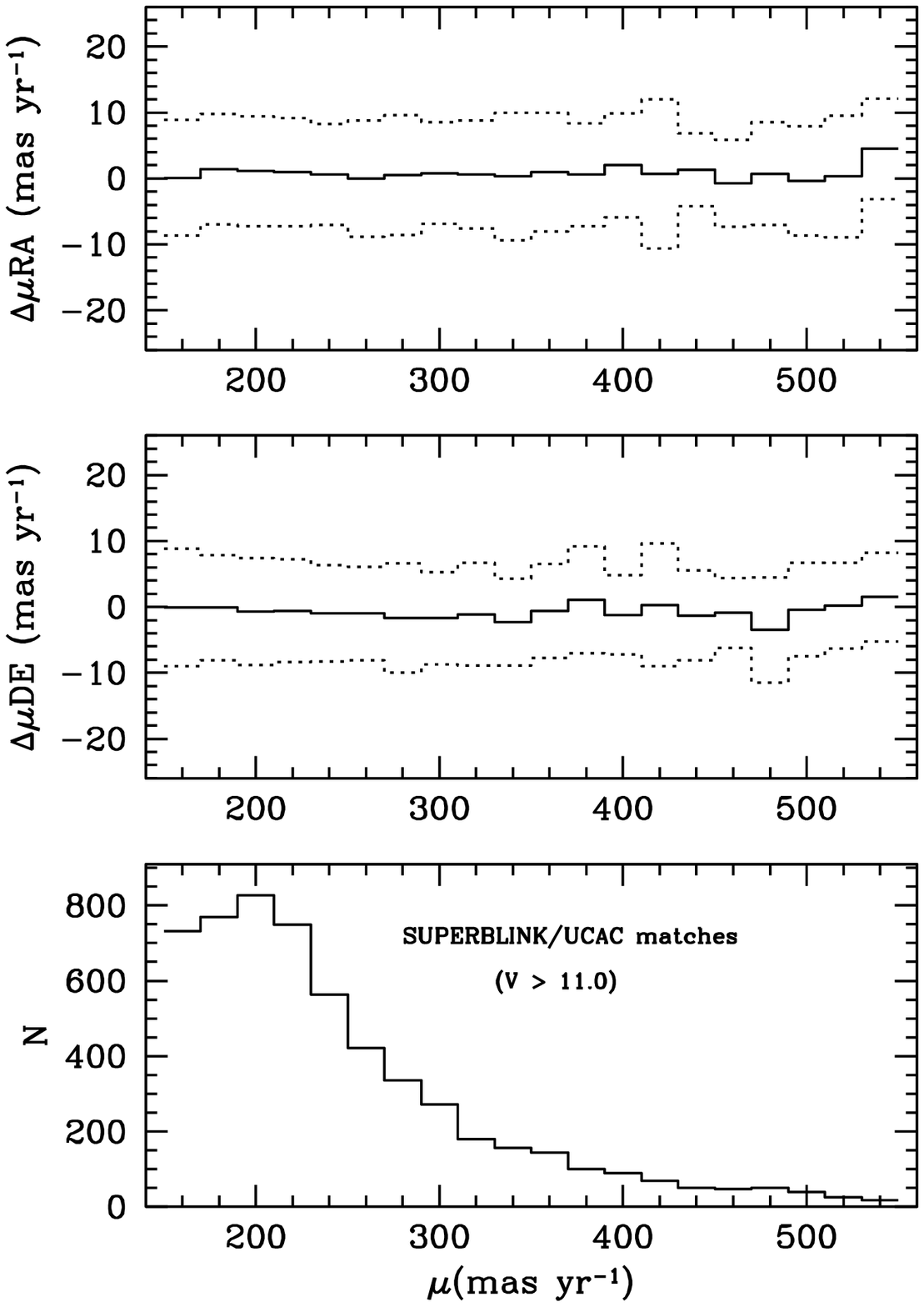}
\caption{\label{fig13} Mean and dispersion in the difference between
SUPERBLINK and UCAC2 proper motions, as a function of proper
motion. The continuous line shows the mean of the difference, and the
dotted lines show the 1$\sigma$ dispersion. The accuracy of the
SUPERBLINK proper motions is largely independent of the magnitude of
the proper motions, as demonstrated here. The fluctuations above 350
mas yr$^{-1}$ are due to small number statistics.}
\end{figure}

\placefigure{fig13}

The proper motion errors from SUPERBLINK are relatively larger than
those quoted for the revised NLTT catalog (rNLTT) of \citet{SG03},
which are claimed to be $\simeq5.5$mas yr$^{-1}$ in both RA and
Decl. It is possible that SUPERBLINK errors are slightly larger
because SUPERBLINK uses photographic plate material for both its first
and second epoch, while \citet{SG03} used data from photographic
plates only for their first epoch (USNO-A catalog, based on POSS-I)
while they used the 2MASS Second Incremental Release as their second
epoch. It is also possible that the larger errors in the SUPERBLINK
proper motions arise from small-scale fluctuations in the systemic
motions of the background stars, which could only be corrected by a
higher-resolution map of zonal corrections. A more likely possibility
is that the SUPERBLINK proper motions are affected by systematic
errors introduced by astrometric magnitude equations (see \S5.6
below).

Readers interested in having more accurate proper motions may
want to check if their star is in the rNLTT catalog \citep{SG03}. The
recovery of rNLTT proper motions is straightforward, since both the
rNLTT and LSPM catalogs provide NLTT identification numbers. One
limitation is that the rNLTT only has data for 15,899 of the northern
NLTT stars, or roughly a quarter of the LSPM stars.

Clearly, there is still room for improvement, and future efforts will
be devoted to obtain more accurate proper motion measurements, which
will be included in future versions of the LSPM catalog. The possibility
of obtaining much more accurate proper motion measurements using data
from CCD-based surveys, such as the Sloan Digital Sky Survey, was
demonstrated recently \citep{GK04,Metal04}. Careful astrometric
calibration using local quasars can yield proper motions with an
accuracy $<4$mas yr$^{-1}$. 

The goal of the current version of the LSPM is to provide the most
complete list of objects possible, with reasonable astrometric accuracy.
Though significant improvements of the proper motion errors are
possible, at least for a fraction of the LSPM stars, this will require
substantial efforts, which are beyond the scope of this paper. In any
case, our positions and proper motions are accurate enough to provide
a solid starting point for future improvements.

\subsection{Accuracy of LSPM 2000.0 positions}

The brighter LSPM stars have their 2000.0 positions extrapolated from
the 1991.25 positions and proper motions of their TYCHO-2
counterparts. Fainter stars with no TYCHO-2 counterparts have their
2000.0 positions extrapolated from the positions of their 2MASS
counterparts and their SUPERBLINK-derived absolute proper
motions.

\begin{figure*}
\plotone{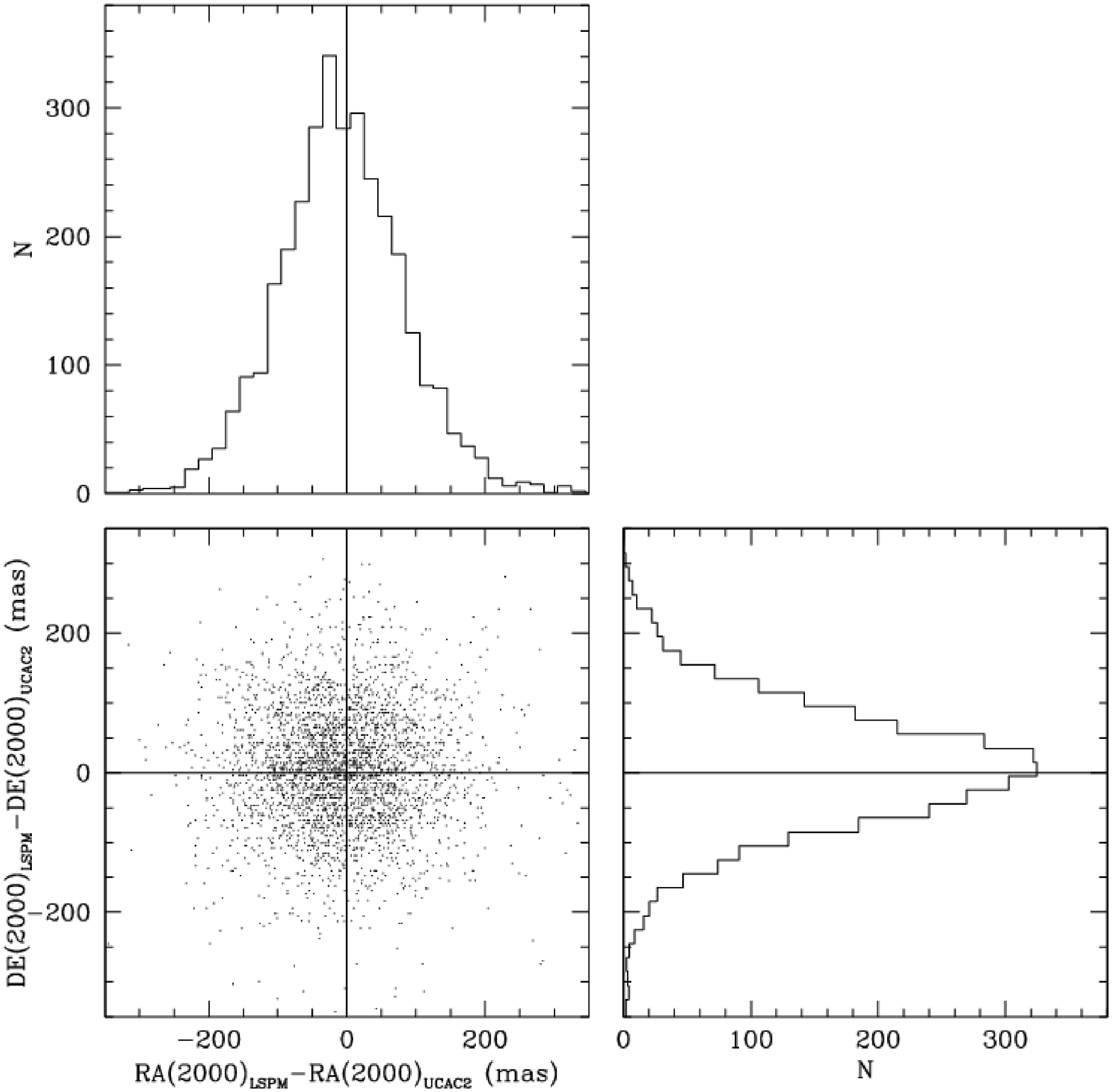}
\caption{\label{fig14}
Difference between the 2MASS-derived 2000.0 positions of 9,105 LSPM
stars with the positions of their UCAC2 catalog counterparts. The
distribution has a dispersion in $[RA,DEC]$ of $[91,88]$
mas. This provides an estimate of the accuracy of the 2000.0 epoch
positions of LSPM stars with no TYCHO-2 counterparts.}
\end{figure*}

\placefigure{fig14}

We estimate the positional accuracy of the fainter ($V>12$) LSPM
objects by comparing the SUPERBLINK-derived positions to those of the
UCAC2 catalog (see \S3.6 above), for those stars that have UCAC2
counterparts (Figure 14). The difference in position has a calculated
dispersion $[91,88]$(mas) in $[RA,DE]$. Note that UCAC2 has a
reported astrometric precision of 20-70 mas in that range of
magnitudes. There is also an offset $[-7.6,6.8]$(mas), which is
small compared to the magnitude of the dispersion, but is statistically
significant. In principle, the offset could be due to small
systematic errors in the SUPERBLINK proper motions, which are used to
extrapolate the 2MASS positions to the 2000.0 epoch. However, the
$[0.9,-0.9]$ mas yr$^{-1}$ systematic offset between the SUPERBLINK
and UCAC2 proper motions is not consistent with the $[-7.6,6.8]$(mas)
positional offset. In any case, we find that our positional errors are
marginally larger than those reported for the entire 2MASS catalog,
which have a dispersion $\approx 80$ mas relative to the UCAC2
catalog.

Most of the brighter LSPM stars ($V<12$) have their 2000.0 positions
extrapolated from the TYCHO-2/ASCC-2.5 positions (epoch 1991.25),
using the TYCHO-2/ASCC-2.5 proper motions. A comparison of the 2000.0
positions with those given in the UCAC2 catalog shows a 
dispersion of $[81,71]$(mas) in $[RA,DE]$. One must note that the
positional errors in the 2MASS catalog are significantly larger for
bright stars ($>120$ mas for $K_s<8$). The TYCHO-2 positions therefore
remain the best choice at this time.

\subsection{Comparison with the NLTT catalog}

The LSPM catalog provides new estimates of the positions and proper
motions of all northern stars in the NLTT catalog. As demonstrated by
\citet{SG03}, the NLTT positions are accurate to no better than a few
arcseconds, with some stars having positional errors of up to a few
arcminutes. The LSPM positions, which are accurate to within one
arcsecond (see above) are a significant improvement. Improved
positions for $\approx$31,000 NLTT stars are also available from
the rNLTT \citep{SG03}, but only for a little more than half the NLTT
stars in the northern sky, while our LSPM is {\em complete} for
northern NLTT stars.

The difference between the NLTT and LSPM proper motions shows a
dispersion $[20.3,18.7]$ mas yr$^{-1}$ in $[\mu RA,\mu DE]$ (after
removal of 3$\sigma$ outliers), with an offset $[1.1,4.4]$ mas
yr$^{-1}$ (Figure 15). This estimate of the NLTT errors is
comparable to the value of $\sigma\simeq20$ mas yr$^{-1}$ estimated in
\citet{SG03}, which was based on a comparison between NLTT and rNLTT
proper motions. The $[2.3,6.6]$ mas yr$^{-1}$ offset is also the
result of NLTT proper motions being relative. Indeed, if we compare
NLTT proper motions to the {\em relative proper motions} measured with
SUPERBLINK, the offset is reduced to $[0.2,-1.6]$ mas yr$^{-1}$.

\begin{figure}
\plotone{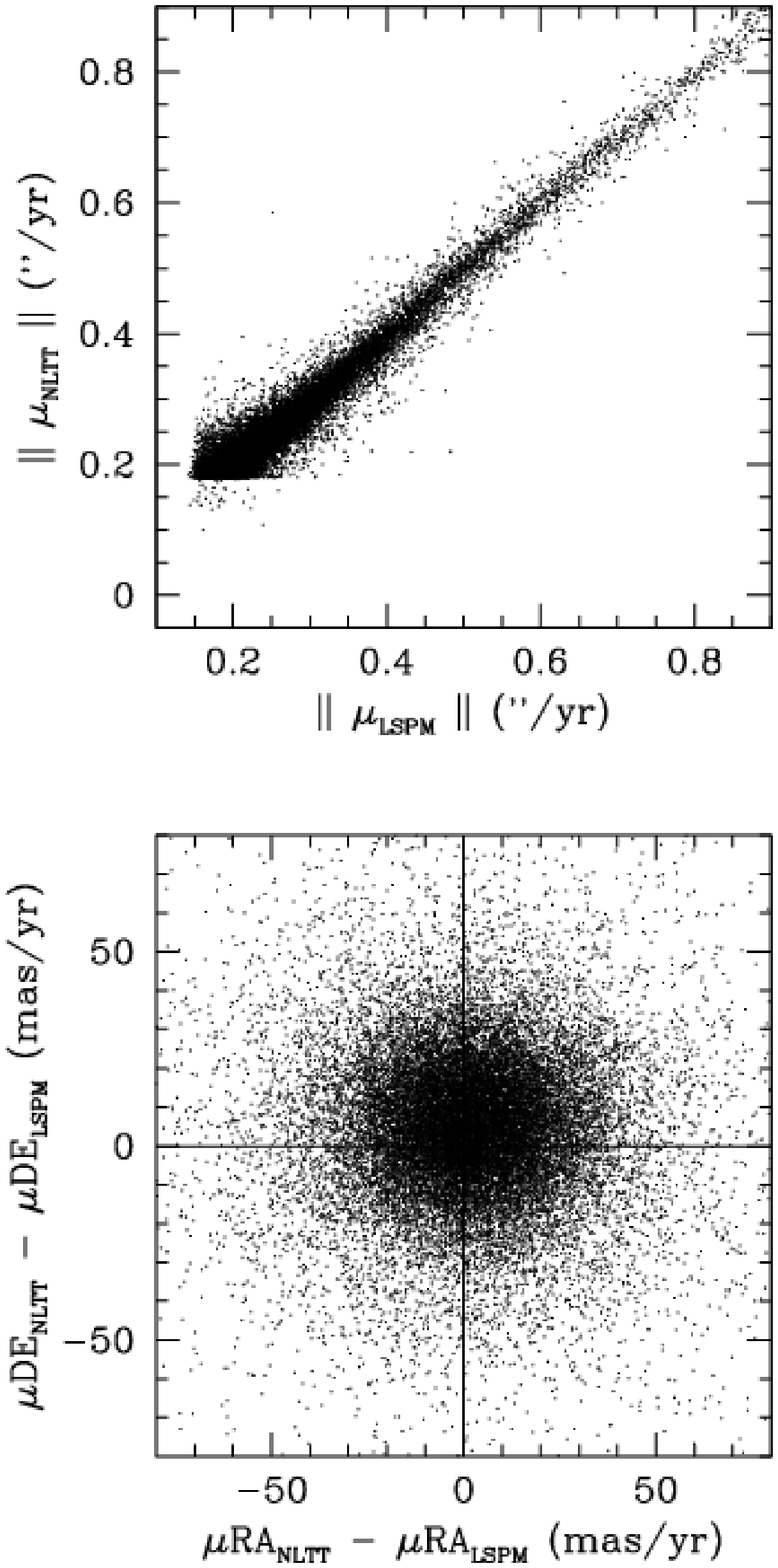}
\caption{\label{fig15}
Comparison between NLTT and LSPM proper motions, for all stars in
common between the two catalogs. Note that errors on $\mu_{LSPM}$ are
$\lesssim 8 \arcsec$ yr$^{-1}$ (see Figure 11); the measurement errors
on $\mu_{NLTT}$ are clearly larger. This accounts for the fact that
$\approx3,000$ NLTT stars have LSPM proper motions below the fiducial
limit of the NLTT catalog ($\mu=0.18\arcsec$ yr$^{-1}$).}
\end{figure}

\placefigure{fig15}

The accuracy of NLTT proper motions is naturally expected to be
larger, since the second epoch of Luyten's survey was only $\approx15$
years after the POSS-I epoch, while the second epoch of the DSS
(POSS-II) is 40 years after that of the POSS-I. The fact that the NLTT
proper motion errors are approximately three times as large as the
SUPERBLINK errors is consistent with the difference in the temporal
baseline. In any case, the LSPM proper motions are a significant
improvement over NLTT proper motions. Not only is the accuracy better
by a factor three, but the LSPM proper motions are absolute, instead of
being relative to the local background stars.

\subsection{Comparison with the USNO-B1 catalog}

Figure 16 compares LSPM proper motions with the proper motions quoted
in the USNO-B1.0. More than 75\% of the stars fall within $30$ mas
yr$^{-1}$ of the $\mu_{USNO-B1.0}=\mu_{LSPM}$ line. The remaining
objects are scattered around, with no obvious correlation with
$\mu_{LSPM}$. In particular, there are 975 stars that have a USNO-B1.0
counterpart with a proper motion of exactly zero. As discussed in
\cite{GK04} these are not stars with measured proper motions of 0.0,
but rather stars for which no significant proper motion was calculated
(to within errors) in the construction of the USNO-B1.0 catalog. In
any case, these zero-proper-motion stars were identified as low-proper
motion objects in the USNO-B1.0.

We emphasize that the high proper motion status of all LSPM stars has
been systematically confirmed by visual inspection. Ambiguous matches
with USNO-B1.0 counterpart have also all been resolved
visually. Furthermore, the comparison between LSPM and NLTT proper
motions shows a very good agreement. Thus, the only explanation for
the outliers in Figure 16 is that their USNO-B1.0 proper motion is in
error.

\begin{figure}
\plotone{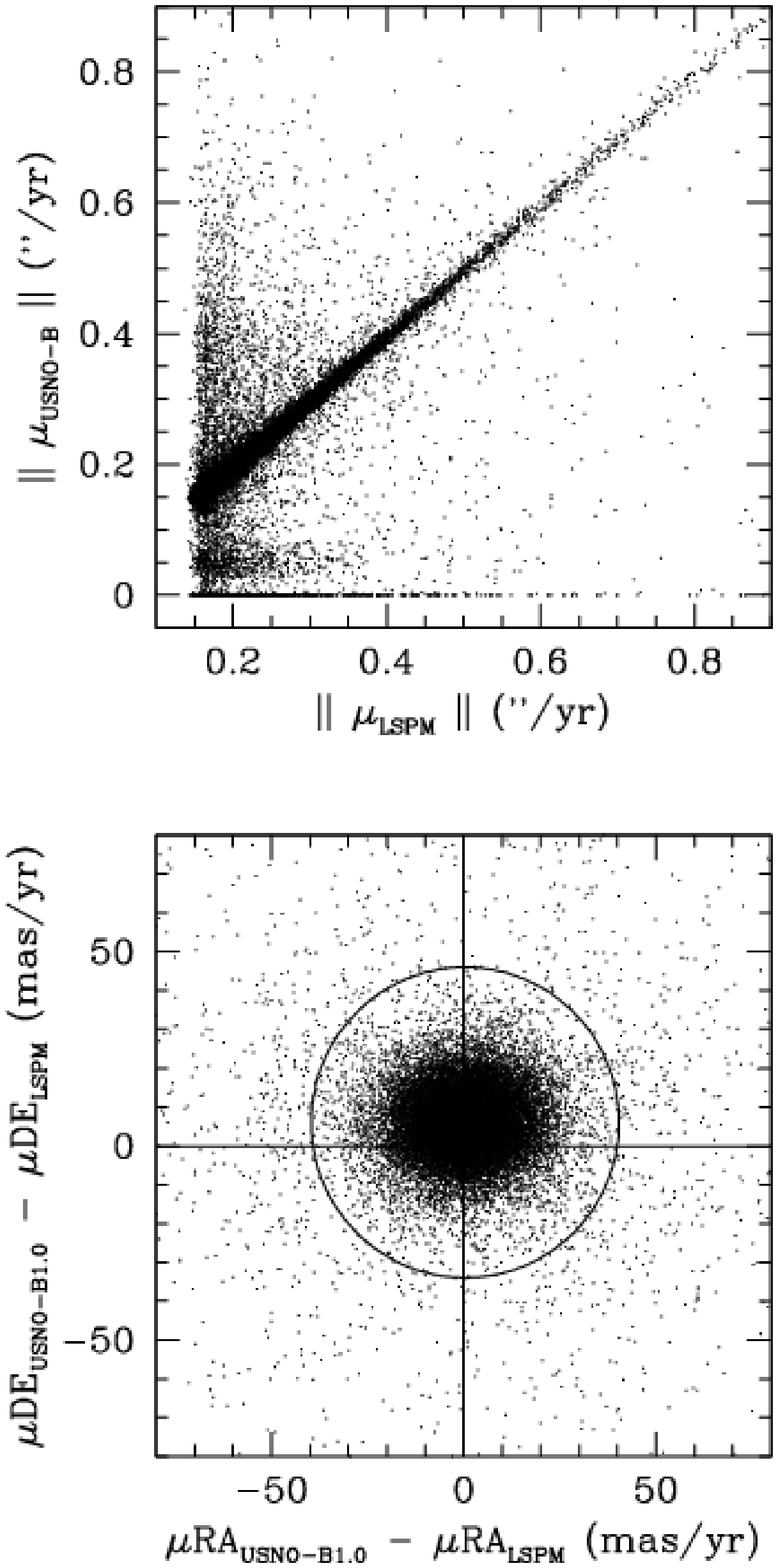}
\caption{\label{fig16}
Comparison between LSPM and USNO-B1.0 proper motions, for LSPM stars
that have USNO-B1.0 counterparts. About 75\% of the stars have
USNO-B1.0 proper motions within 0.02$\arcsec$ yr$^{-1}$ of LSPM proper
motions. Several thousand stars have $\mu_{USNO-B1.0}$ containing
large errors. The proportion of these``problem'' stars, with proper
motion errors exceeding 40 mas yr$^{-1}$ (circle in bottom plot), is
significant (see next figure). There is a mean offset in the proper
motion differences because LSPM lists absolute proper motion while
USNO-B1.0 lists relative proper motions.}
\end{figure}

\placefigure{fig16}

The completeness of the USNO-B1.0 for high proper motion stars was
investigated by \citet{G03b}, from a comparison with the rNLTT. No
USNO-B1.0 counterpart was found for $\approx10\%$ of the rNLTT stars
(``missing'' objects). About $1-2\%$ of the matched objects were also
found to have discrepant (``bad'') USNO-B1.0 proper motions. The
fraction of bad or missing stars (dubbed ``problem fraction'') was
found to be a function of both magnitude and galactic latitude. The
reason for this is that it is both more difficult to pair-up
detections of fast moving stars from different epochs, and to obtain
accurate proper motions of bright stars, which are saturated on the
POSS plates.

We calculate our own ``problem fraction'' for the USNO-B1.0, using all
single stars in the LSPM catalog as a reference. We exclude close
doubles from the analysis, as they are sometimes not resolved in the
USNO-B1.0, and so might tend to overestimate ``problem fraction''. We
calculate the ``problem fraction'' by finding all LSPM stars with no
USNO-B1.0 counterpart (``missing''), or with a USNO-B1.0 counterpart
that has $||\vec\mu_{LSPM}-\vec\mu_{USNO-B1.0}||>$40 mas yr$^{-1}$
(``bad''). We find that most of the problem stars are not ``missing''
objects, like in the \citet{G03b} analysis, but are mostly ($>95\%$)
``bad'' counterparts. Why this difference? \citet{G03b} matched rNLTT
stars to USNO-B1.0 objects within a $5\arcsec$ radius, while our own
search radius was much larger (up to 1$\arcmin$). We have thus
recovered most of those ``missing'' USNO-B1.0 counterparts. It appears
that those stars have large position errors in the USNO-B1.0 because
they also have large proper motion errors. The two values (position
and proper motion) are linked, because the extrapolated 2000.0
positions are dependent on a good estimate of the proper motion. In
other words, the counterparts were ``missing'' because of their
very ``bad'' recorded proper motion. This is why our USNO-B1.0
``problem'' stars are comprised of mostly ``bad'' counterparts.
 
\begin{figure}
\epsscale{2.2}
\plotone{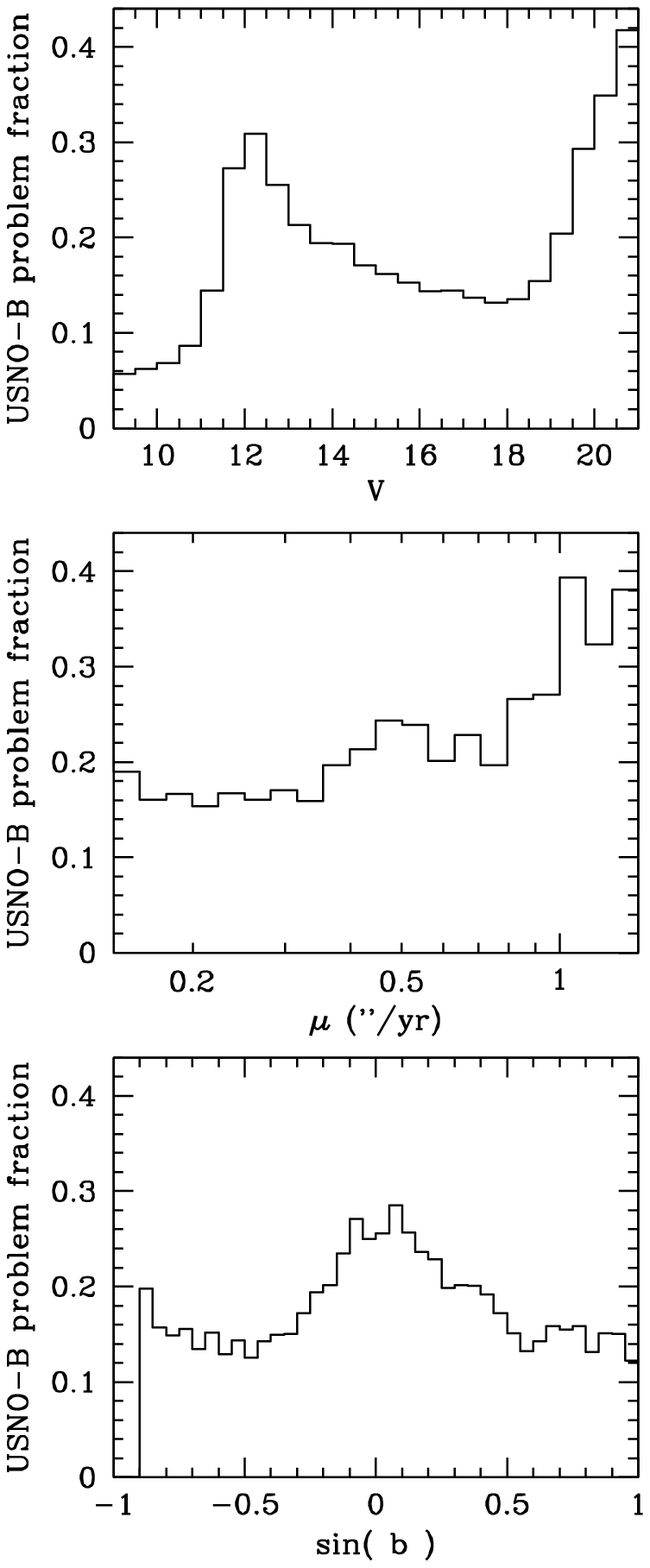}
\caption{\label{fig17}
Fraction of stars in the LSPM catalog with no recorded USNO-B1.0
counterpart, or with a large error in the USNO-B1.0 proper motion
(as compared with the SUPERBLINK proper motion). Close binary stars
are excluded from the analysis, as they may not be resolved in
USNO-B1.0. The calculated USNO-B1.0 ``problem fraction'' is lowest for
moderately faint stars ($14<V<19$) with low proper motions
($\mu<0.3\arcsec$ yr$^{-1}$) at high galactic latitudes ($|b|>20$),
for which it is $\approx10\%$.}
\end{figure}

\placefigure{fig17}

In Figure 17, we plot our USNO-B1.0 problem fraction as a function of
magnitude, proper motion, and galactic latitude. A comparison with
photographic $V$ magnitude shows that large proper motions errors are
more common for USNO-B1.0 counterparts with $11<V<13$ and
$V>19$. The excellent agreement between the USNO-B1.0 and LSPM for the
very brightest stars ($V<10$) reflects the fact that both the LSPM and
USNO-B1.0 use TYCHO-2 positions and proper motions in that range.

Following \citet{G03b}, we interpret the larger problem fraction of
the brighter stars from the fact that these objects are saturated on
the POSS plates, making proper motion determinations
prone to large errors. The large problem fraction of faint ($V>19$)
stars is simply explained by the fact that it is much more difficult
to pair-up faint objects detected in different epochs. This increase
in the problem fraction near the faint star limit doesn't show up in
the \citet{G03b} analysis, because the analysis was restricted to stars
with counterparts in the USNO-A catalog, which has a brighter faint
magnitude limit.

Large proper motion errors on USNO-B1.0 counterparts are also more
frequent for stars with larger proper motions ($\mu_{LSPM}>0.3\arcsec$
yr$^{-1}$), confirming again the analysis of \citet{G03b}. It is
simply more difficult to pair-up stars that have moved very
substantially between different photographic survey epochs. Finally,
crowding is also a major source of confusion, leading to erroneous
USNO-B1.0 proper motions. This is evidenced by the increase in the
problem fraction at low galactic latitude. This is much more
significant than the low-galactic latitude problem fraction of
\citet{G03b}, which was high only because the low-galactic latitude
rNLTT stars contain a larger proportion of bright, saturated
objects. The high problem fraction at low galactic latitude observed
by \citet{G03b} was thus more the result of those stars being dominated
by brighter, saturated objects. Our problem fraction is a more correct
assessment of the completeness of the USNO-B1.0 at low galactic
latitude: the problem fraction reaches $20\%$ near the galactic plane.

The conclusion is that the USNO-B1.0 catalog is at best approximately
$90\%$ complete and accurate at high galactic latitude, for stars with
$\mu<0.5\arcsec$ yr$^{-1}$ and $14<V<19$. Otherwise the completeness
falls to $70\%$. The LSPM is significantly more complete in high
proper motion stars (see \S 7 below). 

\subsection{SUPERBLINK proper motion accuracy and the astrometric
magnitude equations}

The SUPERBLINK proper motions are derived from measurements of
photographic Schmidt plates (POSS-I, POSS-II). One important issue
with these plates is that the measured positions of stars have a
weak dependence on magnitude. The non-linear response of photographic
plates combined with asymmetric stellar images (from, e.g.,
imperfections in the optical system, or inaccurate guiding) causes the
photographic image centroids of the stars to be slightly offset
compared to that on a linear detector. Because they arise from the
non-linear response of the substrate, these offsets are a function of
the magnitude of the star, and they are generally largest for bright,
saturated objects. These offsets, which are also generally dependent
on the position on the plate, are referred to as the ``astrometric
magnitude equations''. Accurate astrometric measurements with
photographic plates require the determination and application of a
magnitude equation correction \citep{G98}.

For the POSS plates, which we use in our SUPERBLINK survey,
\citet{Metal03} have shown the existence of fixed-pattern astrometric
offsets which are on the order of $0.1-0.5\arcsec$ and are strongly
correlated with the XY position on the plate. Plotted for stars of
different magnitudes, these patterns show striking differences, which
indicate the existence of significant astrometric magnitude
equations. The offsets tend to be larger at bright magnitudes, and are
also larger near the edges of the plates. However, the patterns are
relatively regular, and much of the variability occurs on scales
equivalent to at least several minutes of arc.

As described in \S2.1, the SUPERBLINK software intrinsically corrects
for large-scale plate distortions by measuring the relative proper
motions of stars {\em in a frame defined by the local background of
objects}, typically all stars within about $7\arcmin$ of the high
proper motion target. Because SUPERBLINK uses the local backdrop of
{\em stars} to calculate relative proper motions, and not the XY
plate positions, distortions on the photographic plate on scales
$\gtrsim7\arcmin$ do not introduce any significant error on the
SUPERBLINK proper motions. This however, is generally true only if all
the stars (target and background) are locally offset by the same
amount. For example, if the local SUPERBLINK reference frame is
defined by 16th magnitude stars, then the measured relative proper
motions of 16th magnitude stars will not be affected by the
astrometric magnitude equations (since the target and the reference
stars are all offset by the same value). This, of course, is true even
if the value of the offset is different on the first and second epoch
images.

On the other hand, if a proper motion target is significantly brighter
or fainter than the background (reference) stars, the astrometric
magnitude equations will generally introduce systematic errors in the
SUPERBLINK proper motions. One exception to this case is if the {\em
differences} in the offsets between the target and the reference stars
{\em are the same in both the first and second epoch images}. This may
happen, e.g., if the object was recorded at the same plate position
(X,Y) at both of the first and second epoch. Unfortunately, this is
generally not the case for POSS-I and POSS-II, since their plate
centers are on different grids.

Because the systematic proper motion errors introduced by the
astrometric magnitude equations depend on the difference between the
magnitude of the target and the magnitude of the local background
stars, their effect is very difficult to model. One would need to
estimate the local mean offset of the background stars, which have a
variable range of magnitudes, and use estimates of the offsets
expected for the target star given its magnitude. One would need to
make these estimates separately for the first and second epoch images.
Such a procedure would be extremely complex. We have thus made no
attempt to correct the SUPERBLINK proper motions for the effects of
the astrometric magnitude equations. It is very possible that the
larger errors on the SUPERBLINK proper motions ($\approx8$ mas
yr$^{-1}$) compared to the proper motion errors from the rNLTT
($\approx5.5$ mas yr$^{-1}$) are due to the fact that astrometric
magnitude equations have been neglected in SUPERBLINK proper motion
calculations.

\section{The catalog}

\subsection{Format}

The complete catalog in ASCII format is available with the electronic
edition of this paper. The catalog contains 61,977 lines, each 286
characters long. Each catalog entry consists of 29 fields; these are
described in Table 1.

The first field gives the LSPM catalog name. The
next 9 fields provide identifications in the LHS, NLTT, Hipparcos,
Tycho-2, ASCC-2.5, UCAC-2, 2MASS, USNO-B1 catalogs, when these exist
for the star. An additional field gives the original name of the star
in the published literature (e.g. the LSR stars of L\'epine, Shara, \&
Rich). In the current version of the LSPM catalog, however, the
original name is provided {\em only if the star doesn't have a
counterpart in any of the catalogs listed above}. At this point, it is
provided as a means to distinguish ``rediscovered'' LSPM stars from
those that are genuine, ''new'' discoveries. 

\begin{deluxetable}{lllr}
\tabletypesize{\footnotesize}
\tablecolumns{4}
\tablewidth{0pc}
\tablecaption{The LSPM Catalog - field description}
\tablehead{field & datum & units & format }
\startdata
1 & LSPM catalog name       & \nodata               & a16 \\
2 & LHS catalog ID	    & \nodata		    &  a6 \\
3 & NLTT catalog ID	    & \nodata		    &  a6 \\
4 & Hipparcos catalog ID    & \nodata		    &  a7 \\
5 & Tycho-2 catalog ID	    & \nodata		    & a12 \\
6 & ASCC-2.5 catalog ID	    & \nodata		    &  a8 \\
7 & UCAC-2 catalog ID	    & \nodata		    &  a9 \\
8 & Other name 	            & \nodata		    & a31 \\
9 & 2MASS catalog ID	    & \nodata		    & a17 \\
10& USNO-B1 catalog ID	    & \nodata		    & a13 \\
11& R.A.		    & degrees		    & f12.6\\
12& Decl.		    & degrees		    & f11.6\\
13& total relative proper motion     & $\arcsec$ yr$^{-1}$   & f8.3\\
14& relative proper motion in R.A.   & $\arcsec$ yr$^{-1}$   & f8.3\\
15& relative proper motion in Decl.  & $\arcsec$ yr$^{-1}$   & f8.3\\
16& total absolute proper motion     & $\arcsec$ yr$^{-1}$   & f8.3\\
17& absolute proper motion in R.A.   & $\arcsec$ yr$^{-1}$   & f8.3\\
18& absolute proper motion in Decl.  & $\arcsec$ yr$^{-1}$   & f8.3\\
19& astrometric source flag& \nodata		    & a2\\
20& optical $B$ magnitude          & mag            & f6.2\\
21& optical $V$ magnitude          & mag            & f6.2\\
22& photographic blue (B$_{\rm J}$)& mag            & f5.1\\
23& photographic red (R$_{\rm F}$)& mag		    & f5.1\\
24& photographic near-IR (I$_{\rm N}$)& mag         & f5.1\\
25& infra-red J		    & mag		    & f6.2\\
26& infra-red H		    & mag		    & f6.2\\
27& infra-red K$_s$         & mag		    & f6.2\\
28& estimated $V$ magnitude & mag                   & f7.2\\
29& estimated $V-J$ color   & mag                   & f6.2\\
\enddata
\end{deluxetable}

The astrometric information (position, proper motion) is detailed in
the next 9 fields, and includes a flag that gives the origin of the
astrometry. The next 8 fields provide the photometric information,
with optical, photographic, and infrared magnitudes. The last two
fields give the estimated $V$ magnitude and $V-J$ color index.

A sample of the catalog is shown in Tables 2 and 3, in which the first
12 lines are displayed as an example. The full catalog is available
only in electronic format.

\subsection{Names and identifications}

We assign a LSPM name to each star in our catalog, which is based on
the star's R.A. and Decl. at 2000.0 epoch in the International
Celestial Reference System (ICRS, essentially equivalent to
J2000). The first four characters (``LSPM'') are the catalog
identifiers, and stand for Lepine \& Shara Proper Motion. A space then
separates the catalog ID from the positional description. A ``J''
follows, which indicates the equinox of the position. The next four
digits are the hours and minutes of the R.A., then comes the sign of
the Decl. (``+'' for all our stars, since we do not have southern
declinations \--- yet) followed by 4 more digits that represent the
degrees and minutes of Decl. Finally, there is one last character used
to distinguish stars that would otherwise have the same name. Pairs of
stars with the same hours/minutes in R.A. and degrees/minutes in
Decl. have their names appended with an ``N'', ``S'', ``E'', or ``W''
suffix. The choice of suffix depend on the orientation of the
pair. Their separation in both R.A. and Decl. is determined. If the
separation in Decl. is larger, then the stars are given a N/S suffix,
with N (``North'') assigned to the star at higher declination, and S
(``South'') to the other star. If it is the separation in R.A. that is
the largest, then E/W suffixes are used, with E (``East'') assigned to
the star at larger R.A., and W (``West'') to the other one. By no
means are stars with ''NS'' or ''EW'' suffixes necessarily common
proper motion doubles. While this is often the case, there are a
number of chance alignments for which two unrelated high proper motion
stars happen to be in the same arcminute position bin. Conversely, not
all common proper motion doubles have ''NSEW'' suffixes, since it is
often the case that long-period doubles have angular separations large
enough to put them in separate arcminute bins.

Identifications are given for the 2,572 LSPM stars also listed in the
LHS catalog and for the 31,361 stars listed the NLTT. The identification
number for stars listed in the LHS catalog is their LHS \#, which has
been traditionally used in the literature. The identifications for
stars listed in the NLTT catalog are the record \# in the original
NLTT table (``recno'' in the electronic version of the NLTT catalog,
at the VizieR catalog
service\footnote{http://vizier.u-strasbg.fr/}).

We also provide identifiers for the 4,839 stars listed in the
Hipparcos catalog (HIPP number), as well as for 7,943 stars with a
TYCHO catalog number. A total of 4,306 of the HIPP stars also have
data in the TYCHO-2 catalog. We give ASCC-2.5 identification for a
total of 11,430 stars; these include all the TYCHO-2 stars as well as
the HIPP stars that are not in the TYCHO-2. We give UCAC2 catalog
numbers for the 9,137 LSPM stars that are listed in the UCAC2
catalog. Note that since the bulk of the LSPM stars are fainter than
the magnitude limits of these catalogs, the majority of LSPM stars do
not have HIPP, TYCHO-2, ASCC-2.5 or UCAC2 identifiers. 

A few hundred LSPM stars are not listed in any of the catalogs listed
above, but are not entirely ``new'' objects because they have been
previously reported in the literature. An additional column provides
the Simbad\footnote{http://simbad.u-strasbg.fr/} designation for those
objects. About a third of these are the ``LSR'' high proper motion
stars found by our team \citep{LSR02,LSR03}. The other stars are a mix
of objects, some from old proper motion motion catalogs (but that had
not been included in the NLTT or LHS), some identified as M dwarfs or
white dwarfs in field spectroscopic surveys. In particular, stars with
2MASS designations are objects identified recently in various searches
of ultra-cool M and L dwarfs.

Finally, we give identifications for the counterparts of the LSPM
stars in the 2MASS All-Sky Point Source Catalog and in the USNO-B1.0.
Catalog identifiers provided in the LSPM make it easy to retrieve all
relevant information from those catalogs.

\subsection{Positions and proper motions}

Positions in the LSPM catalog are given at the 2000.0 epoch in the
ICRS system, and for the great majority are obtained either by
extrapolating from the TYCHO-2 position using the TYCHO-2 proper
motion (for stars with TYCHO-2 counterparts), or by extrapolating from
the position of the 2MASS counterpart, using the SUPERBLINK-derived
absolute proper motion or the ASCC-2.5 proper motion (see below for
which proper motion is used).

For stars with no 2MASS counterpart (2,345 objects) we used the
coordinates calculated by SUPERBLINK from the DSS scans instead. The
positions of those stars, extrapolated from the position of the star
on the POSS-II scan, are much less accurate than the 2MASS-derived
positions. Unfortunately, the accuracy of the SUPERBLINK-derived J2000
positions is only $\approx0.5\arcsec$ (as estimated from a comparison
with the 2MASS catalog), and future efforts will be devoted to
obtaining more accurate coordinates for those objects.

In the catalog, we list the relative and absolute proper motions in
separate columns. The relative proper motions are always those
determined by SUPERBLINK. This makes it easy to identify stars that
have not been measured with SUPERBLINK: values for their relative
proper motion are set to zero.

For the absolute proper motion, we use either the value derived from
SUPERBLINK, or quote the absolute proper motion from the TYCHO-2 or
ASCC-2.5 catalogs. The order of priority is as follows: (1) TYCHO-2
proper motions, (2) SUPERBLINK proper motion, (3) ASCC-2.5 proper
motion. The quoted TYCHO-2 proper motion errors are $<8$ mas
yr$^{-1}$, for all LSPM stars with a TYCHO-2 counterpart. This
is smaller than the estimated SUPERBLINK proper motion errors, and
justifies that we always defer to the TYCHO-2 proper motion. The
quoted proper motion errors for ASCC-2.5 stars that do not have
TYCHO-2 counterparts are generally $>12$ mas yr$^{-1}$ with a mean
value $\simeq14.5$ mas yr$^{-1}$, larger than the SUPERBLINK errors,
so we use the ASCC-2.5 proper motions only for those stars that have
no SUPERBLINK proper motions. The source of the proper motion is
indicated by the astrometric flag: ``T'' if the TYCHO-2 proper motion
is used, ``S'' if it is the SUPERBLINK-derived proper motion, and
``A'' if the ASCC-2.5 proper motion is quoted.

A total of 508 stars in the LSPM catalog have no TYCHO-2 counterparts
nor ASCC-2.5 counterparts, and their proper motions have not been
measured by SUPERBLINK either. For those objects, we obtain their
proper motion from a variety of sources, mainly from the NLTT and
rNLTT catalogs, and also from the catalog of revised proper motions of
LHS stars of \citet{BSN02}. These objects are denoted by the
astrometric flag ``O''.

Note that for LSPM stars with TYCHO-2 counterparts that have also been
measured with SUPERBLINK, values for the absolute proper motion are
quoted from the TYCHO-2 catalog, but values for the relative proper
motions are those from SUPERBLINK. Note also that zonal corrections
are calculated individually for each star, using all TYCHO-2 objects
within 7$^{\circ}$ of that star. Our map of the zonal corrections can
thus be recovered from the LSPM catalog, by differencing the relative and
absolute proper motions for stars that have their absolute proper
motions derived from the SUPERBLINK values.

\subsection{Magnitudes}

The LSPM catalog lists optical B, V magnitude from the TYCHO-2 and
ASCC-2.5 catalogs. It also gives the photographic blue ($B_J$), red
($R_F$), and near infrared ($I_N$) magnitudes extracted from the
USNO-B1.0 catalog. Stars with no USNO-B1 counterparts are listed with
B$_{\rm J}$ and I$_{\rm N}$ magnitudes of 99.9, indicating these to be
unavailable. When a value for R$_{\rm F}$ could not be obtained from
the USNO-B1.0 catalog, we used the value estimated by SUPERBLINK from
the POSS-II red DSS scans. A value of 99.9 is also used whenever there
is only partial magnitude information from the USNO-B1 counterpart (or
counterparts, see \S3.8 above).

Infrared magnitudes for LSPM stars are obtained from their
counterparts in the 2MASS All-Sky Point Source Catalog
\citep{C03}. The accuracy is about 0.02 mag for $5<J<14$, $5<H<14$,
$4<K_s<13$; it is $\approx0.25$ mags for brighter stars (saturated on
the 2MASS images). For fainter stars, the uncertainty increases with
magnitude. The 2MASS catalog is complete to J$\simeq$16.5,
H$\simeq$16.0, and K$_s\simeq15.5$ \footnote{Detailed documentation
can be found at http://www.ipac.caltech.edu/2mass/}.

Finally, estimated $V$ magnitudes and $V-J$ colors are given in the
last two columns. Values are provided for all but a few entries. These
should be used for quick reference, and for classification of the high
proper motion stars. See \S4.4 for the caveats in using these
estimated values.

\section{Completeness of the LSPM catalog}

\subsection{Comparison with the NLTT catalog}

\begin{figure*}
\plotone{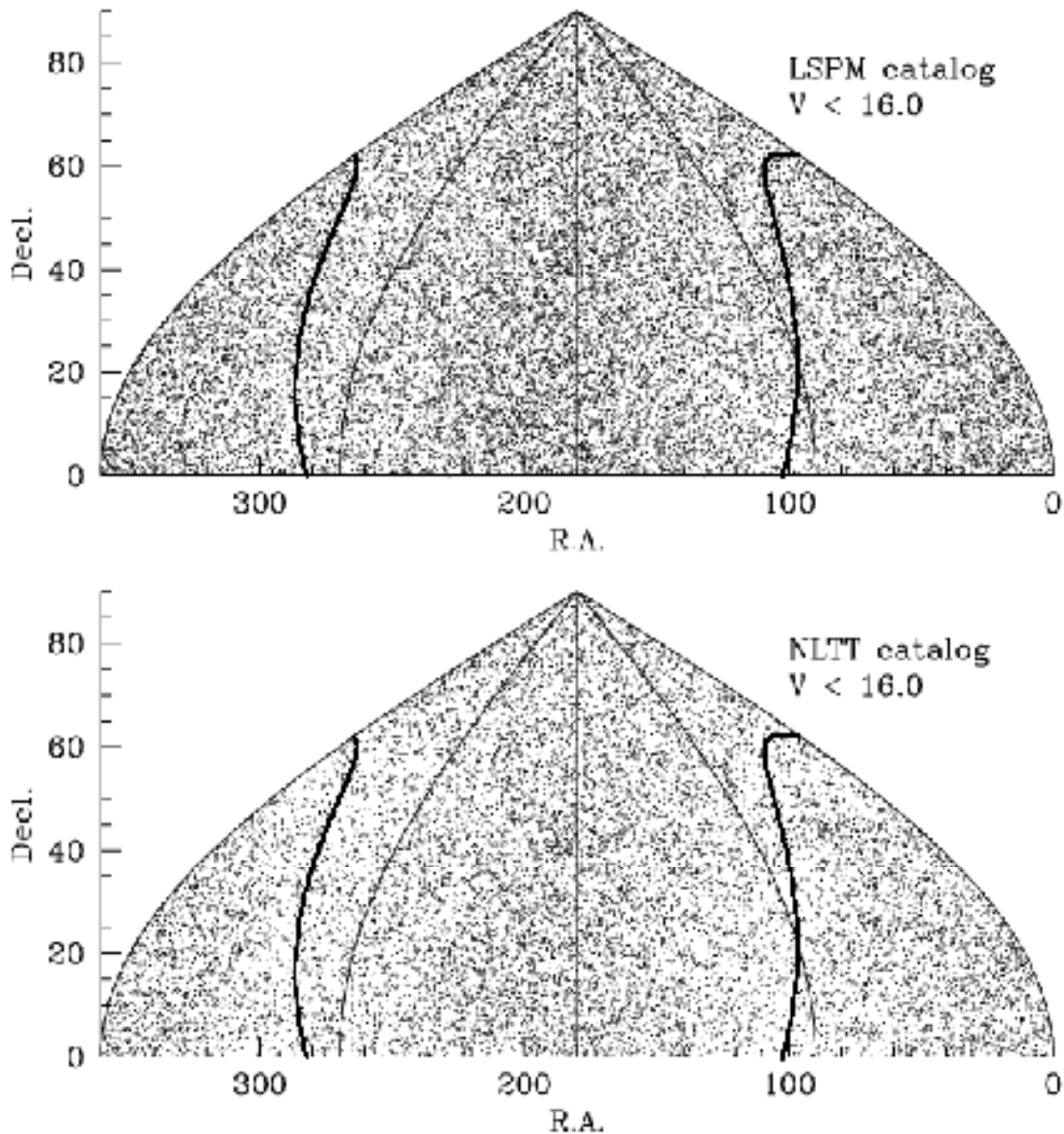}
\caption{\label{fig18}
Distribution of high proper motion stars from our LSPM catalog (top)
compared to the distribution of stars from the NLTT (bottom). Shown
here are stars brighter than $V=16.0$ (see Figure 19 for the
distribution of fainter stars). The Galactic equator is shown (thick
line). Note that even though both catalogs are expected to be
relatively complete ($>90\%$) in that range, it is obvious that the
density of objects is larger at high galactic latitudes. This suggests
that proper motion selected samples are intrinsically
non-uniform. Indeed, one does expect proper motion surveys to be more
sensitive to old disk and halo stars at high galactic, because of the
asymmetric drift.}
\end{figure*}

\placefigure{fig18}

\begin{figure*}
\plotone{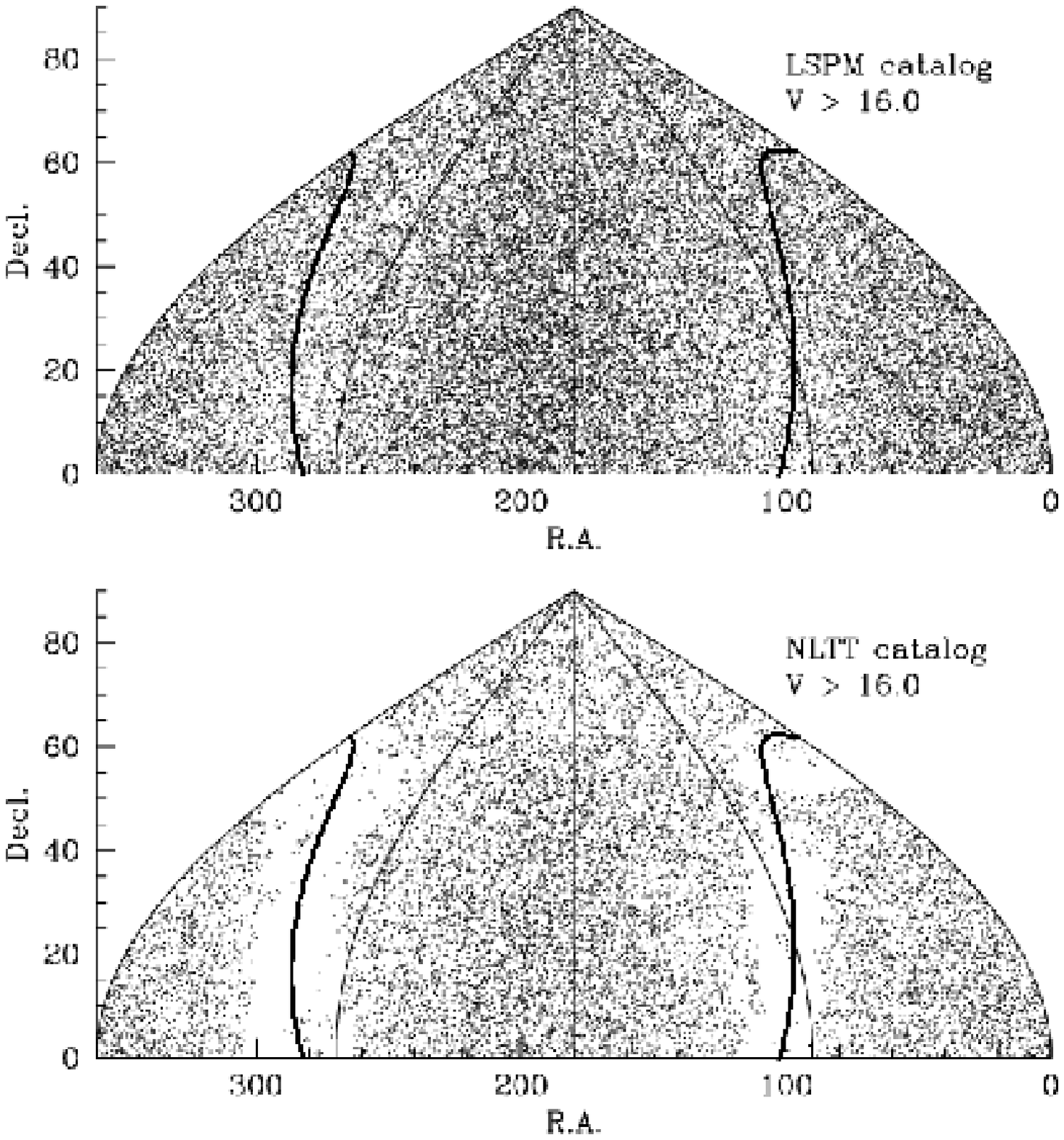}
\caption{\label{fig19}
Same as Figure 18, this time showing stars fainter than $V=16.0$. The
NLTT is dramatically incomplete in low-galactic latitudes. This
situation is much improved with the LSPM catalog. One still observes a
lower density of high proper motion stars at low galactic latitudes.}
\end{figure*}

\placefigure{fig19}

The completeness of a proper motion catalog can be a function of
magnitude, proper motion, and position. The completeness is generally
dependent on how easy it is to identify moving objects against the
backdrop of the ``fixed'' stars. Detection will be more difficult if
the star is fainter, but also if it is moving faster, or if the local
density of background objects is larger. Traditionally, proper motion
surveys have been very incomplete for faint stars at low Galactic
latitudes. As described above, our SUPERBLINK software, with its image
subtraction algorithm, was specifically designed to address this
problem, and detect moving stars in densely populated areas. We
therefore expect the LSPM catalog to be significantly more complete
than the NLTT at low galactic latitudes. The main question is how much
more complete the LSPM is. In particular, we would like to know
whether the LSPM catalog still suffers from some incompleteness near
the Galactic plane.

We first compare the distribution of NLTT and LSPM stars as a function
of position, separating the stars into two groups: stars
brighter than $V=16$, and stars fainter than $V=16$. Figure 18 shows
the distribution of brighter stars. The distributions for both catalogs
are relatively uniform (no sharp discontinuities, or
``holes''). However, the density of objects still appears to be
non-uniform, and is larger in high Galactic latitudes. However, the
decline is very gradual, in sharp contrast to the distributions of
faint NLTT stars, shown in Figure 19. The density of faint NLTT
objects falls very abruptly at low Galactic latitudes: a clear
mark of true incompleteness in the NLTT catalog. On the other hand,
the distribution of faint LSPM objects follows more or less the same
trends as the distribution of bright high proper motion stars.

Two interpretations are possible. One is to say that the distribution
of high proper motion stars should be uniform across the sky, and that
it is the completeness of the catalog that progressively diminishes as
one goes to lower galactic latitudes. With this interpretation, both
the NLTT and LSPM are significantly incomplete, even at moderately
high Galactic latitudes ($20<|b|<60$). The other interpretation, and
the one we favor, is that the distribution of proper motion selected
objects is naturally non-uniform over the sky, and that the
progressive density variations observed in Figure 18, and in the top
panel of Figure 19 have little to do with completeness. Under this
second interpretation, the completeness of the LSPM catalog is
uniformly high, both at high and low galactic latitude. We demonstrate
the truth of this statement as follows.

\begin{figure}
\epsscale{2.25}
\plotone{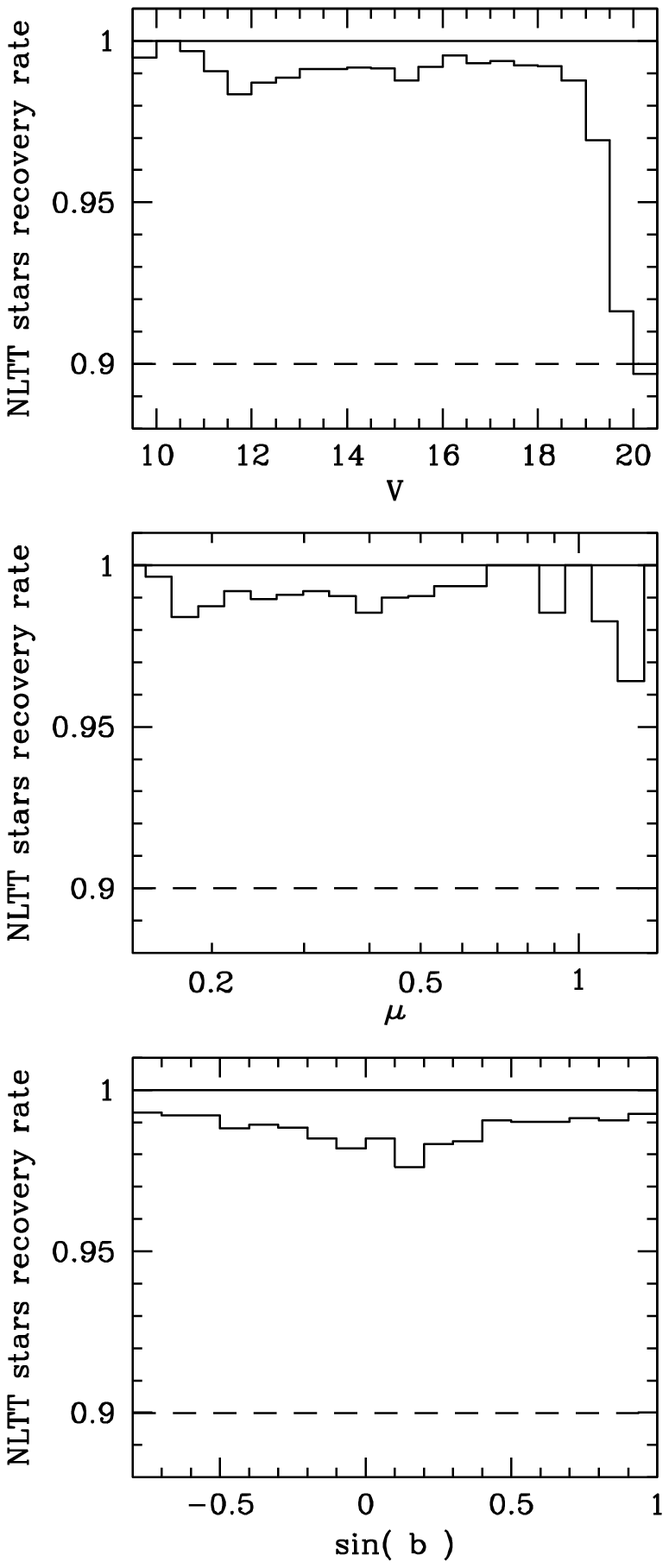}
\caption{\label{fig20} Estimated completeness of the LSPM catalog based
on the recovery rate of NLTT stars. The recovery rate is $>98\%$ for
all values of the proper motion and for magnitudes $V>19.0$. The
recovery rate then drops to $\sim90\%$ at $V=20.0$. The recovery rate
is marginally smaller at low galactic latitudes, but still exceeds
$97\%$.}
\end{figure}

\placefigure{fig20}

Figure 20 lets us estimate the completeness of the LSPM catalog from
the rate at which NLTT stars are recovered by SUPERBLINK and
TYCHO-2. One can guess in advance that the rate is very high, since
very few NLTT stars had to be separately incorporated into the catalog
(see \S 3.4). We exclude from this analysis the NLTT stars that were
missed by SUPERBLINK because they are in areas that were not
processed by the code, since their inclusion would underestimate the
true efficiency of SUPERBLINK. We calculate the fraction of NLTT stars
that have been recovered either by SUPERBLINK or from the
TYCHO-2/ASCC-2.5 catalogs. Results are shown in Figure 20. The
recovery rate is $\approx99\%$ to a magnitude as faint as $V=19$,
falling to $\approx90\%$ at $V=20$. Note also the small dip (to
$\approx98\%$) at the boundary between TYCHO-2 and SUPERBLINK stars
($V=12$), which we investigate further in \S7.2. The recovery does not
vary significantly with the proper motion. There is however the
expected trend that SUPERBLINK misses more stars at low galactic
latitudes, but the recovery rate still exceeds $97\%$ at $|b|<10$.

The recovery rate of NLTT stars by SUPERBLINK/TYCHO-2 is a good
estimate of the completeness of the LSPM catalog, but is valid only in
regions of the [$\mu,V,b$] parameter space that contain a sufficient
number of NLTT stars. We know from Figure 19 that the NLTT catalog is
significantly depleted at $|b|<15$ and $V>16$, which means the completeness
of the SUPERBLINK sample cannot be evaluated for faint stars at low
Galactic latitudes using the above method. Outside of that specific
range, however, we conclude that the completeness of the
SUPERBLINK/TYCHO-2 sample is indeed extremely high.

But the completeness of the LSPM catalog itself is larger than the
completeness of the SUPERBLINK/TYCHO-2 sample, since the missing NLTT
stars have been included in the LSPM. Assuming that the NLTT catalog is
itself more than $90\%$ complete for $|b|>15$ and $V<18$ stars
suggests that $90\%$ of the stars missed by TYCHO-2 and SUPERBLINK
might have been found by Luyten, in which case the LSPM catalog could
up to $\approx99.9\%$ complete down to $V=18$. However, this assumes that
both the SUPERBLINK/TYCHO-2 and the NLTT samples are statistically
independent, an assumption that may not be entirely valid. Indeed, it
is very possible that stars that have been missed by SUPERBLINK have
also been missed by Luyten for the exact same reason. From our
experience, failed detections mostly occur when a faint star moves on
pixels saturated by a bright nearby objects at one of the two epochs
(or worse, at both epochs). But high proper motion stars eventually
move out of the glare, so a star that is hidden at one epoch will
be easy to spot at another. Indeed, this is what happens for several
of the NLTT stars that SUPERBLINK failed to identify: the star is in
the glare of a bright object on the POSS-II image. Because Luyten
used his own 1960s plates as a second epoch, the object was then easy
to identify. As was shown in \citet{LSR03}, many of the new high proper
motion stars found with SUPERBLINK follow the inverse pattern: easy to
find on the POSS-II scans, their trajectory puts then in the glare of
a brighter star in the 1960s. The real problem is for stars hiding on
the POSS-I plates, since both we and Luyten are using it as the first
epoch. Additionally, there might be a few faint stars lost in the
extended, saturated patches from very bright stars at all epochs (if
the total 45 years motion of the star is shorter that the size of the
saturated image), although these cases should be quite rare.

The bottom line is that for every star missed by SUPERBLINK but
found by Luyten, there is probably at least another one that has been
missed by both. Thus, despite the fact that the additional NLTT stars
make the LSPM catalog more complete than the SUPERBLINK/TYCHO-2
sample, we conclude that {\em the LSPM catalog is approximately
$99.0\%$ complete to as faint as $V=19$, and at high ($|b|>15$)
Galactic latitudes.}

\begin{figure}
\epsscale{2.25}
\plotone{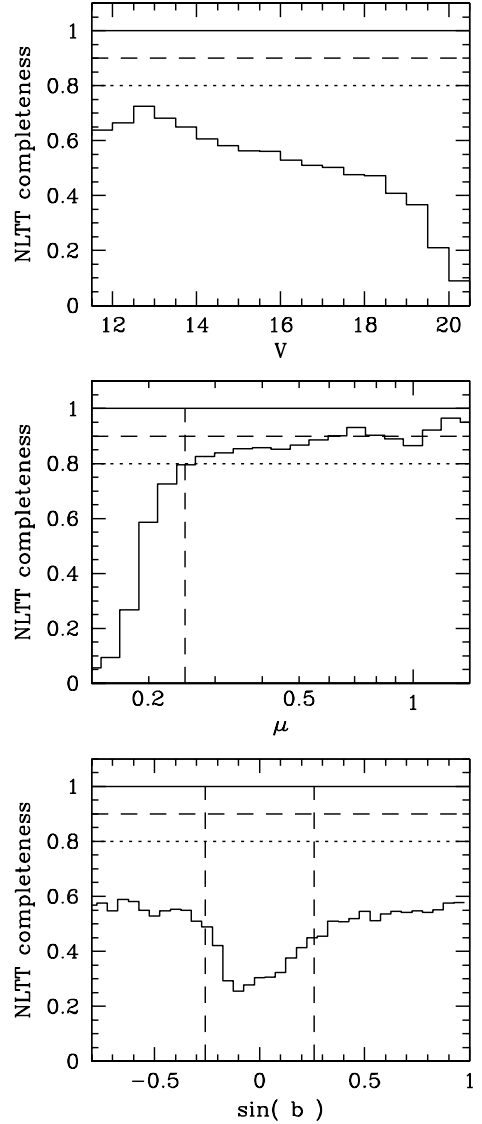}
\caption{\label{fig21} Completeness of the old NLTT catalog, estimated
from the fraction of LSPM catalog objects that are NLTT stars. The
NLTT completeness is a function of both magnitude and proper
motion. While the NLTT was relatively complete for stars with
$\mu>0.3\arcsec$ yr$^{-1}$ and brighter than $R_{\rm F}$=14, its
completeness dropped significantly at fainter magnitudes, and was much
more limited for proper motions near the cutoff at $\mu=0.18\arcsec$
yr$^{-1}$.}
\end{figure}

\placefigure{fig21}

\begin{figure}
\epsscale{2.25}
\plotone{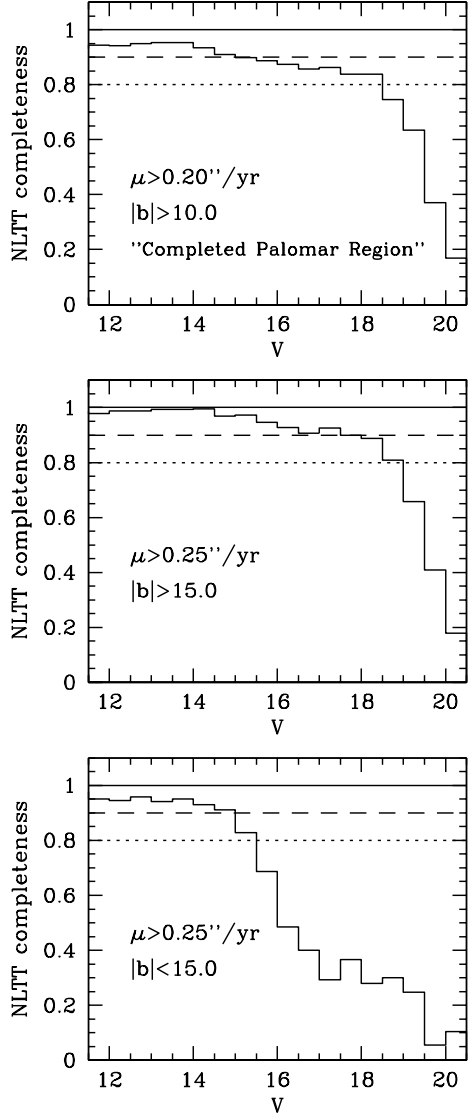}
\caption{\label{fig22} Completeness as a function of magnitude for the
NLTT catalog, based on a comparison with the LSPM, for various ranges
of proper motion and galactic latitudes. The ``Completed Palomar
Region'' is traditionally assumed to be the most complete (top),
although the NLTT is significantly more complete if we exclude the
bands $10<|b|<15$ (middle). At low Galactic latitudes (bottom) the
completeness of the NLTT drops significantly beyond $V=15$.}
\end{figure}

\placefigure{fig22}

We now determine the completeness of the NLTT catalog by comparing it
to the LSPM catalog, assumed to be $99\%$ complete within the limits
quoted above. We first plot the fraction of LSPM stars that are NLTT
objects as a function of magnitude, proper motion, and galactic
latitude (Figure 21). The general completeness of the NLTT over the
whole northern sky for stars with $\mu>0.15\arcsec$ is a little above
$60\%$ down to $V=14$, falls gradually to $40\%$ at $V=19$, and then
drops more abruptly. This is, however, not a fair assessment of the
completeness of Luyten's survey, because the LSPM catalog has a lower
proper motion limit, and because the completeness of the NLTT is
significantly lower at low galactic latitudes.

First we want to determine the effective proper motion limit of the
NLTT, specifically the limit above which the NLTT is most
complete. We find that the completeness of the NLTT catalog decreases
sharply below $\mu=0.20\arcsec$ yr$^{-1}$. This is to be expected from
fiducial limit of $0.18\arcsec$ yr$^{-1}$ and the $0.02\arcsec$
yr$^{-1}$ measurement error. Because the drop in completeness starts a
little above $\mu=0.20\arcsec$ yr$^{-1}$, we define the range of
maximal completeness of the NLTT as $\mu>0.25\arcsec$ yr$^{-1}$.

The NLTT is supposed to be most complete within the Completed Palomar
Region (CPR), as defined by \citet{D86}, which for the northern sky is
simply $|b|>10$, for stars with $\mu>0.2\arcsec$ yr$^{-1}$. However,
we note that the completeness of the NLTT actually starts falling a
few degrees above $|b|=10$. We conclude that the NLTT is most complete
for proper motions $\mu>0.25\arcsec$ yr$^{-1}$ and galactic latitudes
$|b|>15$. 

We now proceed to check the completeness of the NLTT as a function of
magnitude for three regions of parameter space (Figure 22). First we
check the completeness for the CPR. We find that the estimate of
\citet{D86} is accurate, and that the NLTT catalog is indeed more than
$80\%$ complete down to $V=18$. The second region is the one we
identified as the most complete of the NLTT, with the
$\mu>0.25\arcsec$ yr$^{-1}$ stars at Galactic latitudes $|b|>15$.
For that restricted area, we find the NLTT to be $90\%$ complete down
to $V=18.5$. Finally, we check the completeness of the NLTT at low
Galactic latitudes $|b|<15$, again for stars with $\mu>0.25\arcsec$
yr$^{-1}$. We now find that the completeness falls from about $90\%$
at $V=15.0$ to only $30\%$ below $V=17.0$.

Our estimate of a fairly high completeness of the NLTT at high
Galactic latitudes is very significant, because is indicates that the
internal completeness test described by \citet{Fetal2001}
underestimates the completeness of the proper motion sample. The
test of \citet{Fetal2001} suggested that the completeness of the NLTT
in the CPR fell to $80\%$ at $V=15$ and down to $60\%$ at $V=18.5$,
which is significantly smaller than the results of our (external)
test. It appears that the criticism offered by \citet{MFLCHR00} is
legitimate, and that changes in the space density of objects as a
function of distance do lead to an underestimate of the completeness
when applying the internal test of \citet{Fetal2001}.

\subsection{Completeness at the SUPERBLINK/TYCHO-2 boundary}

Completeness problems in the LSPM catalog occur in the magnitude
overlap region between TYCHO-2 and SUPERBLINK stars
($V\approx12$). The problem arises because the completeness of the
TYCHO-2 catalog starts to decrease before SUPERBLINK reaches its full
detection efficiency. As a result, it is very possible that relatively
bright stars have been missed by SUPERBLINK because of plate
saturation, and at the same time are missing from the TYCHO-2 (or
ASCC-2.5) catalogs because they are fainter than these catalogs'
completeness limits.

\begin{figure}
\epsscale{2.15}
\plotone{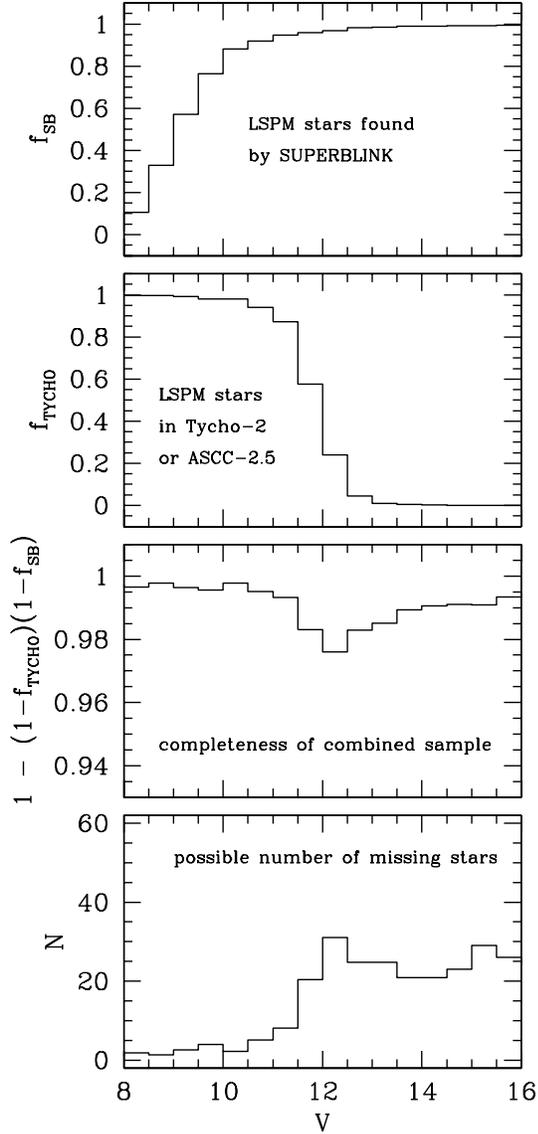}
\caption{\label{fig23}
Estimated completeness of the LSPM catalog at the magnitude boundary
between the TYCHO-2/ASCC-2.5 catalog and the SUPERBLINK stars. The
LSPM is built from the combination of the two samples, which overlap
over a relatively small magnitude range. The coverage appear to be
tight enough that few stars should have been missed at the
boundary; the estimated completeness of the combined sample only drops
to $\approx98\%$ around $V=12.0$, compared to $\approx99.5\%$ at brighter
magnitudes, and $\approx99\%$ at fainter magnitudes. The estimated
number of stars missing from the LSPM because of the boundary effect
should be $\lesssim100$. Note that several of these were recovered as
additional NLTT objects (see Figure 7).}
\end{figure}

\placefigure{fig23}

Fortunately, the TYCHO-2 catalog has a small, but significant overlap
with the SUPERBLINK detections. A large fraction of the brighter
SUPERBLINK stars are in the TYCHO-2 catalog and, likewise, a large
fraction of the fainter TYCHO-2 stars have been recovered by
SUPERBLINK. There is no significant magnitude gap in the combined
TYCHO-2 + SUPERBLINK sample. This is clear from Figure 20 (top panel),
which demonstrates that the combined TYCHO-2/SUPERBLINK sample does
recover more than $98\%$ of the NLTT stars in that magnitude
range. Since the completeness of the full LSPM (including the missed
NLTT stars) is very high, we can use the assumption that it is in fact
complete to estimate the completeness of both the TYCHO-2 and
SUPERBLINK samples as a function of magnitude. The estimated
SUPERBLINK and TYCHO-2 completeness is plotted in Figure 23 (top two
panels). 

We use a probabilistic approach to estimate the completeness of the
combined sample. We can say that the completeness function $f(V)$ is
also the probability that a star of $V$ magnitude will be in the
sample. The probability that a star is in the TYCHO-2 (or ASCC-2.5)
catalog is $f_{T}(V) \simeq N_{T}(V)/N_{LSPM}(V)$, while the
probability for a star to be detected by SUPERBLINK is $f_{S}(V)
\simeq N_{S}(V)/N_{LSPM}(V)$, where $N(V)$ is the number of star in
the sample as a function of magnitude. The probability that a star is
missing from the TYCHO-2 is thus $1-f_{T}$, while the probability that
a star will be missed by SUPERBLINK is $1-f_{S}$. Assuming the two
samples to be independent, we can say that the probability that a star
is missing from {\em both} the TYCHO-2 and SUPERBLINK samples is
$(1-f_{T})(1-f_{S})$. Thus, the completeness of the combined sample
will be $1-(1-f_{T})(1-f_{S})$, which is a function of $V$ magnitude.
From this, we calculate the completeness of the combined
SUPEBLINK+TYCHO-2 list, and the total number of stars that would
presumably be missing from the LSPM because of this incompleteness
(bottom two panels in Figure 23). The effect is very small, and the
completeness only drops to $\approx98\%$ around $V=12$. Overall, we
expect to be missing $\approx50-100$ stars because of the boundary
effect. 

We note that $\approx60$ additional NLTT stars have been found
in the $10<V<14$ magnitude range that were neither in the SUPERBLINK
nor TYCHO-2/ASCC-2.5 lists (Figure 7). The NLTT thus recovers some of
the missing stars. But because the NLTT has a higher proper motion
limit than the LSPM, additional stars in the $0.15\arcsec$
yr$^{-1}<\mu<0.18\arcsec$ yr$^{-1}$ range are probably still
missing. Since about a third of the LSPM stars are in the $0.15\arcsec$
yr$^{-1}<\mu<0.18\arcsec$ yr$^{-1}$ range, we are probably still
missing $\approx30$ stars in the $10<V<14$ magnitude range.

\subsection{Completeness of the LSPM at low galactic latitudes}

A close examination of Figures 18-19 makes it very apparent that the
number density of LSPM stars is not uniform over the sky. Clearly,
there are more LSPM stars at higher galactic latitude than near the
Galactic plane. The lower counts at low galactic latitude are actually
a natural consequence of the low proper motion cut-off of the catalog,
and not a result of decreased completeness in low-galactic latitude
fields.

\begin{figure*}
\plotone{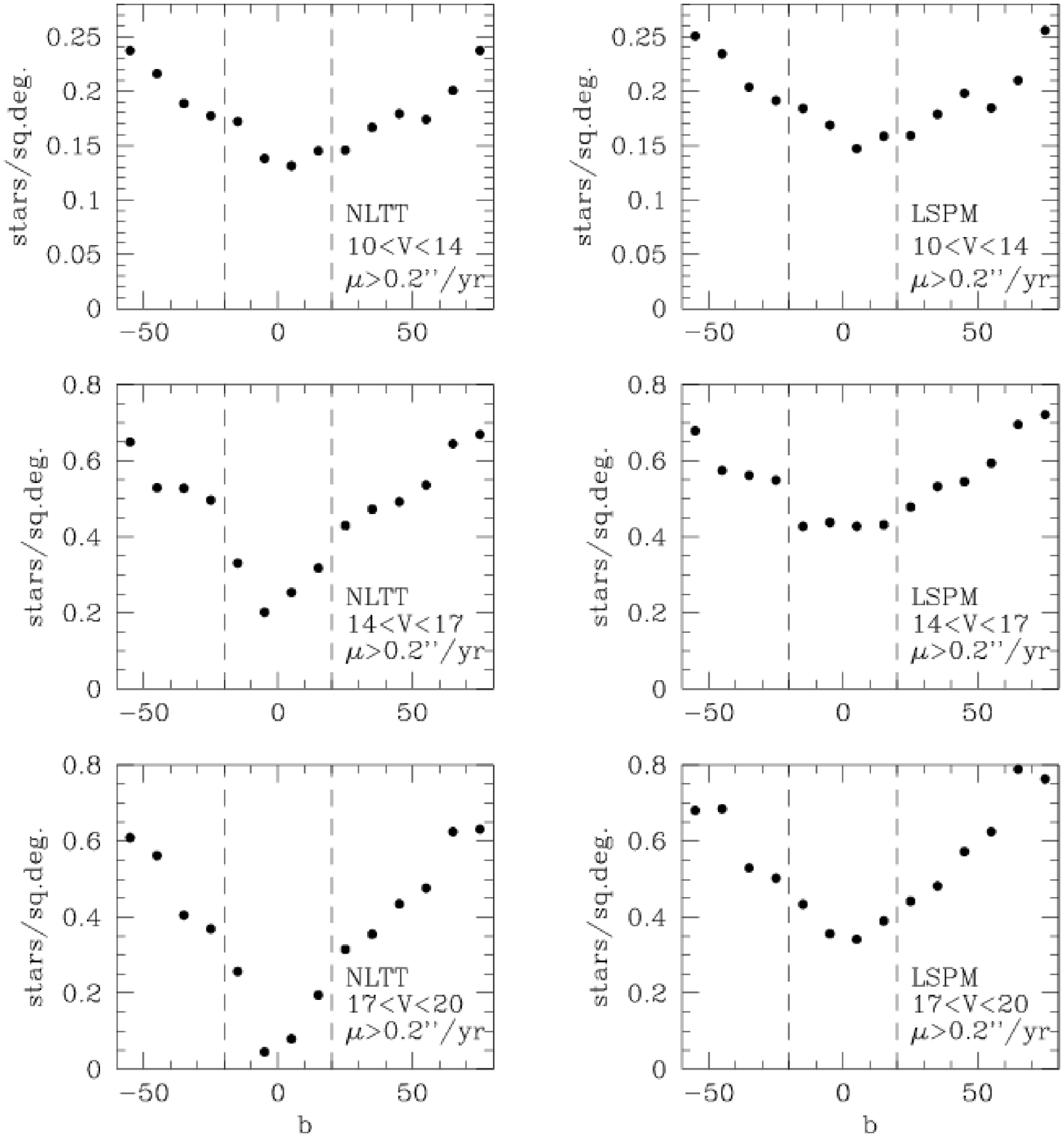}
\caption{\label{fig24}
Surface density of (in stars per square degrees) of the NLTT and LSPM
catalogs as a function of galactic latitude for three groups of
$\mu>0.2\arcsec$ yr$^{-1}$ stars: bright (top), moderately faint
(middle) and faint (bottom). At high galactic latitudes, outside of
the area delimited by the dashed line ($|b|>20$), the NLTT is $>90\%$
complete, and the LSPM $>99\%$ complete, for $V<18$. Both catalogs are
$>90\%$ complete for bright stars ($V<14$) in low galactic latitudes
$(|b|<20)$. The drop in the density of $V<14$ NLTT stars at low
Galactic latitudes is largely due to catalog incompleteness. Not so
for the LSPM catalog, which appears to be largely complete at
$|b|<20$, given the intrinsic trend in number density with $b$.}
\end{figure*}

\placefigure{fig24}

The first line of evidence that this is true is that the lower density
of LSPM stars at low galactic latitude is observed both for bright and
fainter stars. Figure 25 shows that there are $\simeq40\%$ fewer LSPM
stars at low Galactic latitudes than there are at high Galactic
latitudes, and this is independent of the magnitude of the stars. If
there were completeness problems in the LSPM because of crowding, or
other low galactic latitude effects, the proportion of low Galactic
latitude stars would drop with fainter magnitudes, as is observed in
the NLTT catalog (see Figure 24).

\begin{figure*}
\plotone{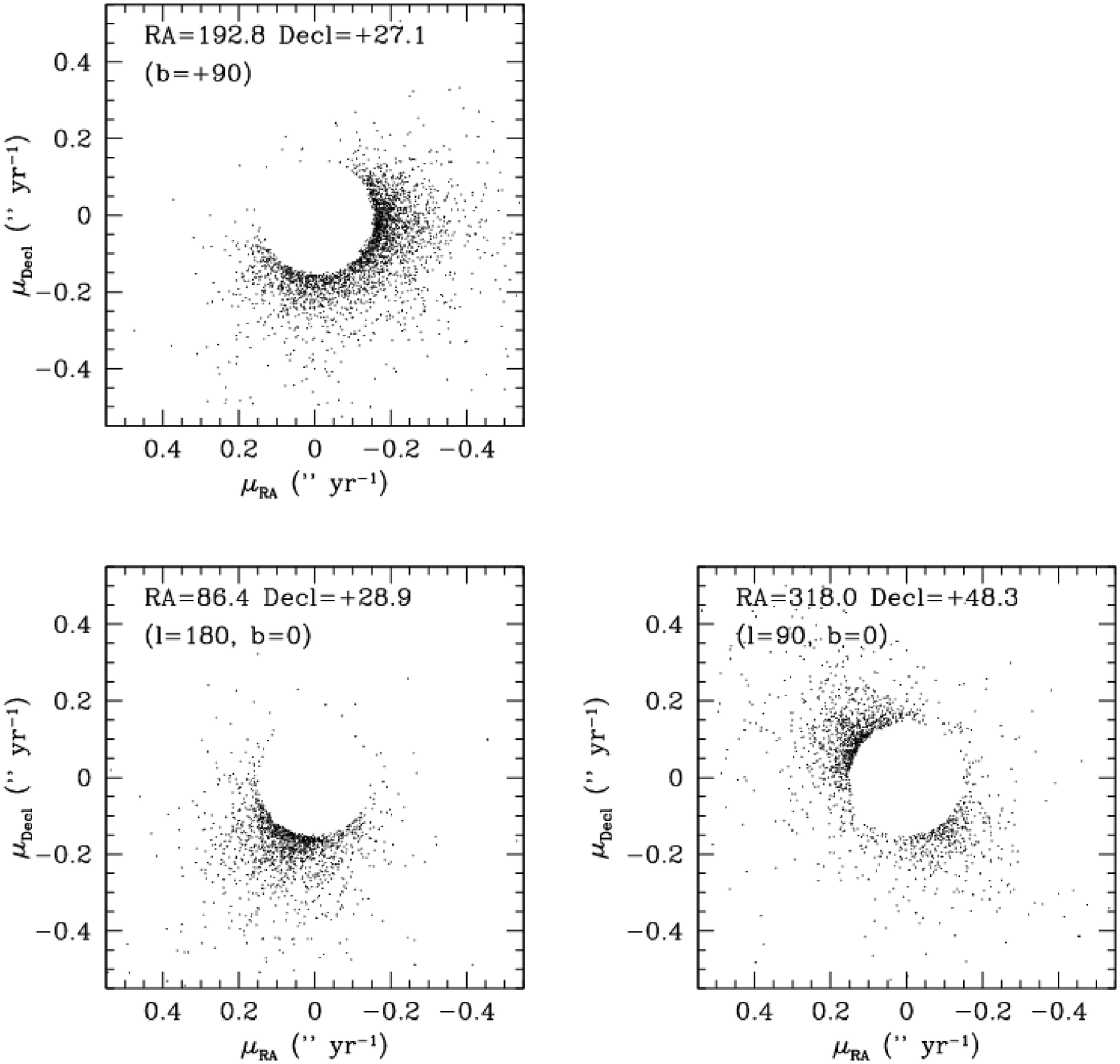}
\caption{\label{fig25}
Distribution of absolute (zonal corrected) proper motions for LSPM
stars within 15$^{\circ}$ of the north Galactic pole (top), the
direction of Galactic rotation (bottom right), and the Galactic
anti-center (bottom left). The distribution of proper motion vectors
is not uniform, and shows a distinct pattern that is clearly dependent
on the position on the sky. The fixed, low-$\mu$ cutoff of the catalog
(empty disks centered on the origin) determines how many stars are
locally included in the LSPM catalog. The larger scatter and larger
offset in the proper motion vectors near the Galactic pole implies
that more stars in that direction will make it into the LSPM. This
explains the higher density of LSPM objects at high galactic
latitudes, near the north galactic pole.}
\end{figure*}

\placefigure{fig25}

If the distribution of LSPM stars is not uniform over the northern
sky, it must be because of proper motion selection effects. The
velocity components of the stars in the vicinity of the Sun are not
isotropic, because of Galactic rotation and because of the Sun's
motion relative to the local standard of rest. The distribution of
stellar proper motion vectors is very much dependent on the position
on the sky, as illustrated in Figure 25. One naturally expects to find
more high proper motion stars at high Galactic latitude because of the
large apparent drift of the halo and old-disk stars in that direction,
as seen from the Sun.

If the distribution of high proper motion stars were uniform over the
sky, estimating the completeness at low galactic latitudes would be
straightforward. Since the LSPM is nearly complete at high Galactic
latitudes (for $V<19$), the completeness at low Galactic latitude
could then be estimated from the ratio between the number density of low
Galactic latitude stars to the number density of high Galactic
latitude stars. This could be calculated for each magnitude bin to
obtain the completeness as a function of magnitude. However, because
the distribution of high proper motion stars is not uniform over the
sky, one has to predict what the density of stars at low Galactic
latitude should be. Estimates of the completeness will be dependent on
the predicted number density of high proper motion stars in the low
Galactic latitude regions.

\begin{figure}
\epsscale{2.2}
\plotone{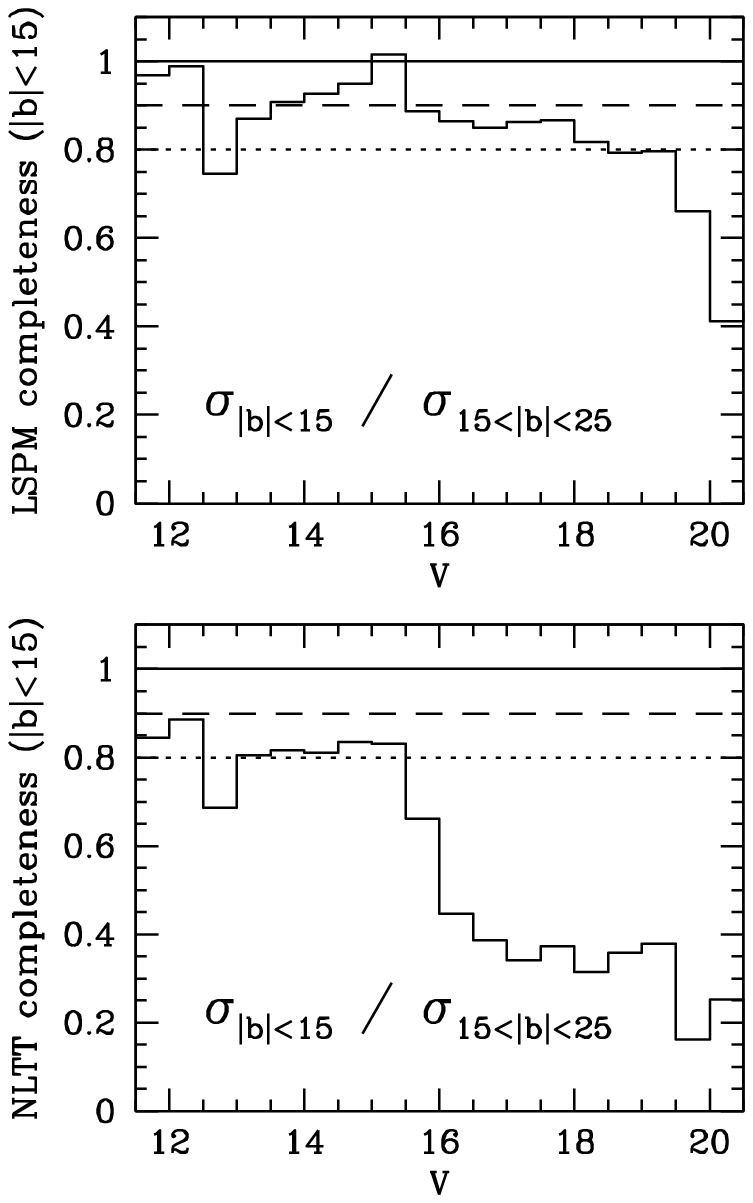}
\caption{\label{fig26} Completeness at low galactic latitude,
estimated from the ratio between the number density of high proper
motion stars at $|b|<15$ and the number density of high proper motion
stars at $15<|b|<25$. This internal test suggests that the LSPM is
$>80\%$ complete at low Galactic latitudes for $V<19$ (top). Applied
to NLTT stars (bottom) this completeness test yields results that are
very similar with the completeness estimate shown in Figure 22, with a
sharp drop at $V=16$. This internal test probably underestimates the
completeness by $5-15\%$, since there are intrinsically fewer high
proper motion stars at lower Galactic latitudes (see Figure 24).}
\end{figure}

\placefigure{fig26}

To estimate the completeness of the LSPM in low galactic latitude
fields, we use the following test. For stars in some magnitude bin
[$V,V+\Delta V$], we first calculate the number density of LSPM
objects (in stars per square degrees) in the area located within 15
degrees of the Galactic plane ($|b|<15$). We then calculate the number
density of [$V,V+\Delta V$] stars in 10 degree bands located above and
below this region ($15<|b|<25$). We assume that the survey is complete
in the $15<|b|<25$ region, and that the distribution of high proper
motion stars should be uniform over the whole $|b|<25$ area. The
completeness in the $|b|<15$ region is thus simply the ratio between
the measured density in $|b|<15$ to the measured density in
$15<|b|<25$. We repeat the calculation for a range of magnitude bins,
to obtain the completeness as a function of $V$. Note that this method
essentially provides an {\em internal} test of completeness. The main
caveat is that the density of high proper motion stars is not uniform
over the sky, as demonstrated above. However, the use of relatively
low Galactic latitude bands ($15<|b|<25$) as a reference should
minimize the effects of the intrinsic non-uniformity. Nevertheless,
since the density of high proper motion stars decreases with Galactioc
latitudes, our internal completeness test is expected to slightly
{\em underestimate} the actual completeness level at low Galactic
latitude, possibly by as much as $\approx5-10\%$, although a detailed
modeling of the distribution of high proper motion stars would be
required to determine the exact value. 

Results of the completeness test are shown in Figure 26, where we have
applied it separately to the complete LSPM catalog (top), and for the
subsample of LSPM stars that are in the NLTT catalog (bottom). Results
for the NLTT stars are consistent with the external completeness test
shown in Figure 22 (which is based on a comparison between NLTT and
LSPM), with a sharp drop in completeness at $V=16$. Note that at
moderately bright magnitudes ($13<V<15$) the internal test suggests
that the NLTT is only $\approx80-85\%$ complete, which is lower than
estimated from the external test ($\approx95\%$, see Figure 22). There
are two different interpretations for the differences in the two NLTT
completeness estimates. First, if the LSPM is only $90-95\%$ complete
at $|b|<15$, $13<V<15$ (as suggested by the internal test), then the
external test overestimates the NLTT completeness, because it assumes
the LSPM to be $100\%$ complete. If indeed the LSPM is only $90-95\%$
complete, then the external test possibly overestimates the
completeness by $5-10\%$, which would bring both values in closer
aggrement. The second interpretation is that the external test is
right, and the LSPM is $100\%$ at $|b|<15$, $13<V<15$, but the
internal test underestimates the completeness level because there are
intrinsically fewer stars at $|b|<15$ than at $15<|b|<15$. If the
second interpretation is valid, then the internal test indeed
underestimates the completeness by as much as $10\%$.

The internal test suggests that the LSPM catalog is at least $80\%$
complete for stars brighter than $V=19$. This should be taken as a the
lowest possible value; as discussed above, the internal test may be
underestimating the completeness by up to $10\%$. Thus, it is possible
that the LSPM is actually $\approx90\%$ in low Galactic latitude
regions. For stars fainter than $V=19$, the internal completeness test
shows a drop to $\approx40\%$ just below $V=20$. The LSPM catalog is
thus definitely shallower at low Galactic latitudes, by 1 to 2
magnitudes. While the LSPM should be regarded as largely complete to
at least $V=20.0$ at high Galactic latitudes, it should be regarded as
relatively complete only down to $V=19.0$ in low Galactic latitude
regions.

Our completeness estimates at low Galactic latitudes are relatively
crude at this point, and we are sorry we cannot provide more accurate
values. More accurate completeness estimates could be obtained only if
we had better estimates of the expected density of high proper motion
stars in the low Galactic latitude regions. Alternatively, a proper
modeling of all the effects that result in SUPERBLINK missing high
proper motion stars might also provide more accurate completeness
estimates. We believe the first option to be beyond the scope of this
paper. The second option, in our opinion, would be extremely difficult
to carry out, because of the complexity of the SUPERBLINK algorithm,
but also because one would need to characterize the combined effects
of all the pairs of POSS-I/POSS-II plates used, with their variety of
saturation levels, point spread functions, and a complete assessment
of plate defects. Ultimately, the most accurate estimates of the
completeness levels will come from better, more accurate proper motion
surveys at low Galactic latitudes (even of limited areas), which will
provide true external tests to the completeness of the LSPM catalog.


\section{Stellar contents of the LSPM catalog}

\subsection{Color-magnitude classification}

\begin{figure*}
\plotone{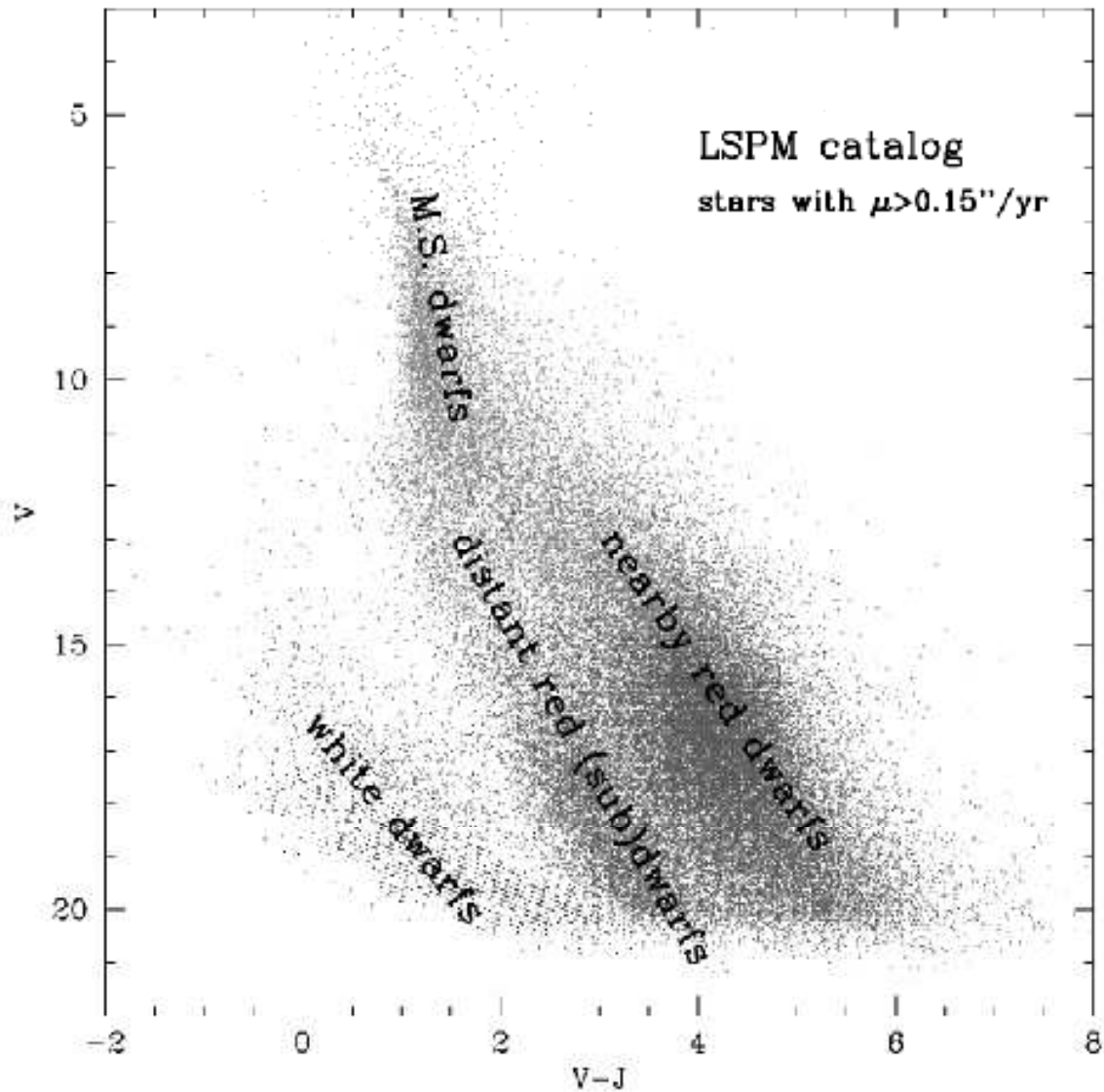}
\caption{\label{fig27} Optical/infrared color-magnitude diagram of
all stars in the LSPM catalog. Stars with TYCHO-2 of
ASCC-2.5 counterparts are plotted in green; these have the most
accurate values of V and V-J. Stars with no 2MASS counterparts are
plotted in blue; most of them are consistent with being white dwarfs.}
\end{figure*}

\placefigure{fig27}

The apparent $V$ magnitude of LSPM stars is plotted in Figure 27 as a
function of the $V-J$ color index. The LSPM stars are nicely clumped
in three main groups. The interpretation is straightforward when one
considers the selection effects implied in a sample of stars selected
for high proper motions. Catalogs of high proper motions stars
essentially contains subsamples of {\em nearby} stars. However, the
spatial extent of the detection range depends on the transverse
velocity of the star. Hence, high velocity (halo, old-disk) stars tend
to be selected from a larger distance. It is thus fair to say that, on first
approximation, high proper motion catalogs combine nearby disk stars
with more distant halo/old-disk stars. Because of the limiting distance, a
diagram of apparent magnitude as a function of color will have the
same general features as a color-magnitude diagram: a main sequence
extending from upper left to lower right, and a white dwarf sequence at the
bottom left. This is indeed what we see in Figure 27. Because the
stars are not exactly at the same distance, all sequences are thicker
and fuzzier, but they are recognizable. Because high velocity stars
are detected to a larger distance, they will form their own sequence,
but {\em shifted down} because or their larger distance and hence
larger mean apparent magnitude. Again this feature is quite apparent
in Figure 27.  

Note in Figure 27 the concentration of stars on the lower left of the
plot, where the nearby white dwarfs are expected to be found. All
stars with a 2MASS counterpart are plotted in red, while stars with no
2MASS counterparts are plotted in blue. It is clear that the majority
of the high proper motion stars with no 2MASS counterparts are white
dwarfs, which confirms the conjecture posed by \citet{SG03}.

\begin{figure*}
\plotone{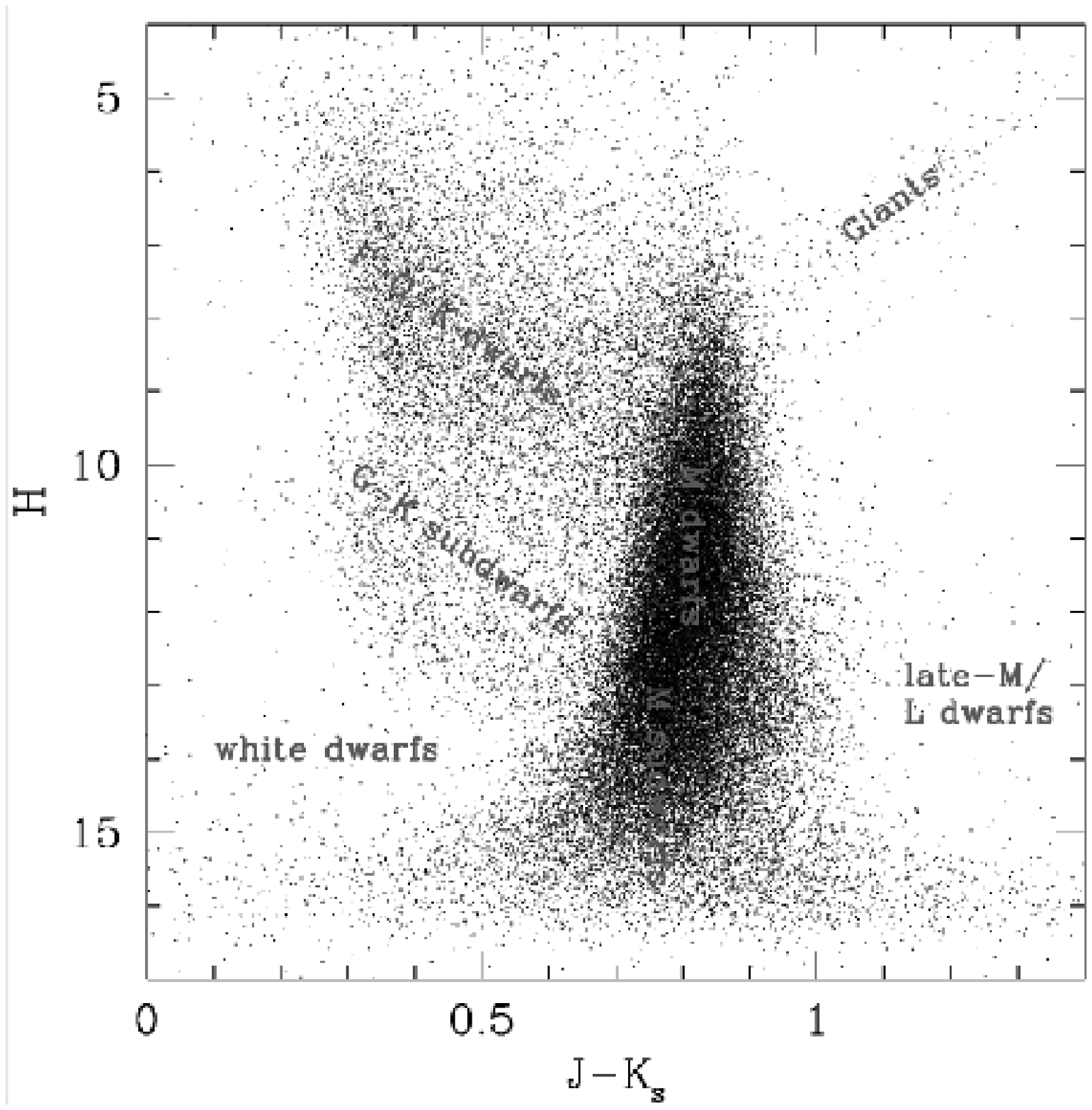}
\caption{\label{fig28} Infrared color-magnitude diagram of stars
in our proper motion catalog, based on JHK$_s$ magnitudes extracted from
the 2MASS All-Sky Point Source Catalog. Main sequence F-G-K stars
populate the upper left of the distribution, while M dwarfs are
clumped around J-K$_s=0.8$. A diffuse wisp of giant stars can be
seen at the upper right. Below H=15, errors in 2MASS magnitudes
increase, which blurs out the color distribution.}
\end{figure*}

\placefigure{fig28}

In the H/J-K color magnitude diagram, the brighter stars are separated
into two main groups: a dense clump of stars around J-K$_s$=0.8, and a
more diffuse one extending from the blue edge of the first one to
around J-K$_s$=0.3 (Figure 28). While the blue clump extends only down
to about H=13, the red clump extends all the way down to the magnitude
limit, where the larger magnitude errors scatter the stars about in
$J-K_s$. The big red clump is populated by M dwarfs and subdwarfs,
which are degenerate in J-K color from M2 to M7 \citep{BB88}. The
blue, diffuse clump consists of F-G-K dwarfs and subdwarfs. The fact
that the distribution of main sequence stars other than M dwarfs ends
(at H=13) well before the 2MASS magnitude limit indicates that the
LSPM is complete for those stellar subtypes. Note also the diffuse
wisp that extends from [J-K$_s$,H]=[0.9,7.5] to [1.5,4.5]; these are
the very few giant stars that have proper motions large enough to be
included in the LSPM.

Dwarfs and subdwarfs are expected to occupy distinct loci on the color
magnitude diagram because these populations have different mean
distances in proper motion catalogs. The proper motion limit
($\mu>0.15\arcsec$ yr$^{-1}$) restricts the detection range of disk
objects, whose transverse velocities are typically $<50$ km s$^{-1}$,
while halo objects (subdwarfs) with typical transverse velocities 2-5
times as large, can find their way into the proper motion catalog from
distances 2-5 times larger. This is illustrated in Figure 28, where we
have noted the probable positions of subdwarfs, 3-4 magnitudes below
the dwarfs.

We also note in Figure 28 the locations expected to be populated
by white dwarfs, and late-type M dwarfs (M7-M9) and L dwarfs. Of
course, one has to look for those objects above the limit (H=15) below
which 2MASS colors become less reliable. We nevertheless find a
significant number of object that are convincing white dwarf
candidates. Late-type M dwarfs (M7-M9) are found in the range
$1.0<$J-K$_s<1.3$, while L dwarfs normally occur beyond J-K$_s>1.3$
\citep{K99}. Several candidates are detected, and they warrant further
investigation. While many might be known objects, chances are high
that at least a few will be new ones, in particular if they occur at
low galactic latitudes, which have largely been avoided by previous L
dwarf surveys \citep{K99,Cr03}.

\begin{figure*}
\plotone{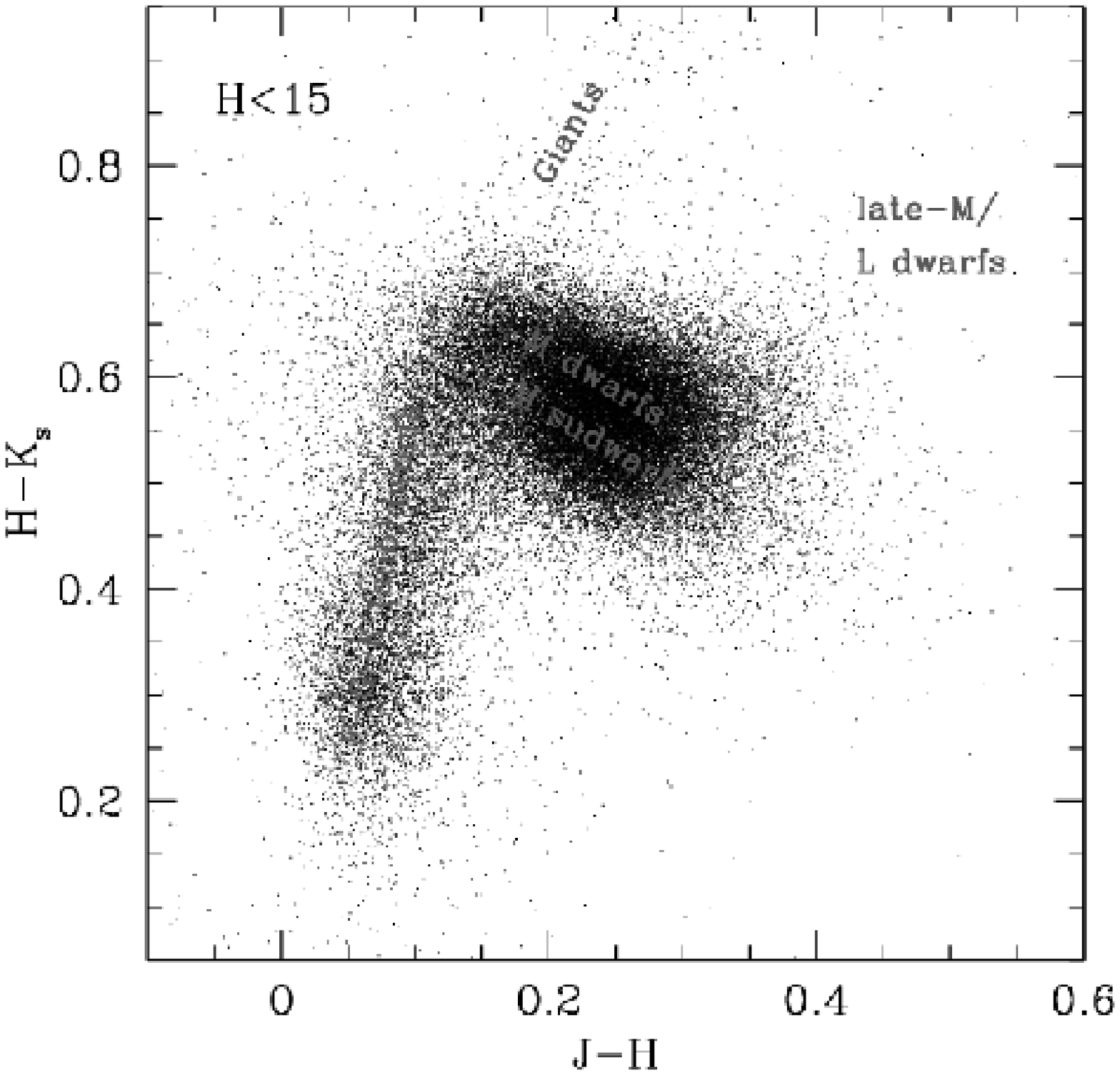}
\caption{\label{fig29} Infrared J-H/H-K$_s$ color-color diagram of
stars in the proper motion catalog. The distribution is consistent with
stars on the main sequence with no reddening.}
\end{figure*}

\placefigure{fig29}

The J-H/H-K color-color diagram is largely consistent with the
vast majority of LSPM objects being main sequence stars with no
significant reddening (Figure 29). The distribution is very similar
to the general color-color diagram of 2MASS sources in
high galactic latitude fields, and follows the standard sequence
determined by \citet{BB88}. The loci of different types of objects are
indicated, as in Figure 28.

Overall, the accuracy of the 2MASS J, H, and K$_s$ magnitudes allows
one to separate different object subtypes and populations (much better
than the photographic magnitudes). They should be very useful in
planning follow-up observations of the LSPM stars.

\subsection{Reduced proper motions}

\begin{figure*}
\plotone{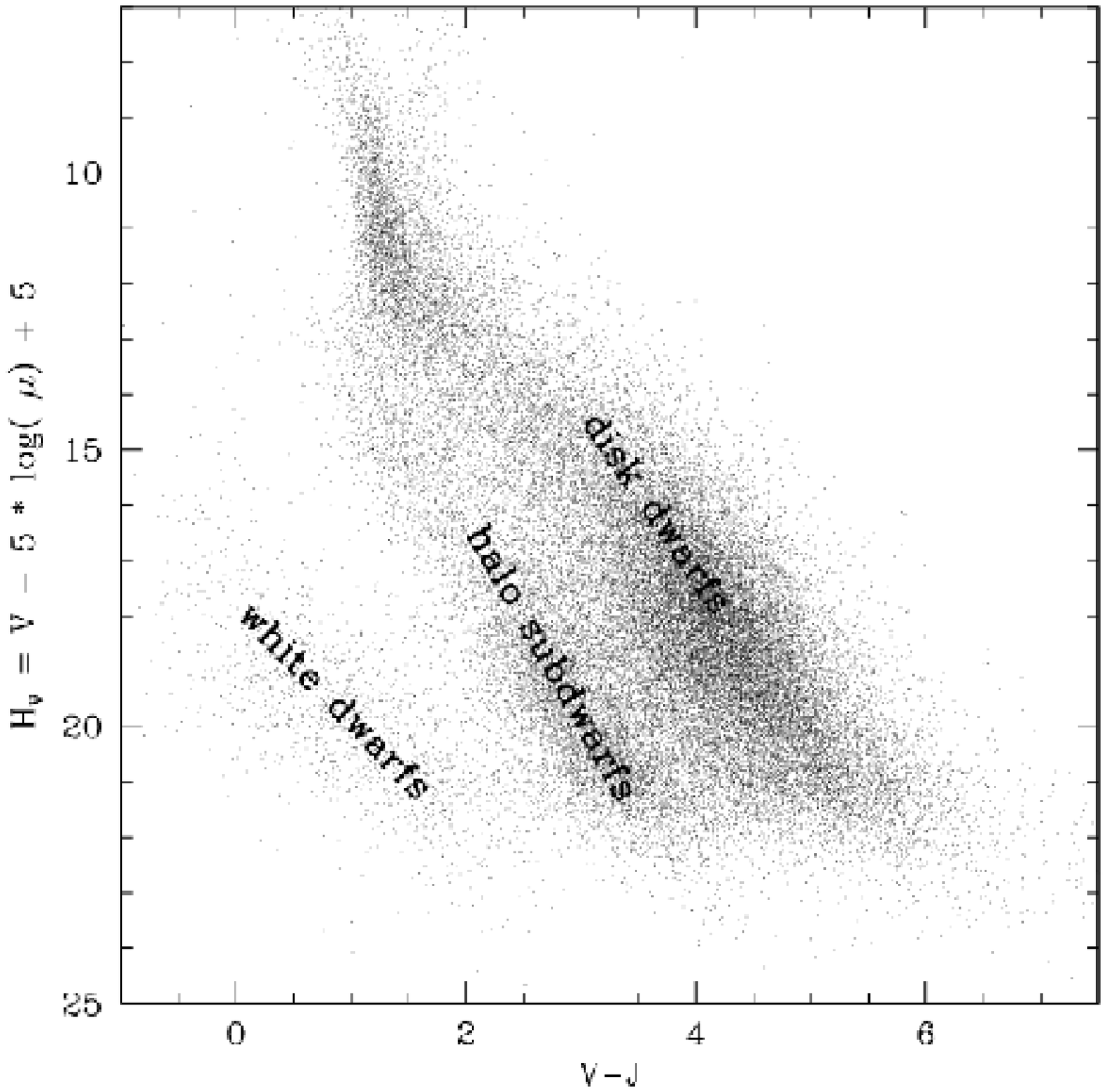}
\caption{\label{fig30}
Reduced proper motion diagram of the LSPM stars. Stars are distributed
in four major groups: brighter stars, cool disk dwarfs, cool halo
subdwarfs, and white dwarfs.}
\end{figure*}

\placefigure{fig30}

Reduced proper motion diagrams are a major tool in the classification
of local stars into different stellar populations
\citep{J72,E92,SG00}. It was recently demonstrated by \citet{SG02}
that {\em optical-infrared} reduced proper motion diagrams are very
efficient in separating samples of high proper motion stars into three
distinct classes: main sequence disk dwarfs, halo subdwarfs, and white
dwarfs. 

Using the proper motions and magnitudes in the LSPM catalog, we build
a reduced proper motion diagram for all but 27 LSPM stars using a
reduced proper motion calculated from the $V$ band ($H_V$) and the
$V-J$ color. The reduced proper motion is analogous to an absolute
magnitude in which the proper motion is used in place of the
parallax. While the absolute magnitude is defined as
\begin{displaymath}
M_V = V + 5 * \log{\pi} + 5,
\end{displaymath}
where $\pi$ is the parallax in seconds of arc, the proper motion is
defined as
\begin{displaymath}
H_V = V + 5 * \log{\mu} + 5,
\end{displaymath}
where $\mu$ is the proper motion in seconds of arc per year. The
reduced proper motion is directly related to the absolute magnitude:
\begin{displaymath}
H_V = M_V + 5 * \log{v_T} - 3.38
\end{displaymath}
where $v_T$ is the projected velocity of the star in the plane of the
sky, in km s$^{-1}$. 

The reduced proper motion diagram is thus analogous to a
color-magnitude diagram, except for the fact that the usual stellar
sequences are ``blurred'' by the $\log{v_T}$ term. But different
populations have different ranges of possible $v_T$. Disk stars, on
the one hand, have a mean transverse velocity $\langle v_T \rangle
\simeq50$km s$^{-1}$, which yields:
\begin{displaymath}
\langle H_V \rangle_{disk} = M_V + 5.1 .
\end{displaymath}
Halo stars, on the other hand, have significantly larger mean
transverse velocities $\langle v_T \rangle \simeq300$km s$^{-1}$, so
their $H_R$ are generally larger:
\begin{displaymath}
\langle H_V \rangle_{halo} = M_V + 9.0 .
\end{displaymath}
The (disk) dwarfs and (halo) subdwarfs thus form distinct sequences in
the reduced proper motion diagram, with halo subdwarfs located well
below the disk dwarfs. White dwarfs occupy a distinct locus on the
lower left of the diagram, a position familiar to users of
color-magnitude diagrams.

We show in Figure 30 the reduced proper motion diagram for LSPM
stars. The corrected, absolute proper motions were used. The
loci of the different stellar classes and populations are
labeled. Note the similarity with Figure 27, which plotted the
apparent magnitude as a function of color. The two plots are
fundamentally very different, however. The reduced proper motion is a
function of {\em luminosity and velocity}, while the apparent
magnitude is a function of {\em luminosity and distance}. The two
plots look similar only because stars are piled up against the LSPM
proper-motion limit of $0.15\arcsec$ yr$^{-1}$, which introduces a
correlation between the velocity and distance. High-velocity stars can
make it into the catalog even at large distances, while only the
nearest of the low-velocity stars have proper motions large enough to
be in the LSPM. Figure 27 thus separates the halo stars based on their
larger average distances, while Figure 30 separates stars based on
their larger transverse velocities. For classification purposes, the
reduced proper motion diagram should be preferred.

The reduced proper motion diagram contains a wealth of information,
and the separation between the different populations warrants a more
detailed analysis of the individual populations represented
here. Such a detailed analysis is beyond the scope of this
paper. Nevertheless, we wish to emphasizes the significant potential
of the LSPM catalog in the study of the local stellar populations.

\section{Conclusions}

We have generated a new catalog of stars with proper motions larger
than 0.15$\arcsec$ yr$^{-1}$ which currently is the most complete of
its kind. We have achieved our initial goals of locating new high
proper motion stars, and redetermining to higher accuracy the
positions and proper motion estimates of previously known objects,
especially the high proper motion stars from the LHS and NLTT
catalogs.

The catalog is limited to northern declinations, and lists 61,977
stars down to a magnitude $V=21$. This essentially doubles the number
of previously cataloged high proper motion stars in this
hemisphere. The catalog is estimated to be $>99\%$ for $V<19$ stars at
high Galactic latitude ($|b|>15$), and $\approx90\%$ for $V<19$ stars
at low Galactic latitude ($|b|<15$). This is a very significant
improvement over previous catalogs.

We provide photometric estimates in the three optical bands of the
POSS-II survey: B$_{\rm J}$ (IIIaJ emulsion with GG385 filter),
R$_{\rm F}$ (IIIaF emulsion with RG610 filter), and I$_{\rm N}$ (IVN
emulsion with RG9 filter). We also provide $B$ and $V$
magnitudes for the brighter stars, from their TYCHO-2/ASCC-2.5 catalog
counterparts. While the $B$ and $V$ magnitudes are very accurate, much
room for improvement remains with the optical photometry of fainter
stars. Infrared photometry, on the other hand, was extracted from the
2MASS All-Sky Point Source catalog, and is accurate down to about 15th
magnitude in each band (J, H, K$_s$). The probability of our stars
having been mismatched in the 2MASS catalog is almost zero. 

To compare all stars in the same color/magnitude system, we provide an
estimate of the apparent $V$ magnitude and the $V-J$ color index for
all but 814 LSPM stars. These are most reliable for the brighter
($V<12$) stars, but should be used with caution for fainter stars, for
which $V$ is estimated from the USNO-B1.0 photographic magnitudes.

The LSPM catalog is a work in progress. An extension to the southern
sky is currently underway. We are also working on an expansion to
lower proper motions, down to 0.10$\arcsec$ yr$^{-1}$. Future plans
include an improvement of the magnitudes, especially in the optical
bands, by using magnitude estimates from a variety of other sources
(SDSS, USNO-A2 catalog, GSC-2.2 catalogs, future versions of the UCAC
catalog).

At this point, the LSPM catalog is ideally suited for follow-up
observations of selected targets of interest. Indeed, we are currently
working on a massive spectroscopic survey of selected LSPM objects,
including all stars with proper motions $\mu>0.45\arcsec$ yr$^{-1}$,
for which spectroscopic observations are now in hand (L\'epine {\it et
al.} 2005, {\it in preparation}).

The LSPM catalog will be updated as new discoveries are made. We
invite investigators who discover new high proper motion objects in
the northern sky to contact the authors so that their discovery can be
included in the catalog. Likewise, investigators who notice that a
known high proper motion star is missing from the LSPM, or who find
errors in our data are invited to communicate with us.


\acknowledgments

{\bf Acknowledgments}

We would like to thank the referee, A. Gould, for invaluable comments
and suggestions.

We are very grateful to Brian McLean, head of the Catalogs and Surveys
Branch of the Space Telescope Science Institute, for allowing us to
massively access the Digitized Sky Surveys. We also thank Michael
Benedetto, of the AMNH IT department, for network connectivity from
AMNH to STScI. MS acknowledges the leadership, support, and friendship
of the late Barry Lasker, who guided the Digitized Sky Surveys project
until his untimely death. Riccardo Giaconni's vision led to the DSS,
and MS acknowledges that seminal idea with gratitude.

R. Michael Rich has been an early and enthusiastic user of this
catalog and we thank him for his support and ongoing collaboration.

This work has been made possible through the use of the Digitized Sky
Surveys. The Digitized Sky Surveys were produced at the Space
Telescope Science Institute under U.S. Government grant NAG
W-2166. The images of these surveys are based on photographic data
obtained using the Oschin Schmidt Telescope on Palomar Mountain and
the UK Schmidt Telescope. The plates were processed into the present
compressed digital form with the permission of these institutions. The
National Geographic Society - Palomar Observatory Sky Atlas (POSS-I)
was made by the California Institute of Technology with grants from
the National Geographic Society. The Second Palomar Observatory Sky
Survey (POSS-II) was made by the California Institute of Technology
with funds from the National Science Foundation, the National
Geographic Society, the Sloan Foundation, the Samuel Oschin
Foundation, and the Eastman Kodak Corporation. The Oschin Schmidt
Telescope is operated by the California Institute of Technology and
Palomar Observatory. The UK Schmidt Telescope was operated by the
Royal Observatory Edinburgh, with funding from the UK Science and
Engineering Research Council (later the UK Particle Physics and
Astronomy Research Council), until 1988 June, and thereafter by the
Anglo-Australian Observatory. The blue plates of the southern Sky
Atlas and its Equatorial Extension (together known as the SERC-J), as
well as the Equatorial Red (ER), and the Second Epoch [red] Survey
(SES) were all taken with the UK Schmidt. 

This publication makes use of data products from the Two Micron All
Sky Survey, which is a joint project of the University of
Massachusetts and the Infrared Processing and Analysis
Center/California Institute of Technology, funded by the National
Aeronautics and Space Administration and the National Science
Foundation.

The data mining required for this work has been made possible with the
use of the SIMBAD astronomical database and VIZIER astronomical
catalogs service, both maintained and operated by the Centre de
Donn\'ees Astronomiques de Strasbourg (http://cdsweb.u-strasbg.fr/).

This research program has been supported by NSF grant AST-0087313 at
the American Museum of Natural History, as part of the NASA/NSF NSTARS
program. MS and SL gratefully acknowledge support from the Cordelia
Corporation, from Hilary Lipsitz, and from the American Museum of
Natural History.

\begin{appendix}

\section{NLTT stars not included in the LSPM-north catalog}

A total of 1,859 objects listed in the NLTT catalog, and claimed to be
located north of the celestial equator, did not make it into the LSPM
catalog. The reasons why these stars were not included in the LSPM
catalog fall into four categories: (1) the object is moving slower
than the 0.15$\arcsec$ yr$^{-1}$ limit of the LSPM catalog, (2) the
object does not show up on the POSS plates, (3) the object is a
duplicate NLTT entry, (4) the object is a high proper motion nebula. 

Most of the slow moving objects are stars that are incorrectly listed
in the NLTT as having large proper motions. A total of 208 such
objects are stars that are listed in the TYCHO-2 catalog; their
TYCHO-2 proper motions unambiguously place them under the LSPM
inclusion limit. An additional 1104 stars had their proper motions
remeasured with SUPERBLINK, and the updated value makes them low
proper motion ($\mu<$0.15$\arcsec$ yr$^{-1}$) stars.

There are also 76 objects which have a quoted NLTT proper motion under
the 0.15$\arcsec$ yr$^{-1}$ limit of the LSPM. While some of these are
possibly bogus (see below), 4 of them have their low proper motion
status confirmed in the TYCHO-2 catalog. Another 63 have their low
proper motion confirmed by SUPERBLINK. Three more stars did not have
their proper motion remeasured by SUPERBLINK but direct examination of
DSS scans confirms they are low proper motion stars.

We have identified 39 objects to be duplicate entries of other NLTT
stars. In all cases, the duplicate entry was listed with a slightly
different position, generally within a few arcminutes of the primary
entry.

The remaining 395 objects could not be recovered on the POSS plates,
and are listed as ``bogus''. An NLTT star can be missing if it
initially was a spurious detection. This is highly probable for stars
initially reported to have a magnitude near the plate limit
(V$>$18). Another possibility is that the quoted NLTT catalog position
was very far off from the actual location of the star. In this case,
it is very likely that the object has been picked up by SUPERBLINK and
is now listed in the LSPM catalog as a ``new'' high proper motion
star. Thus the object is not really missing, but was rather lost, and
has now been rediscovered.

Two high proper motion ``stars'' in the NLTT are found to be small,
compact nebulae with large proper motions. The two objects are NLTT
13414 and NLTT 13424. Moving nebula are not included in the LSPM
catalog at this point. These, and other candidate high proper motion
nebulae found with SUPERBLINK will be discussed in an upcoming paper.

All NLTT objects north of the celestial equator that are not in the
LSPM catalog are listed in the accompanying table (see Table A-1 for a
sample of the first 15 lines). The table lists the NLTT catalog
number, followed by the position, red and blue magnitude, and proper
motion as quoted in the NLTT catalog. The NLTT positions are
transformed into J2000 coordinates from their original values. A flag
value of "B" indicates that the star could not be found by direct
examination  of the Digitized Sky Survey scans of the Palomar Sky
Survey (POSS) photographic  plates. A flag value of "D" means that
the object is a duplicate NLTT catalog entry. A flag value of "L" means
that the star has an NLTT proper motion under the nominal lower limit
($0.15\arcsec$ per year) of the LSPM catalog. A flag value of "N" means
that although the object was found to be real, it is not a star but
rather a proper motion nebula. A flag value of "S" means that the
proper motion of the star was remeasured by SUPERBLINK, and found to
be under the nominal lower limit ($0.15\arcsec$ per year) of the LSPM
catalog. A flag value of "T" means that the star is in the TYCHO-2
catalog, and that its TYCHO-2 proper motion is under the nominal lower
limit ($0.15\arcsec$ per year) of the LSPM catalog. The corrected
position and proper motion is given for all stars that could be
recovered.

We note that this list does not contain a single star from the LHS
catalog. Every one of the northern LHS stars has been accounted for
and is now included in the LSPM catalog.

\end{appendix}


\end{document}